\documentclass[12pt]{article}
\newlength{\abstractwidth}
\usepackage{epsf}

\renewcommand{\thefootnote}{\fnsymbol{footnote}}
\renewcommand{\thanks}[1]{\footnote{#1}} 
\newcommand{\starttext}{
\setcounter{footnote}{0}
\renewcommand{\thefootnote}{\arabic{footnote}}}

\newcommand{\be}{\begin{equation}}
\newcommand{\bea}{\begin{eqnarray}}
\newcommand{\eea}{\end{eqnarray}}
\newcommand{\ee}{\end{equation}}

\newcommand{\<}{\langle}
\renewcommand{\>}{\rangle}
\def\ba{\begin{eqnarray}}
\def\ea{\end{eqnarray}}

\def\v{\vskip .1in}

\def\cB{{\cal B}}

\def\cD{{\cal D}}
\def\cE{{\cal E}}
\def\cF{{\cal F}}
\def\cH{{\cal H}}

\def\cM{{\cal M}}
\def\cN{{\cal N}}
\def\cO{{\cal O}}

\def\R{{\cal R}}
\def\cS{{\cal S}}

\def\cV{{\cal V}}

\def\cZ{{\cal Z}}

\def\sdet{{\rm sdet}}

\def\z{{\bf z}}
\def\bZ{{\bf Z}}

\def\chiz{{\chi _{\bar z}{} ^+}}

\def\chix{{\chi _{\bar x}{} ^+}}
\def\chiy{{\chi _{\bar y}{} ^+}}

\def\p{\partial}

\def\ff{ P}
\def\FF{F}

\def\chiz{{\chi (z)}}

\def\chix{{\chi (x)}}
\def\chiy{{\chi (y)}}

\def\half{ {1\over 2}}

\def\no{\nonumber}
\def\sm{\smallskip}

\def\a{\alpha}
\def\b{\beta}

\def\e{\epsilon}

\def\l{\lambda}

\def\o{\omega}
\def\f{\varphi}

\def\t{\theta}
\def\z{\zeta}
\def\ep{\varepsilon}

\def\ep{\varepsilon}

\def\half{ {1\over 2}}

\def\z{{\bf z}}

\def\chiz{{\chi _{\bar z}{} ^+}}

\begin{document} \starttext \baselineskip=15pt
\setcounter{footnote}{0} \newtheorem{theorem}{Theorem}
\newtheorem{lemma}{Lemma} \newtheorem{corollary}{Corollary}
\newtheorem{definition}{Definition}

\rightline{UCLA/07/TEP/25} \rightline{Columbia/Math/07}

\bigskip

\begin{center} {\Large \bf TWO-LOOP SUPERSTRINGS VII}
\\
\bigskip
{\Large\bf Cohomology of Chiral Amplitudes} \footnote{Research supported
in part by National Science Foundation grants PHY-01-40151,
PHY-04-56200,  DMS-02-45371 and DMS-05-14003}

\bigskip

\bigskip

{\large Eric D'Hoker$^*$ and
D.H. Phong$^{\dagger}$} \\

\bigskip

$^*$ {\sl Department of Physics and Astronomy} \\
{\sl University of California, Los Angeles, CA 90095}\\

\v

$^{\dagger}$ {\sl Department of Mathematics} \\
{\sl Columbia University, New York, NY 10027}

\end{center}
\v \v\v\v \begin{abstract}

The relation between superholomorphicity and holomorphicity 
of chiral superstring $N$-point amplitudes for NS bosons on a genus 2
Riemann surface is shown to be encoded in a hybrid cohomology theory,
incorporating elements of both de Rham and Dolbeault cohomologies.
A constructive algorithm is provided which shows that, for arbitrary $N$ 
and for each fixed even spin structure, the hybrid cohomology classes 
of the chiral amplitudes of the $N$-point function on a surface of genus 2 
always admit a holomorphic representative. Three key ingredients in the 
derivation are a classification of all kinematic invariants for the $N$-point
function, a new type of 3-point Green's function, and a recursive construction 
by monodromies of certain sections of vector bundles over the moduli space 
of Riemann surfaces, holomorphic in all but exactly one or two insertion points.

\end{abstract}

\newpage

\baselineskip=15pt
\setcounter{equation}{0}
\setcounter{footnote}{0}

\section{Introduction}
\setcounter{equation}{0}

A basic feature of superstrings in the Ramond-Neveu-Schwarz (RNS)
formulation \cite{RNS1}, is that the space-time theory arises from a
two-dimensional $\cN=1$ supergravity theory on the string
worldsheet \cite{RNS2}. The fields of this supergravity are the worldsheet
metric $g_{mn}$ and gravitino field $\chi_{m\alpha}$. Each
equivalence class under all local symmetries - reparametrizations,
local supersymmetry, Weyl, and super-Weyl invariance in the critical
space-time dimension 10  - defines a supergeometry, and hence a
super-complex structure on the worldsheet. The space of equivalence
classes of supergeometries is supermoduli space, which is itself
endowed with a supercomplex structure  \cite{dp87,superg}. The
Chiral Splitting Theorem \cite{dp89} guarantees that superstring amplitudes
arise from a pairing of superholomorphic (right movers) and
anti-superholomorphic (left movers) chiral amplitudes.

\sm

The notion of superholomorphicity on a worldsheet and
its associated supermoduli space, and the notion of holomorphicity on the
underlying bosonic worldsheet and its associated moduli space, are two
different things, which do not coincide.
Key physical properties of superstring amplitudes, such as the absence of
unphysical singularities in the $S$-matrix, are intimately tied in with the
{\sl holomorphic} structure on an underlying bosonic worldsheet.
As a consequence, the inherent {\sl holomorphic} structure of strings
has to be recovered from the distinct {\sl superholomorphic} structure
defined by the two-dimensional supergeometry. This latter notion
(see \cite{dp87,superg} and references therein) is still relatively unexplored,
and the problem of recovering all the required holomorphic information
from the superholomorphic structure has yet to be fully resolved.

\sm

Integrating out the Grassmann odd supermoduli \cite{supermoduli}, which are
encoded in $\chi_{m\a}$, provides a projection from a supergeometric structure
to a purely bosonic geometry on the worldsheet. If this projection
$(g_{mn}, \chi _{m\a}) \to g_{mn}$ is carried out naively, by simply integrating
over $\chi _{m \a}$ at fixed $g_{mn}$, then the resulting {\sl holomorphic structure}
will fail to be invariant under local supersymmetry, and will thus fail to be
defined intrinsically \cite{dp88,dp89Rome}.
The difficulties resulting from the naive projection
$(g_{mn},\chi_{m\a})\rightarrow g_{mn}$ were explored
from many points of view in early studies of superstring
perturbation theory \cite{vv1, ambiguities}.

\sm

In \cite{dp88, dp89Rome}, it was proposed instead to fix a homology basis on the worldsheet, and to project
the supergeometry $(g_{mn},\chi_{m\alpha})$ onto a bosonic geometry $\hat g_{mn}$,
whose complex structure is defined by the super period matrix $\hat\Omega_{IJ}$.
Since $\hat \Omega _{IJ}$ is invariant  under local supersymmetry, this projection is
well-defined for worldsheets of genus 2 throughout moduli space, and for arbitrary
genus $h\geq 3$, away from a lower-dimensional subvariety. The super period matrix
projection provides a consistent way of identifying the correct holomorphic  structure
defined by a supergeometry. This projection was applied successfully
to the genus 2 superstring measure at fixed even spin structures in \cite{I,II,III,IV,dp02}
producing the measure as a modular form on which the physical properties
of proper factorization and finiteness could be checked explicitly.

\sm

Consistent projection of the measure should be viewed, however, only as a first step.
Superstring amplitudes must similarly be projected onto holomorphic blocks with
respect to $\hat\Omega_{IJ}$, and the major problem is to extract these from the
superholomorphic amplitudes of the two-dimensional supergeometries. In \cite{V,VI},
this was solved for the the $N$-point function for massless NS bosons, for $0\leq N\leq 4$,
by exploiting the crucial simplification provided by the Gliozzi-Scherk-Olive (GSO)
projection \cite{GSO}. The GSO projection allows the extraction process to be carried
out after averaging over spin structures, which resulted in many cancellations for
$0\leq N\leq 4$, and gave proofs of various non-renormalization theorems
at two-loop order \cite{VI} and \cite{D'Hoker:2005ht}, whose validity had been
conjectured earlier on general grounds \cite{old1,Green:1998by}, and 
on the basis of string dualities, such as in
\cite{Green:1997as,Berkovits:1997pj,Pioline:1998mn}.

\sm

The purpose of the present paper is to develop a concrete and systematic
theory of the relation between the superholomorphic structures of
chiral amplitudes defined by supergeometry data $(g_{mn}, \chi _{m\a})$
and the holomorphic structures of the chiral amplitudes defined
by the super period matrix $\hat \Omega_{IJ}$.
Recall from \cite{dp89} that
the chiral $N$-point amplitude $\cF [\delta]$ is a {\sl superholomorphic} form
in each of its vertex operator insertion points ${\bf z}_r=(z_r,\theta_r)$ 
and in supermoduli.
However, as was shown in \cite{V}, it incorporates forms of type
\be
\label{01forms}
(0,1)_r\,\otimes\,(0,1)_s
\ee
in up to two insertion points $z_r$
and $z_s$. Thus the notion of $\partial_{{\bar z}_r}$ is not well-defined on
$\cF[\delta]$, and it does not even make sense to speak of the holomorphicity of
$\cF[\delta]$. Nevertheless, we shall show in the present paper that
$\cF[\delta]$
admits the
following decomposition,
\bea
\cF [\delta ] = \cZ [\delta ] + \cD [\delta ]
\eea
modulo Dirac $\delta$ functions at coincident points
\footnote{By the ``cancelled propagator argument", in presence of the
factor $E(z_i,z_j)^{k_i\cdot k_j}$,
such Dirac $\delta$ functions do not contribute to the physical amplitudes.
Mathematically, this follows from analytic continuation in the
Mandelstam variables $s_{ij}$, e.g. as was carried out explicitly for one-loop
amplitudes in \cite{dp92}.}.
Here, schematically, $\cZ [\delta ]$ is a {\sl holomorphic} $(1,0)$ form in each
vertex insertion point, and $\cD [\delta]$ is {\sl exact in one or two vertex insertion points},
and holomorphic in all other vertex insertion points.
The blocks $\cF[\delta]$, $\cZ [\delta]$, and $\cD[\delta]$ are subject
to certain monodromy properties, already familiar from \cite{V}, and which will be
explained in detail in this paper. The exact piece $\cD [\delta]$ will
be immaterial in any physical superstring amplitude, and may be ignored.
In this sense, the chiral amplitudes $\cF[\delta]$ can be identified with
the holomorphic amplitudes $\cZ[\delta]$, and the holomorphic structure of
superstrings has now been recovered.

\sm

As was already apparent in \cite{VI}, the key to the extraction of the holomorphic
blocks $\cZ [\delta]$ is a new cohomology theory, in which certain cohomology
classes would admit holomorphic representatives. We shall refer to it as the
{\sl hybrid cohomology} of chiral blocks, as the objects of interest are the chiral blocks
of correlation functions in two-dimensional supergravity, and as the
cohomology theory incorporates elements of both de Rham and Dolbeault cohomologies.

\sm

In this paper, we shall show that, for genus $h=2$, the chiral amplitudes $\cF [\delta]$
do admit holomorphic representatives, for arbitrary number $N$ of external
massless NS strings, and for each individual even spin structure $\delta$.
We shall show that, although intermediate calculations
involve complicated combinatorics, the final result is remarkably simple.
In fact, the structure of the various holomorphic blocks that enter may be
schematically understood in terms of just a single fundamental building block.
In the present paper, we develop an algorithm for the calculation of these
blocks, and work out in detail the cases of $\cD [\delta]$, which allows us
to develop a constructive proof that each chiral amplitude does indeed
admit a holomorphic representative with the correct monodromy properties.
In the subsequent  paper, we shall apply the same algorithm
for computing explicitly also the  holomorphic representatives $\cZ [\delta]$.
We turn next to a more precise description.

\subsection{Hybrid cohomology of chiral blocks}

The essentials of the hybrid cohomology theory which we need can be 
defined as follows.

\medskip

Let $\Sigma$ be a Riemann surface and let $\hat\Sigma$ be the quotient
of its universal cover by the commutator subgroup
of the fundamental group $\pi_1(\Sigma)$ of $\Sigma$.
For each fixed integer $N$, let $\hat\Sigma^N$ be the product of $N$
copies of $\hat\Sigma$, and $\Gamma(\hat\Sigma^N,\Lambda^k)$,
be the space of $k$-forms on $\hat\Sigma^k$, which are smooth
on $\hat\Sigma^N$
away from the diagonal $z_r=z_s$.
The de Rham exterior differential $d_r=dz_r\p_{z_r}+d\bar z_r\p_{\bar z_r}$
operates on $\Gamma(\hat\Sigma^N,\Lambda^k)$,
\be
\label{derham}
d_r\ :\ \Gamma(\hat\Sigma^N,\Lambda^k)\ \longrightarrow\
\Gamma(\hat\Sigma^N,\Lambda^{k+1})
\ee
A form $\cF[\delta]$ in this cohomology theory is said to be {\sl closed} if
\be
\label{closed}
\cF[\delta] \, \in \, \bigcap_{r=1}^N {\rm Ker}\,d_r
\ee
A form $\cD[\delta]$ in this cohomology theory is said to be {\sl exact}
if it is a closed form expressible as, modulo Dirac $\delta$ functions
supported only on the diagonal $z_r=z_s$,
\bea
\label{exact}
&&
\cD[\delta]=\sum_{r_1<\cdots<r_\ell}d_{r_1}\cdots 
d_{r_\ell}\,S_{r_1\cdots r_\ell}(z_1,\cdots,z_N),
\hskip 0.6in \ell \geq 1
\nonumber\\
&&
{\rm with} \qquad 
S_{r_1\cdots r_\ell}\,\in\,\bigcap_{r\not\in\{r_1,\cdots,r_\ell\}}{\rm Ker}\,d_r.
\eea
The cohomology class of a closed form $\cF[\delta]$ is its equivalence class modulo
exact forms.

\sm

Given a complex structure on $\Sigma$ and a closed form $\cF[\delta]$,
the main question is then whether the equivalence class
of $\cF[\delta]$, in terms of this cohomology theory, admits a representative
of pure type $(1,0)$ in each insertion point $z_r$. Since a closed
$(1,0)$ form on a Riemann surface is automatically holomorphic,
the fact that $\cF[\delta]$ is a closed
form implies then that this representative must be holomorphic in all insertion points,
away from the diagonal $z_r=z_s$.
In practice, given a closed form $\cF[\delta]$, the existence of a pure $(1,0)$
representative implies that the
leading $(0,1)_{r_1}\otimes \cdots \otimes (0,1)_{r_\ell}$ obstruction
in $\cF[\delta]$ is of the form
$\bar\partial_{r_1}\cdots\bar\partial_{r_\ell}\cS_{r_1\cdots r_\ell}[\delta]$
for some $\cS_{r_1\cdots r_\ell}[\delta]$. Replacing $\cF[\delta]$ by 
$\cF [\delta] -d_{r_1}\cdots d_{r_\ell}\cS_{r_1\cdots r_\ell}[\delta]$,
we can iterate the process and recognize the problem as equivalent
to all the $(0,1)$ obstructions being successively in the range
of a product of $\bar\partial_r$ operators.
In this sense, the problem of finding a holomorphic representative in the
present cohomology theory is a hybrid mixture of de Rham and Dolbeault cohomologies.

\sm

Our main result is that the answer is affirmative for the chiral amplitudes $\cF[\delta]$
of the $N$-point function, when the complex structure of $\Sigma$ is defined by the
super period matrix $\hat\Omega_{IJ}$. In fact, both the exact differential $\cD[\delta]$ and
the resulting holomorphic representative $\cZ[\delta]$ exhibit a rich structure
that can be summarized in the following theorem:

\bigskip
\noindent
{\bf Main Theorem}

\medskip

{\it Let the worldsheet $\Sigma$ be of genus $h=2$, equipped with
a supergeometry $(g_{mn},\chi_{m\a})$. For any fixed even spin structure
$\delta$ on $\Sigma$, and any positive integer $N$,
let ${\cal F}[\delta]$ be the chiral amplitude
of the $N$-point function
for scalar superfields, as given in (\ref{Ys})
in the section below. Let the complex structure on $\Sigma$
be defined by the super period matrix $\hat\Omega_{IJ}$,
as given explicitly in (\ref{Omegahat2}).
Then there exist forms
\be
\cS_r[\delta]\in \bigcap_{t\not=r}{\rm Ker}\,d_t,
\qquad
\cS_{rs}[\delta]\in \bigcap_{t\not=r,s}{\rm Ker}\,d_t,
\ee
smooth away from the diagonal $z_r=z_s$,
such that, modulo Dirac $\delta$ functions supported only on the diagonal,
we have
\be
\cF[\delta]
=
\cZ[\delta]
+
\sum_rd_r\cS_{r}[\delta]
+\sum_{rs} d_rd_s\cS_{rs}[\delta],
\ee
where $\cZ[\delta]$ is a holomorphic
form of pure type $(1,0)$ in each insertion point $z_r$.
More precisely, the forms $\cS_r[\delta]$ and
$\cS_{rs}[\delta]$ are linear combinations of certain basic forms
\be
\Pi^{(n+2)},\qquad \Pi_I^{(n+1)},
\qquad
\Pi_{\pm}^{(m+1|\ell+1)},
\ee
with coefficients given
by certain basic kinematic invariants
\be
K_{[t_1\cdots t_n]}^{\mu\nu},
\qquad
K_\pm^{(m+1|\ell+1)}
\ee
(see sections \S 4 and \S 11 below). In particular, the forms $\cS_r[\delta]$ and
$\cS_{rs}[\delta]$ are actually $(1,0)$ forms, and hence holomorphic,
in all insertion points $z_t$ for $t$ different from $r$ and $r,s$ respectively.}

\subsection{Differential structure of the chiral amplitudes $\cF [\delta]$}

In this subsection, we shall review the key results of \cite{V}.
The $N$-point chiral amplitudes $\cF [\delta]$ are naturally differential forms
on $\Sigma ^N$, and more precisely they are a 1-form in each insertion point $z_i$.
This may be traced back to the fact that the expression for the
un-integrated vertex operator \cite{dpvertex} for a massless NS-NS  string of momentum
$k^\mu$ and polarization  tensor $\e^\mu \bar \e ^\nu$ is given
in terms of the matter scalar superfield
$X^\mu=x^\mu+\theta\psi_+^\mu+\bar\theta\psi_-^\mu$
as follows,
\bea
V(z,\bar z, \theta, \bar \theta, \e, k) = \e^\mu \bar \e ^\nu | dz d\theta |^2 (\sdet E_M {}^A)
\cD _+ X^\mu \cD _- X^\nu e^{i k \cdot X}
\eea
Chiral splitting of this vertex must be carried out with care, since $(\sdet E_M {}^A)$
is non-trivial on a genus 2 worldsheet, and depends on the
underlying supergeometry $(g_{mn},\chi_{m\a})$. The resulting  chiral vertex is the sum of
three contributions,
\bea
\label{vertexdecomp}
\cV^{(0)}(z; \epsilon, k)
& = &
\epsilon^\mu\,dz(\p_zx_+^\mu-ik^\nu\psi_+^\mu\psi_+^\nu)(z)\,
e^{ik \cdot x_+ (z)}
\no \\
\cV^{(1)}(z; \epsilon, k)
&=&
-\half \e^\mu d\bar z\chiz\psi_+^\mu(z) \, e^{ik \cdot x_+ (z)}
\no \\
\cV^{(2)}(z; \epsilon, k)
&=&
- \epsilon^ \mu\hat\mu_{\bar z}{}^zd\bar z(\p_zx_+^\mu
-ik^\nu\psi_+^\mu\psi_+^\nu)(z)\,e^{ik \cdot x_+ (z)}
\eea
Here, $\cV ^{(0)}$ is the familiar vertex operator ($x_+(z)$ and $\psi _+(z)$
denote respectively the effective chiral scalar boson and the chiral fermion
components of the superfield $X^\mu$, as used, for example, in \cite{V}),  
and $\cV^{(1)}$ and $\cV^{(2)}$ are
corrections which depend both on the gravitino slice $\chiz$ and the Beltrami
differential $\hat\mu_{\bar z}{}^z$ for the passage from
period matrix to super period matrix. In all these expressions,
$z$ is a holomorphic coordinate for the complex structure defined by the
super period matrix $\hat \Omega _{IJ}$, 
and not for the complex structure defined by the
original bosonic metric $g_{mn}$ in the supergeometry $(g_{mn},\chi_{m\a})$.

\sm

We stress that $\cV^{(0)}$ is a $(1,0)$ form, but that $\cV^{(1)}$ and $\cV^{(2)}$
are $(0,1)$  forms, and hence the full chiral vertex is a differential 1-form in the
vertex insertion point $z$, containing forms of both types $(1,0)\oplus (0,1)$.
Although $\cV^{(1)}$  and $\cV^{(2)}$ are the source of complications,  their
omission would certainly lead to unacceptable gauge-dependent  results for the
final superstring amplitudes.

\sm

The chiral amplitudes $\cF [\delta]$ are correlators of the full chiral vertex operators,
and thus  receive contributions from 3 different types of differential
forms,\footnote{The correlators on the first line of $\cF [\delta]$ are connected,
as indicated by the subscript $\< \cdots \> _{(c)}$, in the following sense.
To be excluded are all self-contractions of $T_m$, as well as the contributions
in which both $x_+$ and $\psi _+$ are contracted between the two
supercurrents $S_mS_m$. To be included are all the contractions of only
a single field between the two $S_m$-operators, with the remaining
operators contracted elsewhere.}
\bea
\label{Ys}
\cF [\delta ]
& = &
\left \< Q(p_I)\,  \left ( {1 \over 8 \pi ^2}  \int \! \chi S_m ~ \int \! \chi S_m
+ {1 \over 2 \pi }  \int \! \hat \mu T_m \right )
\prod _{t=1}^N \cV_t ^{(0)} \right \> _{(c)}
\no \\
&& +
 \sum _{r=1} ^N \left \< Q(p_I)\, \left ( {1 \over 2 \pi } \int \! \chi S_m ~ \cV ^{(1)} _r
 +  \cV ^{(2)} _r  \right )
~ \prod _{t \not= r}^N  \cV_t ^{(0)}  \right \>
\no \\
&& +
\half \sum _{r \not= s} ^N \left \< Q(p_I) ~
 \cV^{(1)} _r  ~ \cV^{(1)} _s  ~ \prod _{t \not= r,s}^{N}  \cV_t ^{(0)} \right \>
\eea
Here, $S_m$ is the worldsheet supercurrent, and $T_m$ is the stress tensor
for the matter fields $x_+$ and $\psi _+$. The chiral amplitude is evaluated
at fixed even spin structure $\delta$ and fixed internal loop momenta $p_I$,
as guaranteed by the insertion of the operator
\be
Q(p_I) = \exp \left \{ i p_I ^\mu \oint _{B_I} dz \p_z x_+^\mu(z) \right \}.
\ee
All quantities above are expressed with respect to the  super periods
$\hat \Omega _{IJ}$. We recall the relations between the period
matrix $\Omega _{IJ}$, the super period matrix $\hat \Omega _{IJ}$
and the Beltrami differential $\hat \mu _{\bar z} {}^z$ which provides the
complex structure deformation between $\Omega _{IJ}$ and $\hat \Omega _{IJ}$,
\bea
\label{OmegaDef}
\Omega _{IJ} - \hat \Omega _{IJ}
=
i \int \! d^2 z \hat \mu _{\bar z} {} ^z  \omega _I (z) \omega _J (z)
\eea

\sm

The full chiral amplitude  ${\cal B}[\delta]$ for $N$ massless NS bosons,
incorporating the chiral measure, ghost, and superghost
contributions was derived in \cite{V}, and is given by\footnote{Strictly speaking, 
of the integral of $\cF[\delta]$
with respect to all $\theta_r$ variables. To lighten the terminology,
we shall not insist on this distinction when there is no possibility of confusion.}
\bea
\label{BandF}
{\cal B}[\delta] (z; \e ,k ,p_I)
=
d \mu _2 [\delta]   ~ \left \< Q(p_I)  \prod _{t=1}^N \cV _t ^{(0)}  \right \>
+ d \mu _0 [\delta ]   \int _\zeta     \cF [\delta ]
\eea
The integration $\int _\zeta$ is over the odd super-moduli $\zeta ^\alpha, ~ \alpha =1,2 $;
the prefactors  $d \mu _0 [\delta]$ and $d \mu _2 [\delta]$
are components of the chiral measure,  defined and evaluated in \cite{II};
their explicit form will not be needed here.
The chiral blocks $\cF [\delta]$ and ${\cal B}[\delta]$ are 1-forms (including both $(1,0)$
and $(0,1)$ components) in each vertex point $z_r$ with the following monodromy,
\bea
\label{monodromy}
\cF [\delta] (z_r+\delta_{rs} A_K; \e_r,k_r,p_I)
& = &
\cF [\delta] (z_r; \e_r,k_r,p_I)
\no \\
\cF [\delta] (z_r+\delta_{rs}B_K; \e_r, k_r,p_I)
& = &
\cF [\delta] (z_r; \e_r,k_r,p_I- 2 \pi \delta_{IK}k_s)
\eea
Thus they should be viewed as sections of a flat vector bundle
over the moduli space of Riemann surfaces with $N$-punctures.

\medskip

The following results on the differential structure of the  chiral amplitudes $\cF [\delta]$
and  $\cB [\delta]$ were proven in \cite{V},

\medskip

{\bf  (a)  Closedness}: ~
The forms $\cF [\delta]$ and $\cB [\delta]$  are closed in each variable $z_j$;

\medskip \smallskip

{\bf  (b) Slice-change}: ~
Under infinitesimal changes of either the gravitino slice $\chi$ or the
Beltrami differential $\hat\mu$, the forms $\cF [\delta]$ and $\cB [\delta]$ change by
terms which are de~Rham $d$-exact in one variable and de~Rham
$d$-closed in all other variables,
\bea
\cB [\delta] (z; \e ,k ,p_I)
\to
\cB [\delta] (z; \e, k,p_I)
+
\sum_{r=1}^Nd_r{\cal R}_r [\delta] (z; \e , k,p_I)
\eea
Specifically, $\R_r [\delta]$ is a form of weight $(0,0)$ in $z_r$,
and a form of weight $(1,0)\oplus (0,1)$, which is de~Rham closed,
in each $z_s$ for $s\not=r$;
Finally, ${\cal R}_r [\delta] $ has the same monodromy as $\cB [\delta] $.

\sm

Henceforth, we concentrate on the problem of finding a holomorphic representative
within the cohomology classes of $\cB[\delta]$ and $\cF[\delta]$
in the sense of section \S 1.1.
Since the first expression on the right hand side of the equation
(\ref{BandF}) is manifestly a holomorphic $(1,0)$-form in each insertion point
(away from the diagonal), $\cB[\delta]$ and $\cF[\delta]$
have the same non-holomorphic terms, and we can speak interchangeably
of the existence of a holomorphic representative in either class.
In \cite{VI}, this problem was solved for $0\leq N\leq 4$,
and after summation over the even spin structures $\delta$.
Here we shall treat the case of general $N$, and for each fixed
spin structure $\delta$ separately.

\subsection{The main $(0,1)_r\otimes (0,1)_s$ obstruction}

The central result of the present paper is the development of an
explicit algorithm for the construction of the differential blocks $d_r \cS_r [\delta]$
and $d_rd_s\cS_{rs}[\delta]$
for a general chiral $N$-point amplitude $\cF [\delta]$,
so that the difference
\be
\label{difference}
\cF[\delta]-\sum_r d_r\cS_r[\delta]-\sum_{[rs]}d_rd_s\cS_{rs}[\delta]
\ee
is a pure $(1,0)$ form in all insertion points.
A crucial requirement of the algorithm is to make sure
that $\cS_r[\delta]$ and $\cS_{rs}[\delta]$ are closed
forms in any insertion point $z_t$ for $t$ different from $r$ and $r,s$ respectively.
It is as a consequence of this requirement
that the difference (\ref{difference})
is closed and a pure $(1,0)$-form, and thus automatically holomorphic in all insertion
points.

\sm

We establish in this manner the existence of the holomorphic representative 
$\cZ[\delta]$ for the blocks $\cF[\delta]$, and $\cH [\delta]$ for the blocks 
$\cB[\delta]$. These holomorphic representatives have themselves
a very rich structure, the derivation of which
is postponed until a subsequent paper.

\sm

Although the intermediate calculations to isolate the blocks $d_r \cS_r [\delta]$,
$d_r d_s \cS_{rs} [\delta]$, and $\cH [\delta]$ will be quite involved, the final
results are remarkably simple, and may all be related to one fundamental block.

\sm

To see how this comes about, we start from the chiral amplitude $\cF [\delta]$
in (\ref{Ys}). Its first line is a $(1,0)$ form in each insertion point, and these
terms will not contribute to blocks of the type $d_r \cS _r [\delta]$ and
$d_r d_s \cS_{rs} [\delta]$, since they have no $(0,1)$ components. Its second
line is a sum of forms which is of type $(1,0)$ in all but one insertion point, while
the third line is a sum of forms of type $(1,0)$ in all but two insertion points.
The last term is  the top obstruction term, in that it exhibits the highest degree in
$(0,1)$ forms. This term will determine $d_r d_s \cS_{rs} [\delta]$.
It is natural to start by examining this term, and investigate how its
$(0,1)_r \otimes (0,1)_s$ component can be recast in the form of an exact
differential. The correlator in question is
\bea
\left \< Q(p_I) \, \cV_r ^{(1)} \cV _s ^{(1)} \prod _{l \not= r,s} ^N \cV _l ^{(0)} \right \>
\eea
The proportionality of $\cV_r ^{(1)}$ to $\chi (r) \psi _+ (r)$ forces this correlator
to always contain a linear chain of RNS fermion $\psi _+$ contractions.
Therefore, the above correlator will always contain a linear chain of
Szeg\"o kernels $S$, arranged as follows,
\bea
{ 1 \over 4} d\bar r \, \chi (r) S(r,t_1) S(t_1,t_2) \cdots
S(t_{n-1} , t_n) S(t_n, s)  \chi (s) d\bar s
\eea
Here, the points $t_1,t_2, \cdots, t_{n-1}, t_n$ are all distinct from one another
and distinct from  $r$ and $s$, as is guaranteed by the structure of the Wick contractions
for the free field $\psi _+$. The remainder of the correlator which multiplies
each such linear chain is manifestly holomorphic in all vertex insertion points.

\sm

This form in $(0,1)_r \otimes (0,1)_s$ can be recast in terms of a double
exact differential in $r$ and $s$ provided there exists a function
$\Pi ^{(n+2)} (r;t_1, \cdots , t_n;s)$, which is
a $(0,0)$ form in $r$ and $s$, and a $(1,0)$ form in $t_1,\cdots, t_n$, and is
such that
\bea
\p_{\bar r} \p_{\bar s} \Pi ^{(n+2)} (r;t_1, \cdots , t_n;s)
= {1 \over 4} \chi (r) S(r,t_1) S(t_1,t_2) \cdots
S(t_{n-1} , t_n) S(t_n, s)  \chi (s)
\eea
The blocks $\Pi ^{(n+2)} (r;t_1, \cdots,  t_n;s)$ must be holomorphic
in each of the points $t_1, \cdots, t_n$, away from coincident points $t_i= t_j$
for $i \not= j$, and away from the points $r$ and $s$. Finally, the blocks
must have the same mirror symmetry as the correlator, so that we require
\bea
\label{symmPi}
\Pi ^{(n+2)}  (r;t_1, \cdots , t_n;s)
= (-)^n \Pi ^{(n+2)}  (s;t_n, \cdots , t_1;r)
\eea
Although one might initially have hoped that blocks $\Pi ^{(n+2)}$ would exist
without monodromy in $r,s$, it turns out to be impossible to achieve without
introducing new  singularities at extraneous points. Instead, we may require
that its monodromy in $r$ is independent of $r$, and the monodromy in $s$
is independent of $s$. This non-trivial monodromy of $\Pi ^{(n+2)}$,
instead of causing a problem,  will  combine precisely with the
monodromies of the $(0,1)$ form obstructions and help lift those as well.

\sm

The fundamental blocks $\Pi ^{(n+2)} (r;t_1, \cdots,  t_n;s)$
have the connectivity of a {\sl linear chain}, and will be constructed
explicitly in terms of bosonic and fermionic worldsheet Green functions in the
body of the paper.

\subsection{Recursive relations through monodromy}

Under monodromy in their endpoints $r,s$, the fundamental
blocks $\Pi ^{(n+2)} (r;t_1, \cdots,  t_n;s)$ produce new linear chain blocks
$\Pi ^{(n+1)}_I$ and $\Pi ^{(n)} _{IJ}$,  whose roles will be as follows,
\bea
\Pi ^{(n+2)} (r;t_1, \cdots,  t_n;s) &  & {\rm contributes ~ to} ~ \cS _{rs} [\delta]
\no \\
r \to r+ B_I ~ {\rm monodromy} ~   ~ \bigg \downarrow \hskip 0.5in &  &
\no \\
\Pi ^{(n+1)}_I (t_1; \cdots,  t_n;s) \hskip 0.05in & & {\rm contributes ~ to} ~ \cS _s [\delta]
\no \\
s \to s+ B_J ~ {\rm monodromy} ~   ~ \bigg \downarrow \hskip 0.5in & &
\no \\
\Pi ^{(n)}_{IJ} (t_1; \cdots;  t_n) \hskip 0.1in &  & {\rm contributes ~ to} ~ \cH [\delta]
\eea
These linear chain blocks are schematically  depicted in Figure 1 below.

\begin{figure}[tbph]
\begin{center}
\epsfxsize=5.6in
\epsfysize=2.7in
\epsffile{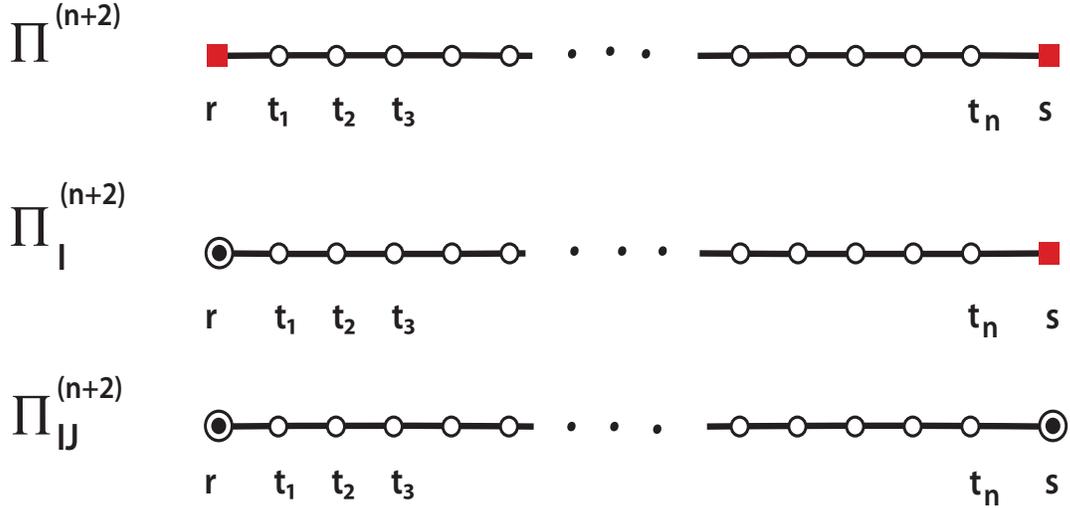}
\label{figure1}
\caption{Connectivity of the linear chain blocks
$\Pi ^{(n+2)}$, $\Pi ^{(n+2)}_I $, $\Pi ^{(n+2)}_{IJ}$;
they are holomorphic at all points, except at the points represented by red squares.}
\end{center}
\end{figure}

\sm

The fundamental blocks $\Pi ^{(n+2)} (r;t_1, \cdots,  t_n;s)$ may also be
linked, by letting one endpoint, say $r$, coincide with the other endpoint $s$,
or with one of the midpoints $t_1,\cdots, t_n$ of the linear chain. This produces
{\sl singly linked chains}, consisting of a single loop with $\ell +1$ points
$r,u_1, \cdots, u_\ell$ connected,  at the point $r$, to a linear chain
with $m+2$ points $s, t_1, \cdots, t_m, r$.

\sm

By themselves, the singly linked blocks obtained this way will not be holomorphic
in the  point $r$, but we shall show that suitable ``counter-terms" may be added to restore
holomorphicity at $r$, while maintaining holomorphicity in all $u$- and $t$-variables.
It will be natural to ``symmetrize" the resulting blocks, so that
\bea
\label{linkblocks1}
\Pi _\pm ^{(m+1|\ell+1)}  (s;t_1,\cdots, t_m,r,u_1,\cdots, u_\ell)
=
\pm  (-)^\ell \Pi _\pm  ^{(m+1|\ell+1)}  (s;t_1,\cdots, t_m,r,u_\ell,\cdots, u_1) \quad
\eea
From the singly linked blocks $\Pi ^{(m+1|\ell+1)} _\pm$, we obtain new
holomorphic blocks by taking the monodromy in $s$. This process may be
schematically represented as follows,
\bea
\Pi ^{(m+\ell +3)} (s;t_1, \cdots,  t_m,r, u_1, \cdots, u_\ell; v) & & {\rm linear ~ chain}
\no \\
{\rm linking} ~ v ~ {\rm to} ~ r ~ \bigg \downarrow \hskip 1in &  &
\no \\
\Pi _\pm ^{(m+1 | \ell +1)} (s;t_1, \cdots,  t_m,r, u_1, \cdots, u_\ell) &&
{\rm contributes ~ to} ~ \cS _s [\delta]
\no \\
s \to s+ B_I ~ {\rm monodromy} ~   ~ \bigg \downarrow \hskip 1in & &
\no \\
\Pi _{ \pm I} ^{(m | \ell +1)} (t_1; \cdots,  t_m,r, u_1, \cdots, u_\ell) &&
{\rm contributes ~ to} ~ \cH [\delta]
\eea
The connectivity of the resulting singly linked chains is depicted in Figure 2 below.

\begin{figure}[tph]
\begin{center}
\epsfxsize=5in
\epsfysize=3.2in
\epsffile{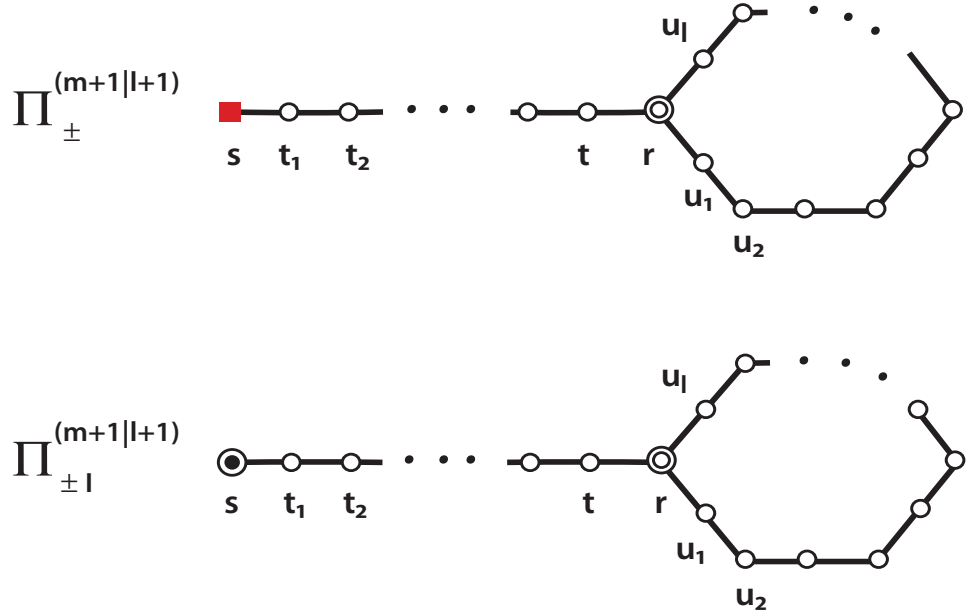}
\label{figure2}
\caption{Connectivity of the singly linked chain blocks
$\Pi _\pm ^{(m+1|\ell+1)}$, $\Pi _{\pm I} ^{(m+1|\ell+1)} $;
they are holomorphic at all points, except at the points represented by red squares.}
\end{center}
\end{figure}

Finally, the fundamental blocks $\Pi ^{(n+2)} (r;t_1, \cdots,  t_n;s)$ may also
be linked by letting both end points coincide with distinct midpoints. These
blocks only contribute to $\cH [\delta]$ and will be calculated and studied
in the subsequent paper.

\subsection{Kinematic invariants}

A fundamental component of the algorithm for finding $\cS_r[\delta]$ and
$\cS_{rs}[\delta]$ is the identification of all kinematic invariants
arising in the $N$-point function. These are discussed in detail in
section \S 4, but we would like to single out some particularly
important features in this Introduction.

\sm

The key quantity is the following odd Grassmann valued differential 
form of even degree, which is vector valued in Minkowski space-time, 
\be
K_r^\mu \equiv\ep_r^\mu\wedge dz_r+
ik_r^\mu \theta_r
=\epsilon^\mu d\theta_r\wedge dz_r+ik_r^\mu \theta_r
\ee
Here $\e^\mu _r$ is the physical polarization vector associated
with external state $r$, and $k_r ^\mu$ is the momentum of that state.
In the body of the paper, we shall often use the composite 
$\ep ^\mu _r = \e^\mu _r d \theta _r$. As before, the coordinates 
of the vertex insertion point $r$ are given by $(z_r, \theta_r)$ and its
complex conjugate.

\sm

The bilinears in $K_r$ give rise immediately to the gauge-invariant field strengths,
\be
K_r^\mu K_r^\nu=i(\ep_r^\mu k_r^\nu-\ep_r^\nu k_r^\mu)dz_r \theta_r
=i(\epsilon_r^\mu k_r^\nu-\epsilon_r^\nu k_r^\mu)d\theta_r\wedge dz_r \theta_r.
\ee
Then all kinematic invariants are given either by {\sl linear chains}
in $K_r^\mu$ of the form
\be
K_{[t_1t_2\cdots t_{n-1}t_n]}^{\mu\nu}
=
K_{t_1}^\mu \bigg(\prod_{k=1}^{n-1}K_{t_k}^{\rho_k}K_{t_{k+1}}^{\rho_k}\bigg)K_{t_n}^\nu
\ee
(contracted with either $p_I^\mu$, $ik_r^\mu$, or $\ep_r^\mu\theta_r$),
or {\sl singly linked chains}, of which there are two types
$K_{\pm}^{(m+1|\ell+1)}$, symmetric and anti-symmetric,
\bea
K_+^{(m+1|\ell+1)}
&=&\half \ep_s^\mu\theta_s\, \bigg(K_{[t_1\cdots t_mru_1\cdots u_\ell]}^{\mu\rho}
+
(-)^\ell K_{[t_1\cdots t_mru_\ell\cdots u_1]}^{\mu\rho}\bigg)ik_r^\rho
\nonumber\\
K_- ^{(m+1|\ell+1)}
&=&
\ep_s^\mu\theta_s
K_{[t_1\cdots t_m]}^{\mu\rho}
ik_r^\rho K_{[ru_1\cdots u_\ell]}^{\sigma\sigma}
\eea
Thus, the full chiral amplitude becomes a sum of such kinematic
factors, with coefficients which are the holomorphic or differential 
blocks $\Pi$, $\Pi_I$ or $\Pi _{\pm}$. 

\sm

This decomposition into sums of kinematic factors times basic 
Lorentz scalar amplitudes seems  reminiscent, to some degree, of 
the color and helicity decomposition of scattering amplitudes 
in massless Yang-Mills theory (see e.g. \cite{Parke:1986gb},
as well as the developments  in \cite{Bern:2002wt,Cachazo:2004kj}, 
and \cite{Bern:2007dw} where the subject is reviewed.
While the Yang-Mills  amplitudes, stripped of their color and helicity 
dependence, are found to obey certain analyticity and recursion
relations in momentum space, our two-loop  chiral amplitudes,
stripped of their polarization tensor kinematic factors, are found to 
obey certain holomorphicity and recursion relations on the string worldsheet. 
It is tempting to imagine that this similarity has a real, and perhaps
practical, significance.

\newpage

\subsection{Organization of the paper}

The remainder of the paper is organized as follows.

\medskip

Section \S 2 is devoted to a brief summary of two-dimensional supergeometry,
the chiral splitting theorem, and the function theory which we
need, including super Abelian differentials, the super period matrix, and
the super prime form.

\medskip

In section \S 3, we discuss simple examples of how the cohomology class
of a superholomorphic object can contain a holomorphic representative.
The most basic case is that of a superholomorphic Abelian differential,
which already played a major role in the derivation in \cite{I,II,III,IV}
of the superstring measure. More complicated examples are the super prime
form and its derivatives.

\medskip

In section \S 4, all the kinematic invariants are derived.

\medskip

In section \S 5, we provide a first splitting of the chiral amplitudes
$\cF[\delta]$ into differentials and remaining terms. The underlying
combinatorics are explained in some detail. At this time, the differentials
are not holomorphic in the remaining insertion variables, and are not yet
the terms $d_r\cS_r[\delta]$ and $d_rd_s\cS_{rs}[\delta]$ that we seek.
In order to eliminate the $(0,1)$ terms, we shall need to solve $\bar\partial$
equations, with right hand side having non-trivial monodromies.

\medskip

In section \S 6, we discuss such equations systematically. It turns out
that all the equations we encounter can be solved using a unique function
$Q_0(s;x,y)$, which can be viewed as a kind of Green's function in 3 variables
$s$, $x$, and $y$. In preparation for the construction of the differentials
$\cS_r[\delta]$ and $\cS_{rs}[\delta]$, we construct all the differential blocks
that we need: first, the linear chain blocks in section \S 7,
and then the singly linked chain blocks in section \S 8. As explained previously,
the key starting block is $\Pi$, from which all other blocks descend by
recursive monodromy relations. 

\medskip

In sections \S 9 and \S 10, the differentials obtained in the preliminary
decomposition of section \S 5 can then be re-assembled, after solving suitable
$\bar\partial$ equations, into terms of the form $d_r\cS_r[\delta]$ and
$d_rd_s\cS_{rs}[\delta]$, this time with forms which are $(1,0)$ and holomorphic
in any insertion point $z_t$ with $t$ different from $r$ and $r,s$ respectively.
Thus these are the hybrid exact differentials that we seek.

\medskip

In section \S 11, we present a summary of the resulting formulas for 
$\cS_r[\delta]$ and $\cS_{rs}[\delta]$.

\medskip

Finally, we should note that there is a very extensive literature on superstring
perturbation theory. Some representative papers are the early works in 
\cite{old0, bis2, bis1}, with the case of genus 2 considered in e.g. \cite{bis3}, 
and more recently in \cite{iz,zwzI,zwzII,zwzIII,Cacc}. The pure spinor approach
to multiloop superstring amplitudes is developed in \cite{berkovits}. Recent proposals
for  the superstring measure and scattering amplitudes 
beyond two-loop order may be found in \cite{Russo:1997fi}, 
\cite{D'Hoker:2004xh}, and \cite{Matone}.
We refer to these papers, as well as to the earlier papers \cite{II,V,VI}
of this series, for a fuller list of references on the subject.

\newpage

\section{Chiral $N$-point Amplitudes for Massless NS Strings}
\setcounter{equation}{0}

In this section, we present the basic facts about two-dimensional supergeometries
and the correlators of chiral scalar superfields. The corresponding results were
originally derived in \cite{dp89}. Our goal here is to provide a succinct summary,
in a form convenient for the extensive study of the chiral $N$-point amplitudes that
will follow in the sequel of this paper. The main point to be brought forward in this
section is that a supergeometry defines a notion of superholomorphicity, and that
the $N$-point chiral amplitudes for massless NS strings are superholomorphic
and expressible in terms of superholomorphic quantities which are analogous
to standard objects from the theory of Riemann surfaces such as
period matrices, Abelian differentials, and prime forms.

\subsection{Function theory on a Riemann surface}

The starting point is an orientable compact Riemann surface $\Sigma$ of genus 2
and with fixed even spin structure $\delta$. Let $(A_I,B_I)$, $I=1,2$, be a fixed
choice of canonical homology basis for $\Sigma$, with intersection numbers
$\#(A_I\cap A_J)=\#(B_I\cap B_J)=0$, $\#(A_I\cap B_J)=\delta_{IJ}$. The
worldsheet metric $g_{mn}$, defines a complex structure on $\Sigma$. The key
ingredients of the associated complex function theory are the Abelian differentials
$\o_I(z) $, the period matrix $\Omega_{IJ}$, the Szeg\"o kernel $S_\delta(z,w) $,
the prime form $E(z,w)$, and the Green's function $G(z,w)$.

\sm

The Abelian differentials $\o_I$, $I=1,2$, are the basis of holomorphic
$(1,0)$-forms with respect to the metric $g_{mn}$ dual to the cycles $A_I$, and
the period matrix $\Omega_{IJ}$ is the matrix of their periods around the
$B_J$ cycles,
\bea
\label{Abelian1}
\oint_{A_J}\o_I=\delta_{IJ},
\hskip 1in
\oint_{B_J}\o_I=\Omega_{IJ}
\eea
The three periods $\Omega _{IJ}= \Omega _{JI}$ are local coordinates on the
genus two moduli space $\cM_2$ of compact Riemann surfaces.

\sm

The Szeg\"o kernel $S_\delta(z,w)$ is the Green's function for the $\bar \partial$
operator on spinors of  weight $(1/2,0)$, and spin structure $\delta$, and obeys,
\bea
\p_{\bar z} S_\delta (z,w) = 2 \pi \delta (z,w)
\eea
Throughout this paper, the even spin structure is fixed to be $\delta$, and we
shall drop the subscript $\delta$ from the Szeg\"o kernel, and simply denote
it by $S(z,w) \equiv S_\delta (z,w)$.

\sm

The prime
form $E(z,w)$ is a holomorphic function of weight $(-1/2,0)$ in each variable $z$
and $w$ on the universal cover of $\Sigma$, which behaves as $E(z,w)\sim z-w$
for $z$ close to $w$, and has the following monodromy,
\bea
E(z+A_K,w)&=&E(z,w)\nonumber\\
E(z+B_K,w)&=&E(z,w)\,{\rm exp}\, \left (-i \pi -\pi i\Omega_{KK}- 2\pi i\int^z_w \o_K \right )
\eea
The Green function $G(z,w)$ is a  meromorphic $(1,0)$ form in $z$,
which may be defined by
\bea
G(z,w) = \p_z \ln \left ( {E(z,w) \over E(z,w_0)} \right )
\eea
It has simples poles at $w$ and $w_0$ with residues $\pm 1$, so that
\bea
\p _{\bar z} G(z,w) & = & +2 \pi \delta (z,w) - 2 \pi \delta (z,w_0)
\no \\
\p _{\bar w} G(z,w) & = & -2 \pi \delta (z,w)
\eea
The point $w_0$ will be considered  fixed throughout,
and its dependence will not be exhibited. $G(z,w)$ is single-valued in $z$; it
coincides with the third Abelian differential with vanishing $A$-periods in $z$;
but has non-trivial monodromy in $w$, given by
\bea
G(z,w+A_K) & = & G(z,w)
\no \\
G(z,w+B_K) & = & G(z,w) + 2 \pi i \o_K(z)
\eea
Note for later use that $\p_w G(z,w) = \p_z G(w,z) = \p_z \p_w \ln E(z,w)$ is
symmetric in $z$ and $w$.

\subsection{Two-dimensional supergeometry}

The worldsheet $\Sigma$ of a supergeometry is equipped with a worldsheet metric
and gravitino field $(g_{mn},\chi_m{}^\a)$, which in the superfield formalism
may be parametrized by a superframe $E^A= dz^M E_M{}^A$ and a superconnection
$\Omega = dz^M \Omega _M$. Here, $z^M=(z,\bar z,\theta,\bar\t)$ denote local
coordinates for the super Riemann surface associated with $\Sigma$, and
$A=(a,\alpha)$ labels the corresponding frame index. The torsion constraints on $E^A$,
and $\Omega$
allow one to relate the superfields to the component fields by\footnote{Throughout,
auxiliary fields will play no role and will not be exhibited. Their explicit dependence
may be found in \cite{superg, dp88}.}
\bea
E_m{}^a=e_m{}^a+\theta\gamma^a\chi_m, \hskip 1in
g_{mn}=e_m{}^ae_n{}^b\delta_{ab}.
\eea
Supergeometry in the superfield formalism is invariant under super diffeomorphisms,
and super Weyl transformations. Upon decomposition into components, these symmetries reduce to the
customary diffeomorphism and Weyl invariance, as well as local supersymmetries.
In local complex coordinates $z, \bar z$ on $\Sigma$, the bosonic frame indices $a$
and Einstein indices $m$ may be identified, and we use the frame labels
$A= (z, \bar z, +,-)$.
Infinitesimal diffeomorphisms $\delta _v$ and supersymmetries $\delta _\xi$ are
generated by vector field $v^z$ of weight $(-1,0)$ and the field $\xi ^+$ of weight
$(-1/2,0)$ respectively and  their action is given by
\bea
e_{\bar z} {}^m \delta_v e_m {}^z = \p_{\bar z} v^z
& \hskip 1in &
e_{\bar z} {}^m \delta_\xi \chi _m {}^+ = - 2 \p_{\bar z} \xi ^+
\no \\
e_{\bar z} {}^m \delta_v \chi _m {}^+ = 0 \hskip 0.2in
& \hskip 1in &
e_{\bar z} {}^m \delta_\xi  e_m {}^z = \xi ^+ \chi _{\bar z} {}^+
\eea
Super Weyl invariance of the critical superstring allows us to restrict
to vanishing even Weyl changes $e_z {}^m \delta e_m {}^z=0$,
and vanishing odd Weyl changes $\chi _z ^+=\chi _{\bar z} ^-=0$.
The action of local supersymmetry on $\chiz$ leaves two odd supermoduli,
which we denote by $\zeta ^1, \zeta ^2$, and which may be used to
parametrize the worldsheet gravitino field as follows,
\bea
\chiz = \chi (z) = \zeta ^1 \chi _1(z,\bar z) + \zeta ^2 \chi _2(z, \bar z)
\eea
Henceforth, we shall use the notation $\chi (z)$ to denote this two-dimensional
parametrization. To simplify notation, we indicate the $z$ and $\bar z$ dependence
of $\chi$ only by $z$, but it must be kept in mind that the slice functions
$\chi_{1,2}(z,\bar z)$, and thus also $\chi (z)$ are generally allowed to be
arbitrary smooth functions of $z$ and $\bar z$.

\sm

The super covariant derivative $\cD_-^{(n)}$ on a superfield $V(z,\bar z, \theta,\bar\theta)
=V_0+\t V_++\bar\t V_-$ of weight $(n,0)$ is given by
\bea
\cD_-^{(n)}V=
V_- + \bar \t \left ( \p_{\bar z}V_0+{1\over 2}\chi  V_+ \right )
-\t \bar \t \left ( \p_{\bar z}V_+ + {1\over 2}\chi  \p_z V_0
+n\p_z\chi  V_0-{1\over 4}\chi  \overline{\chi } V_- \right )
\eea
where $V_0, V_+$ and $\chi$ are all functions of $z,\bar z$.
A superfield $V$ of weight $(n,0)$ is said to be superholomorphic if
\bea
\label{superholomorphicity1}
\cD_-^{(n)}V=0.
\eea
We consider only superfields with a definite $\bZ_2$-grading, which is
defined to be the grading of the lowest component $V_0$.
The components of a super holomorphic form $V$ are then given by $V_-=0$, and
$V_0, V_+$ obeying the equations,
\bea
\label{superholomorphicity2}
\p_{\bar z}V_0+{1\over 2}\chi (z) V_+=0
\hskip 1in
\p_{\bar z}V_+ + {1\over 2}\chi (z) \p_z V_0+n\p_z\chi (z) V_0=0
\eea
Solutions to these equations may be viewed as $\chi$-deformations of
holomorphic $(n,0)$-forms $V_0$ which lead to even-grading (or simply ``even")
superholomorphic $V$, or $\chi$-deformations of holomorphic $(n+1/2,0)$-forms
$V_+$ which lead to odd-grading (or simply ``odd") superholomorphic $V$.

\subsection{Super-Abelian differentials and super prime forms}

The notions of holomorphic Abelian differentials generalize
to supergeometry. Super-holomorphic Abelian differentials $\hat\o=\hat\o_0+\t\hat\o_+$
are forms of weight $(1/2,0)$ satisfying
\be
\cD _- ^{(1/2)} \hat \o=0.
\ee
On a genus 2
super Riemann surface with even spin structure $\delta$,
there exist no even superholomorphic Abelian differentials, since there
exist no holomorphic $(1/2,0)$ forms. The space of odd superholomorphic Abelian
differentials is 2-dimensional, and admits a basis
$\hat\o_I({\bf z})=\hat\o_{I0}(z)+\t\hat\o_{I+}(z)$, given by
\bea
\label{omegahat}
\hat\o_{I+}(z) &=&
\o_I(z) - {1\over 16\pi^2} \int _x \int _y
\p_z \p_x \ln E(z,x)\,\chi (x) S (x,y)\chi (y) \o_I (y)
\no \\
\hat\o_{I0}(z)
&=&
-{1\over 4\pi}\int _x \,S (z,x)\,\chi (x) \o_I(x)
\eea
Super holomorphic $(1/2,0)$ forms $\hat\o=\hat\o_0+\t\hat\o_+$
support the notion of a line integral, which may be defined by
\bea
\label{lineintegral}
\int_{(w,\t_w)}^{(z,\t_z)} \hat \o
=
\int_w^z\, \left ( dz\,\hat\o_+ - {1\over 2} d\bar z\,\chi (z) \hat\o_0 \right )
+ \theta_z \hat\o_0(z)-\theta_w\hat\o_0(w).
\eea
The basis $\hat\o_I$ is dual to the basis $A_I$ of homology cycles in the sense of
line integrals (\ref{lineintegral}), and we define its $B_J$ periods to be the super
period matrix $\hat \Omega_{IJ}$,
\bea
\label{Omegahat1}
\oint_{A_J} \hat\o_I = \delta_{IJ},
\hskip 1in
\oint_{B_J} \hat\o_I = \hat \Omega_{IJ}
\eea
By construction, both $\Omega_{IJ}$ and $\hat\Omega_{IJ}$ are invariant
under diffeomorphisms, but while $\hat\Omega_{IJ}$ is also invariant under local
supersymmetry transformations, $\Omega_{IJ}$ is not. Thus $\hat\Omega_{IJ}$
can be viewed as the supersymmetric completion of $\Omega_{IJ}$.
From (\ref{omegahat}) for $\hat\o_I$, it follows that
\bea
\label{Omegahat2}
\hat\Omega_{IJ} = \Omega_{IJ}
- {i\over 8\pi} \int _x \int _y  \o_I(x)\chi(x) S (x,y)\chi(y) \o_I (y)
\eea
Finally, we can also construct a super prime form $\cE_\delta({\bf z},{\bf w})$ by
\bea
\label{superprimeform1}
\ln \cE_\delta({\bf z},{\bf w})
=
- F_0(z,w)   - \theta_z  \FF (z,w)-\theta_w  \FF (w,z)
-\theta_z\theta_w F_1 (z,w)
\eea
with the functions $F_0, F_+$, and $F_1$ defined  by
\footnote{Throughout this paper, conditionally convergent integrals
of the form $\int_{\bf C}{f(z)\over (z-w)^2}d^2z$ are regulated as
$\p_w\int_{\bf C}{f(z)\over z-w}d^2z$, for $f$ smooth function with compact
support.}
\bea
\label{superprimeform2}
F_0(z,w)  & = &
- \ln E(z,w) + {1 \over 16 \pi ^2} \int _x \int _y G(x,z) \chi (x) S (x,y) \chi (y) G(y,w)
\\
F_+ (z,w) & = &
{1 \over 4 \pi} \int _x  S (z,x) \chi (x) G(x,w)
\no \\
F_1 (z,w) & = &
S (z,w) - {i\over 16\pi^2} \int _x \int _y S(z,x)
\chi (x) \p_x \p_y \ln E(x,y) \chi(y) S(y,w)
\no \eea
The super prime form $\cE_\delta({\bf z},{\bf w})$ has properties similar to
those of the customary prime form $E(z,w)$: it is superholomorphic in 
${\bf z}$ and ${\bf w}$, and $\cE_\delta({\bf z},{\bf w})=
\cD_+^{\bf z}\cE_\delta({\bf z},{\bf w})=0$ if and only if 
the superpoints coincide ${\bf z}={\bf w}$ \cite{dp89}.

\subsection{The chiral splitting theorem}

We consider a theory of  scalar multiplets $(x^\mu,\psi_+^\mu,\psi_-^\mu)$,
$0\leq \mu\leq 9$, on the surface $\Sigma$ with a supergeometry
$(e_m{}^a, \chi_m{}^\a)$. In the superfield formalism,
the multiplet $(x^\mu,\psi_+^\mu,\psi_-^\mu)$ corresponds to a scalar superfield
$X^\mu(z,\theta,\bar\theta)=x^\mu(z)+\theta\psi_+^\mu+\bar\theta\psi_-^\mu$
whose action is
\bea
I_m(E_M{}^A,X^\mu)
=
{1\over 4\pi}
\int d^{2|2}{\bf z}\, ({\rm sdet}\,E_M{}^A) \, {\cal D}_+X^\mu{\cal D}_-X^\mu
\eea
where $d^{2|2}{\bf z}=d^2z d\theta d\bar\theta$, and ${\cal D}_\pm = \cD _\pm ^{(0)}$
are the covariant derivatives with respect to the supergeometry $(g_{mn},\chi_m{}^\a)$.
The generating vertex for massless NS strings is given by \cite{dpvertex}
\bea
\label{vertex1}
V({\bf z},\bar{\bf z};\e^\mu_o,\bar\e^\mu_o;k^\mu)
=
{\rm exp} \Big ( ik^\mu X^\mu({\bf z},\bar{\bf z})
+
\e^\mu_o {\cal D}_+X^\mu+\bar\e^\mu _o {\cal D}_-X^\mu \Big )
\eea
where $k^2=k\cdot\e_o =k\cdot\bar\e_o =0$. The coefficients $\e_o$ and
$\bar \e_o$ must be Grassmann odd (as will be indicated by the subscript $o$)
in order to obtain a superfield
vertex $V$ that has definite (even) grading. In the next subsection, we shall
explain more in detail how $\e_o$ and $\bar \e _o$ are determined.
In particular, the physical massless NS-NS string vertex operators will
correspond to retaining precisely the term linear in $\e_o$ and linear
in $\bar \e_o$ in the expansion of $V$ and discarding all other terms.

\sm

The matter part of the string amplitudes for the scattering of $N$ massless
NS superstrings is derived from the correlator  of $N$ vertex operators $V$
for various polarizations $\e_{or}$, momenta $k_r$ and vertex insertion points $\z_r$,
with $r=1,\cdots, N$. Up to a factor of the vacuum amplitude (which is
independent of $\e_{or},k_r,\z_r$), this correlator is given by the functional
integral over the superfield $X$,
\bea
\left \<\prod_{r=1}^NV({\bf z}_r,\bar{\bf z}_r;\e_{or}^\mu,\bar\e_{or}^\mu;k_r^\mu) \right \>
\sim
\int DX^\mu e^{-I_m(E_M{}^A,X^\mu)}
\prod_{r=1}^NV({\bf z}_r,\bar{\bf z}_r;\e_{or}^\mu,\bar\e_{or}^\mu;k_r^\mu)
\eea
The {\sl Chiral Splitting Theorem} of \cite{dp89} states that this correlator
is chirally split at fixed internal loop momenta $p_I$, $I=1,2$. More
precisely, it was shown in \cite{dp89} that we have
\bea
\label{chiralsplitting}
\left \<\prod_{r=1}^NV({\bf z}_r,\bar{\bf z}_r;\e_{or}^\mu,\bar\e_{or}^\mu;k_r^\mu) \right \>
=
\int dp_I^\mu\ \bigg |{\cal F}[\delta](\z_r;\e_{or}^\mu ;k_r^\mu;p_I^\mu) \bigg |^2
\eea
where $p_I^\mu$ are parameters which can be interpreted as
internal momenta circulating in the loop $B_I$. (Such parameters had
been introduced in \cite{vv2} for the bosonic string.) The chiral blocks
${\cal F}[\delta](\z _r ;\e_{or}^\mu ;k_r^\mu;p_I^\mu)$ are given explicitly
by\footnote{Throughout, a factor of $2 \pi$ in the internal loop momenta will be
scaled out as compared to the conventions of \cite{dp89}. Also,
the dependence on super moduli of the chiral blocks
$\cF [\delta]$ will not be exhibited.}
\bea
\label{FE}
\cF [\delta]
& = &
\exp \biggl \{
{i \over 4 \pi } p_I ^\mu \hat \Omega _{IJ} p_J ^\mu
+  p_I ^\mu \sum _{r=1} ^N \left ( \e ^\mu _{or} \hat \omega _I (\z _r)
+ ik ^\mu _r \int ^{\z _r} _ {\z _0} \hat \omega _I  \right ) \biggr \} \times
\no \\
&&
\exp \half \sum _{[r,s]} ^N \biggl [
k_r ^\mu k_s ^\mu \ln \cE _\delta (\z _r, \z _s)
+ \epsilon _{or} ^\mu \epsilon _{os} ^\mu \p_+ ^r \p_+ ^s \ln \cE _\delta (\z _r, \z _s)
\no \\ && \hskip 0.7in
- i k _r ^\mu  \epsilon _{os} ^\mu \p_+ ^s \ln \cE _\delta (\z _r, \z _s)
- i k _s ^\mu  \epsilon _{or} ^\mu \p_+ ^r \ln \cE _\delta (\z _s, \z _r)
\biggr ]
\eea
where $\p_+=\p_\theta+\theta\p_z$, and the super period matrix $\hat \Omega$,
the super Abelian differentials $\hat \o_I$, and the super prime form $\cE_\delta$
have been defined and calculated earlier in this section.\footnote{We shall use the
summation subscript $[r,s]$ for a summation
over both $r$ and $s$ with $r\not= s$, while the subscript $r\not= s$
will be reserved to denote a summation over $r$ only for fixed $s$.}

\sm

The monodromy properties of $\cF [\delta]$ are as follows,
\bea
\cF [\delta ] (\z_r + \delta _{rs} A_K, \e_{or}^\mu ;k_r^\mu;p_I^\mu)
& = & \cF [\delta ] (\z_r , \e_{or}^\mu ;k_r^\mu;p_I^\mu)
\no \\
\cF [\delta ] (\z_r + \delta _{rs} B_K, \e_{or}^\mu ;k_r^\mu;p_I^\mu)
& = & \cF [\delta ] (\z_r , \e_{or}^\mu ;k_r^\mu;p_I^\mu - 2 \pi \delta _{IK} k_s)
\eea

\subsection{The use of Grassmann odd polarization vectors}

The Grassmann odd nature of the coefficients $\e_o$ and $\bar \e_o$
was used originally in \cite{dp89} only as a formal device to recover the
physical vertex operators for massless NS states via a generating
function. Actually, it is possible to trace the odd Grassmann nature
of $\e_o, \bar \e_o$ back to the structure of the integration measure
that occurs in the physical vertex operator, and then use this
structure beyond the formal level to our advantage.

\sm

The key observation is that the unintegrated vertex operator in the
superfield formalism should naturally be viewed as a differential form
on the super Riemann surface $\Sigma$, generalizing the differential
form structure of the component formulation of the vertex operators
of \cite{V} to superspace. The unintegrated physical vertex
operator  for massless NS strings is given by
\bea
\label{uninvertex}
d^{2|2}{\bf z}\, ({\rm sdet}\,E_M{}^A)
\, \e^\mu \, \bar \e ^\nu \, \cD_+ X^\mu \, \cD_- X^\nu \exp ( i k \cdot X )
\eea
Here, $k$ is again the momentum of the string, but $\e$ and $\bar \e$
are now Grassmann even, or simply, complex numbers-valued polarization
vectors, satisfying $k^2=k\cdot\e =k\cdot\bar\e =0$. Additional insight is
gained by using an observation made in \cite{V}, namely that the super
volume form admits a natural chiral splitting,
\be
\label{chiralvolume1}
d^{2|2}{\bf z}\, ({\rm sdet}\,E_M{}^A) =
e^z\wedge  e^{\bar z} \wedge d\theta \wedge d\bar \theta
\ee
with the chiral frame $e^z$ given explicitly in Wess-Zumino gauge,
and local complex coordinates $z$ and $\bar z$, by
\be
\label{chiralzweibein1}
e^z=dz-{1\over 2}\theta \, \chi (z) \, d\bar z,
\qquad
e^{\bar z}=d\bar z-{1\over 2}\bar \theta \, \bar \chi (z) \, dz
\ee
Therefore, it is clearly natural to rearrange  the unintegrated vertex
operator of (\ref{uninvertex}) by grouping together the chiral
polarization vector $\e$ and the chiral differentials $d \theta$ and $e^z$
(and similarly for $\bar \e, d \bar \theta$ and $e ^{\bar z}$),
\bea
\label{uninvertex1}
\Big ( \e^\mu d \theta \wedge e^z \cD_+ X^\mu \Big ) \wedge
\Big ( \bar \e^\mu d \bar \theta \wedge e^{\bar z} \cD_- X^\mu \Big )
 \exp ( i k \cdot X )
\eea
The  generating function from which these vertex operators may
be derived is given by
\bea
\label{vertex2}
{\rm exp} \Big ( ik^\mu X^\mu({\bf z},\bar{\bf z})
+ \e^\mu d \theta \wedge e^z {\cal D}_+X^\mu
+\bar \e^\mu d \bar \theta \wedge e^{\bar z}  {\cal D}_-X^\mu \Big )
\eea
which coincides with (\ref{vertex1}) provided we make the identifications,
\bea
\e _o ^\mu & = & \e^\mu d \theta \wedge e^z
= \e^\mu d \theta \wedge \left  ( dz-{1\over 2}\theta \, \chi (z) \, d\bar z \right )
\no \\
\bar \e _o ^\mu & = & \bar \e^\mu d \bar \theta \wedge e^{\bar z}
= \bar \e^\mu d \bar \theta \wedge
\left ( d\bar z-{1\over 2}\bar \theta \, \bar \chi (z) \, dz \right )
\eea
Given that the polarization vectors $\e^\mu , \bar \e^\mu$ are complex numbers,
we see that $\e_o ^\mu$ and $\bar \e_o^\mu$ are Grassmann odd,
in view of the fact that $\theta, \chi, \bar \theta, \bar \chi$, and $d \theta, d \bar \theta$
are all Grassmann odd. Moreover, $\e_o ^\mu$ and $\bar \e_o^\mu$ are
differential forms of even degree, since $d\theta \wedge dz$,
$d\theta \wedge d\bar z$, $d\bar \theta \wedge dz$, and $d\bar \theta \wedge d\bar z$
are of even degree. The above set-up will be very useful in the sequel.
In particular, $\cF[\delta]$
should be viewed as a differential form in each of the vertex insertion points $\z_r$.

\newpage

\section{Holomorphicity and Super-Holomorphicity}
\setcounter{equation}{0}

The key to the approach to superstring perturbation theory developed in \cite{I,II,III,IV,V,VI}
is a deformation of the complex structure on the worldsheet defined by the metric $g_{mn}$
to the complex structure defined by the super period matrix $\hat\Omega_{IJ}$.

\medskip
The deformation from $g_{mn}$ to $\hat\Omega_{IJ}$ was required in order to
preserve local supersymmetry on the worldsheet, and produce a well-defined
projection from supermoduli space to moduli space. But it turned out also that
fundamental superholomorphic objects, such as the super Abelian differentials
$\hat\o_I(\z)$, were related to the complex structure defined by
$\hat\Omega_{IJ}$ rather than the one defined by the original metric $g_{mn}$
(\cite{II}, \S 5.1).

\medskip
This relation between superholomorphicity with respect to $(g_{mn},\chi_m{}^\a)$ and holomorphicity
with respect to $\hat\Omega_{IJ}$ is the main theme of this paper. The purpose of this
section is to present the simplest examples where this correspondence can be established,
as a warm-up before the considerably more complicated treatment for the general case.

\subsection{Deformations of complex structures and metrics}

From the point of view of moduli, a deformation of complex structures
can just be viewed as a deformation of period matrices (after a choice of canonical homology basis)
\be
\Omega_{IJ}\, \to \, \hat\Omega_{IJ}
\ee
However, the objects of interest to us are tensors on the worldsheet,
and we need to realize this
deformation of complex structures as a deformation of metrics
\be
g_{mn}\, \to \, \hat g_{mn}
\ee
As in \cite{II}, we proceed as follows. Let $\hat g_{mn}$ be a metric on the
worldsheet $\Sigma$ whose period matrix in $\hat\Omega_{IJ}$.
In genus $h=2$, the dimension of odd supermoduli is 2, and thus we may
assume that $\hat g_{mn}=g_{mn}+\cO(\zeta^1\zeta^2)$, where $\zeta^1$ and
$\zeta^2$ are two odd supermoduli parameters. Let $z$ denote now complex
coordinates\footnote{We stress that this is a change of notation from the
previous section, where $z$ denoted complex coordinates for the
metric $ds^2$ associated with $g_{mn}$. This change is motivated by the fact
that the ultimate goal is to express all correlation functions in
terms of the $\hat\Omega_{IJ}$ complex structure.}
for the metric $d \hat s^2$ associated with $\hat g_{mn}$. The deformation
from the metric $d\hat s^2$ to the metric $ds^2$
\bea
d\hat s^2 = 2\hat g_{\bar zz}dzd\bar z,
\hskip 1in
ds^2= 2g_{\bar zz}dzd\bar
z+\delta g_{zz}dz^2+\delta g_{\bar z\bar z}d\bar z^2
\eea
is characterized by the Beltrami differential
\bea
\label{beltrami}
\hat \mu_{\bar z}{}^z= \mu (z) = - {1\over 2}\,\hat g^{z\bar z}\delta g_{\bar z\bar z}.
\eea
The condition that the period matrix for the metric $\hat g_{mn}$ is the super period
matrix $\hat\Omega_{IJ}$ is then equivalent to
\bea
i\int_\Sigma \o_I(z)\o_J(z)\,  \mu (z)
= \Omega_{IJ}- \hat\Omega_{IJ}.
\eea
This equation determines the Beltrami differential $\mu (z) $ only up to a
reparametrization $\mu (z) \to \mu (z) + \bar \p v(z)$, where $v(z)$ is a
smooth vector field on the surface $\Sigma$. Thus we can view the
equivalence  class of $\mu(z) $ as the intrinsic object, and a choice of
$\mu(z)$ within this equivalence class as just a gauge choice. Obviously,
this gauge choice did not enter the starting formulas
for the superstring amplitudes, and it should cancel out at the end.

\subsection{Super-holomorphic $\half$-forms and holomorphic $(1,0)$-forms}

The relation between superholomorphic differentials of  weight $(1/2,0)$ and
holomorphic differentials (with respect to $\hat\Omega_{IJ}$) of weight $(1,0)$
was discovered in \cite{II}. If $\hat\o (\z) =\hat\o_0 (z)+\theta\hat\o_+(z)$ is a
superholomorphic differential of  weight $(1/2,0)$, then there is a smooth,
single-valued scalar function $\lambda(z)$ and a 1-form $\o_z (z) dz$, holomorphic
with respect to $\hat g_{mn}$, so that the integral in $\theta$ alone yields,
\bea
\label{example1}
\int d\theta\wedge \hat e^z\,\hat\o (\z) = \o_z(z) dz + d\lambda(z)
\eea
We re-derive this result here, in a formalism suitable for later extensions.

\sm

A first important observation \cite{V} is that the superholomorphicity of $\hat\o$
implies that the above left-hand side is a closed 1-form. Closedness is a
property independent of the complex structure, and can be established using
the isothermal coordinates $z$ defined by the metric $g_{mn}$. Thus
$e^z$ is given by the expression (\ref{chiralzweibein1}), and we have
\be
\int d\theta \wedge e^z\,\hat\o (\z)
=
dz\,\hat\o_+(z)  -{1\over 2} d\bar z \, \chi (z) \hat\o_0(z)
\ee
In particular,
\be
d \left ( \int d\theta \wedge e^z\,\hat\o \right )
= -\, \left ( \p_{\bar z}\hat\o_++{1\over 2}\chi (z) \hat\o_0 \right )\,
dz\wedge d\bar z=0
\ee
in view of the superholomorphicity of $\hat\o$ and hence of (\ref{superholomorphicity2}).

\sm

The next observation is that the $d\bar z$ component of the left-hand side
of (\ref{example1}), {\sl with respect to the complex structure defined by
$\hat g_{mn}$}, is a $\bar\partial$-exact $(0,1)$-form. It suffices to show
this for each element $\hat\o_I$ of the basis of superholomorphic differentials
given by (\ref{omegahat}). Under the deformation of metrics
$g_{mn}\to \hat g_{mn}$, the forms $dz$ and $d\bar z$ get deformed as,
\be
\label{dzdeformation}
dz \to dz- \mu (z)  dz
\hskip 1in
d\bar z \to d\bar z- \bar \mu (z) dz,
\ee
where the $z$ coordinates on the left hand side of the arrows are holomorphic coordinates
with respect to $g_{mn}$, while on the right hand side, they are holomorphic coordinates
with respect to $\hat g_{mn}$. Thus, with $z$ holomorphic coordinates for $\hat g_{mn}$,
we have
\bea
\hat e^z & = & dz-\left ( \mu (z)+{1\over 2} \theta \chi (z)  \right ) d\bar z,
\no \\
\hat e^{\bar z} & = & d\bar z- \left ( \bar\mu(z)+{1\over 2}\bar\theta{\chi_z{}^-} \right ) dz
\eea
and hence
\be
\int d\theta\wedge \hat e^z\,\hat \o_I (\z)
=
dz\, \hat\o_{I+} - d\bar z\, \left (\mu (z) \hat \o_{I+} (z) + {1\over 2} \chi (z) \hat\o_{I0}(z) \right )
\ee
We claim that the $d\bar z$ component is orthogonal to any holomorphic differential.
Indeed, since it is already $\cO(\zeta^1\zeta^2)$ in its dependence on supermoduli,
in the pairing with an arbitrary holomorphic differential $\o_J(z)$, we can ignore the
distinction between holomorphic
forms with respect to $g_{mn}$ and $\hat g_{mn}$. Thus we have,
\bea
\int_\Sigma
d^2z\, \o_J  \left (\mu \hat \o_{I+} + {1\over 2}\chi \hat\o_{I0} \right )
=
\int_\Sigma d^2z \, \o_J \o_I \mu + {1\over 2}\int_\Sigma d^2z \, \o_J\chi \hat\o_{I0} =0
\eea
since the first term on the right hand side equals $\Omega_{IJ}-\hat\Omega_{IJ}$
by the definition of $\mu$, and the second term
equals $\hat\Omega_{IJ}-\Omega_{IJ}$ in view of the formula (\ref{Omegahat2})
for $\hat\Omega_{IJ}$. Thus there exists a smooth function $\lambda_I (z)$,
defined uniquely up to an additive constant, so that
\be
\mu (z) \o_I (z) +{1\over 2}\chi (z) \hat\o_{I0} (z) =-\p_{\bar z} \lambda_I(z).
\ee
We can write then
\be
\int d\theta \wedge \hat e^z\,\hat\o_I (\z)
=
\left ( \int d\theta \wedge \hat e^z\,\hat\o_I (\z) - d \lambda_I (z) \right )
+ d\lambda_I(z)
\ee
The term between parentheses on the right hand side is by construction a $(1,0)$-form,
which is closed, since both $d\lambda_I$ and the left-hand side are closed.
Thus it must be holomorphic. By examining its periods, we can easily recognize
it as $dz\,\o_I (z)$, and thus we obtain the desired formula,
\be
\int d\theta \wedge \hat e^z\,\hat\o_I (\z)
=
dz\,\o_I(z)+ d\lambda_I(z).
\ee
We observe that the function $\lambda_I$ depends on the
choice of metric $\hat g_{mn}$ within all metrics with period matrix
$\hat\Omega_{IJ}$. Under a gauge change $\mu (z) \to \mu (z) + \bar \p v(z)$,
the function $\lambda_I$ changes by $\lambda_I (z) \to \lambda_I (z) -v(z) \o_I (z) $,
up to a $z$-independent additive term. The form of this additive term will be determined
by the definition of the arbitrary additive constant in $\lambda _I(z)$, and we shall
take this to be $\lambda _I(w_0)=0$, where $w_0$ is the point on $\Sigma$
where the Green function $G$ is defined to vanish, $G(z,w_0)=0$ for all $z \in \Sigma$.

\subsection{The super prime form in $\hat g_{mn}$ conformal coordinates}

The next illustrative example, closer to the full chiral blocks,
is that of the super prime form $\cE_\delta({\bf z},{\bf w})$.
The super prime form $\cE_\delta({\bf z},{\bf w})$ was defined by (\ref{superprimeform1})
and (\ref{superprimeform2}).
It is important to recall that, in those formulas, the coordinates $z$ and $w$
are holomorphic coordinates with respect to the metric $g_{mn}$. As in the previous
example, to discuss holomorphicity, we need to express $\cE_\delta({\bf z},{\bf w})$
with respect to holomorphic coordinates $z$ and $w$ with respect to $\hat g_{mn}$.
It is easy to see that the prime form $E$ and the Szeg\"o kernel $S$
with respect to $g_{mn}$ can be expressed as
\bea
\label{hatco1}
\ln  E (z,w) & \to &
\ln\,E(z,w) + {1\over 2\pi} \int _x \mu(x) G(x,z) G(x,w)
\\
S (z,w) & \to &
S(z,w) +{1\over 4\pi}
\int _x \mu(x) \Big ( S(x,z)\p_xS(x,w)-S(x,w)\p_xS(x,z) \Big )
\no \eea
where all expressions on the right hand side are defined with respect to the metric
$\hat g_{mn}$. Taking into account the fact that $\mu=\cO(\zeta^1\zeta^2)$,
we can rewrite the super prime form as
\bea
\label{suprime2}
\ln\,\cE_\delta({\bf z},{\bf w})
=
- \hat F_0(z,w) - \theta_z \hat F_+(z,w)
- \theta_w \hat F_+(w,z) - \theta_z\theta_w \hat F_1 (z,w)
\eea
with the component functions defined by
\bea
\label{suprime3}
\hat F_0(z,w)&=&-\ln E(z,w)+ \ff (z,w)
\no \\
\hat F_+(z,w) &=& \FF (z,w)
\nonumber\\
\hat F_1 (z,w)
&=& S(z,w)+\Psi (z,w)
\eea
where we use the following abbreviations,
\bea
\label{hatcoordinates2}
\FF (z,w) & = & {1\over 4\pi}\int _x S(z,x)\chi(x) G(x,w)
\no \\
\ff (z,w)
&=&
{1\over 16\pi^2} \int _x \int _y G(x,z) \chi(x) S(x,y) \chi(y) G(y,w)
- {1\over 2\pi} \int _x \mu(x)G(x,z) G(x,w)
\no \\
\Psi (z,w)
&=&
- {1\over 16\pi^2}
\int _x \int _y S(z,x) \chi(x) \p_x\p_y\ln E(x,y) \chi(y) S(y,w)
\no \\
&&
\qquad
+ {1\over 4\pi} \int _x \mu(x) \Big ( S(x,z)\p_xS(x,w)
- S(x,w)\p_xS(x,z) \Big )
\eea
The functions $\FF, \ff, \Psi $ as well as $\hat \o _{I0}$ and
$\lambda_I$ are the basic constituents of the holomorphic blocks
that will be obtained in this paper.

\subsection{Basic properties of the functions $\hat \o _{I0}, \lambda _I, \FF, \ff$
and $\Psi$}

Here, we shall collect all the basic properties of the functions
$\hat \o _{I0}, \lambda _I, \FF, \ff$ and $\Psi$ that will be required in the sequel.
They are differentiation, monodromy and variation formulas under
changes of slice for $\chi $ and $\mu$. Note that we have the following symmetry properties,
\bea
\ff (z,w) & = & + \ff (w,z)
\no \\
\Psi (z,w) & = & - \Psi (w,z)
\eea

\subsubsection{Differential formulas}

They are all written in holomorphic coordinates for $\hat g_{mn}$,
\bea
\label{differentialrelations}
\p_{\bar z}\hat\o_{I0}(z)
&=&
-{1\over 2}\chi (z)  \o_I(z)
\\
\p_{\bar z}\lambda_I(z)
&=&
-\mu(z) \o_I(z)-{1\over 2} \chi (z) \hat\o_{I0}(z)
\no \\
\p_{\bar z} \FF (z,w)
&=&
+{1\over 2} \chi (z)  G(z,w)
\no \\
\p_{\bar w} \FF (z,w)
&=&
-{1\over 2} S (z,w)\chi (w)
\no \\
\p_{\bar z} \ff (z,w)
&=&
-{1\over 2} \chi (z)  \FF (z,w) + \mu(z) G(z,w)
\no \\
\p_{\bar z} \Psi (z,w)
&=&
-{1\over 2} \p_z \FF (w,z)\chi (z) - \mu(z)\p_z S(z,w)
- {1\over 2}\p_z \mu(z) S(z,w)
\no \eea

\subsubsection{Monodromies}

All monodromies on $A_K$-cycles vanish, while the monodromies on $B_K$-cycles are
given as follows for $z'=z+ B_K$,
\bea
\hat \omega _{I0} (z') & = & \hat \omega _{I0} (z)
\no \\
\lambda _I(z') & = & \lambda _I (z)
\no \\
\FF (z',w) & = & \FF (z,w)
\no \\
\FF (z,w') & = & \FF (z,w) - 2 \pi i \, \hat \omega _{K0} (z)
\no \\
\ff (z',w) & = & \ff (z,w) - 2 \pi i \, \lambda _K(w)
\no \\
\Psi (z',w) & = & \Psi (z,w)
\eea

\subsubsection{Variations of $\mu$-Slice}

The basic transformation is under a vector field $v^z= v(z)$ which is {\sl
second order} in $\zeta$,
\bea
\delta _v \chi (z) & = & 0
\\
\delta _v \mu (z) & = & \p _{\bar z} v (z)
\no \\
\delta _v \hat \omega _{I0} (z) & = & 0
\no \\
\delta _v \lambda _I (z) & = & - v(z) \omega _I (z)
\no\\
\delta _v  \FF (z,w) &=& 0
\nonumber \\
\delta _v  \ff (z,w) &=& v(z) G(z,w) + v(w) G(w,z) + \p _{w_0} v(w_0)
\no \\
\delta _v \Psi (z,w) & = & - v(z) \p_z S (z,w) - \half \p_z v(z) S (z,w)
- (z \leftrightarrow w)
\eea
All other quantities, $\omega _I$, $S$, $\ln E(z,w)$ are untransformed as
long as they are evaluated with respect to the metric $\hat g_{mn}$. (For simplicity, we
assume that $v(w_0)=0$.)

\subsubsection{Variations of $\chi$ and $\mu$ by local supersymmetry}

The basic transformation is under a spinor field $\xi ^+= \xi (z)$ which is
{\sl first order} in $\zeta$,
\bea
\delta _\xi \chi (z)   & = &  -2 \p _{\bar z} \xi  (z)
\no \\
\delta _\xi  \mu (z) & = & \xi (z) \chi (z)
\no \\
\delta _\xi \hat \omega _{I0} (z) & = & \xi (z) \omega _I(z)
\no \\
\delta _\xi \lambda _I (z) & = & \xi (z) \hat \omega _{I0} (z)
\nonumber \\
\delta _\xi  \FF (z,w) &=&   - \xi  (z) G(z,w)  + \xi  (w) S (z,w)
\nonumber \\
\delta _\xi  \ff (z,w) &=&  + \xi  (z)  \FF (z,w) + \xi (w)  \FF (w,z)
\no \\
\delta _\xi \Psi (z,w) & = & - \xi (z) \p_z \FF (w,z) - \p_w \FF (z,w) \xi (w)
\eea
(For simplicity, we have assumed that $\xi (w_0)=0$.)

\subsubsection{Dependence on $w_0$}

The Green function $G (r,s)$ is defined so that
$G(r,w_0)=0$ from which it follows that $ \lambda _I (w_0) = \ff (r,w_0)=
\ff (w_0,s)=  \FF (r,w_0) =0$. Denoting the corresponding quantities
defined with respect to a new point $w_0'$ by $G'(r,s)$, we have
the following relations
\bea
G' (r,s) & = & G (r,s) - G (r,w_0')
\no \\
\lambda _I ' (w) & = & \lambda _I (w) - \lambda _I (w_0')
\no \\
 \ff ' (r,s) & = &  \ff (r,s) -  \ff (r,w_0 ') -  \ff (w_0' ,s) +  \ff (w_0',w_0')
\nonumber \\
\FF  ' (r,s) & = &  \FF (r,s) -  \FF (r,w_0')
\eea

\subsection{Holomorphic blocks from $\p_+^z\ln\cE_\delta$ and
$\p_+^w\p_+^z\ln\cE_\delta$}

The superholomorphic chiral amplitudes $\cF [\delta]$ of (\ref{FE}) involve,
besides the differentials $\hat \o _I$ and the super prime form $\cE_\delta$
also first and second derivatives of $\ln \cE_\delta$, multiplied by certain
factors involving the frame $e^z$. We shall now show that these
superholomorphic objects also produce holomorphic forms on $\Sigma$,
up to exact differentials. All our considerations are modulo $\delta$-functions
supported at coincident vertex insertion points; these terms do not
contribute in a suitable definition of the full physical amplitude,
obtained by analytic continuation in the external momenta.
In the old dual model language, this fact is often referred to as
the {\sl cancelled propagator argument}.

\sm

Using the component form of $\ln \cE _\delta$ of (\ref{suprime2}),
the  derivatives (with respect to complex coordinates $z$ and $w$ on $\Sigma$
associated with the complex structure of $\hat \Omega _{IJ}$) are given by
\bea
\label{derprime}
\p_+^z\ln \cE_\delta({\bf z},{\bf w})
& = &
- \hat F_+(z,w) -\theta_w \hat F_1 (z,w)-\theta_z
\Big ( \p_z \hat F_0(z,w)+\theta_w \p_z \hat F_+(w,z) \Big )
\\
\p_+ ^w \p_+^z\ln \cE_\delta({\bf z},{\bf w})
& = &
- \hat F_1 (z,w) + \theta _z \p_z \hat F_+(w,z)
- \theta _w \Big ( \p_w \hat F_+(z,w) + \theta _z \p_w \p_z \hat F_0(z,w) \Big )
\no \eea
where all the partial derivatives on the right hand side can be taken to
be partial derivatives with respect to the holomorphic coordinates $z$
of the metric $\hat g_{mn}$, the difference between these and the original
derivatives with respect to $g_{mn}$ having cancelled because $\FF$ and $\ff$ are
respectively of orders $1$ and $2$ in $\zeta^\a$, and $\ln E(z,w)$ is holomorphic.
We shall now prove the following formulas, in which the integrations are
understood to be carried out over the $\theta _z$ and $\theta _w$ variables only,
\bea
\label{spint}
\int d\theta_z  \hat e^z \p_+^z\ln \cE_\delta({\bf z},{\bf w})
& = &
dz \p_z\ln E(z,w) - d_z \ff (z,w) - \theta_w d_z \FF (w,z)
\no \\
\int \! \! \int d\theta_w \hat e^w
 d\theta_z \hat e^z\,\p_+^w\p_+^z \ln \cE_\delta({\bf z},{\bf w})
& = &
dz dw \, \p_w \p_z \ln E(z,w) + d_w d_z \ff (z,w)
\eea
On the right hand side of the first line above, we have omitted a term
$-d\bar z \mu(z)\p_z \! \ln E(z,w_0)$, which is independent of $w$, and
will cancel out of the chiral amplitude thanks to momentum conservation.
The forms $dz \p_z \ln E(z,w)$ and $dz  dw \, \p_w \p_z \ln E(z,w)$
are holomorphic forms on $\Sigma$ with respect to the complex
structure of $\hat \Omega _{IJ}$. The remaining terms on the right
hand side of both formulas are exact differentials either in $z$, in $w$
or in both.

\sm

The proof of both formulas in (\ref{spint}) is obtained by first carrying out
explicitly these $\theta_z$ and $\theta _w$ integrations. We begin with
the first formula,
\bea
\int d\theta_z \hat e^z \p_+^z \ln \cE_\delta({\bf z},{\bf w})
&=&
-dz \p_z \hat F_0(z,w)- \theta_w dz \p_z \hat F_+(w,z) - d\bar z \mu(z) \p_z\ln E(z,w)
\no \\ &&
+ \half d\bar z \bigg ( \chi (z) \hat F_+(z,w) + \theta_w \hat F_1(w,z) \chi (z) \bigg )
\eea
Using the differential relations (\ref{differentialrelations}), we can essentially
absorb the $d\bar z$ terms in an exact de Rham differential, and recover
the first line of (\ref{spint}). As mentioned above, we omit an irrelevant
term of the form $-d\bar z \mu(z)\p_z \ln E(z,w_0)$.

\sm

To prove the second line in (\ref{spint}), we apply the operation
$\int d\theta _w \wedge e^w \p_+ ^w$ to the first line of (\ref{spint}), and obtain,
\bea
\int \! \! \int d\theta_w  \hat e^w  d\theta_z  \hat e^z\,
\p_+^w\p_+^z \ln \cE_\delta({\bf z},{\bf w})
&=&
dz  dw \,\p_w\p_z \ln E(z,w) + dw \,\p_w d_z \ff (z,w)
\no \\
&&
+ d\bar w dz \, \mu(w) \p_w\p_z \ln E(z,w)
\no \\
&&
- {1\over 2}d\bar w \, \chi (w) \, d_z \FF (w,z).
\eea
Here we note that the shift of $\p_z$, from holomorphic coordinates $w$ with respect to
$g_{mn}$ to holomorphic coordinates $w$ with respect to $\hat g_{mn}$,
results only in additional $\delta(z,w)$ terms, when applied to $\p_z\ln E(z,w)$.
In view again of the relations (\ref{differentialrelations}), we can write
\bea
dw\,\p_w d_z \ff (z,w)
&=&
d_w d_z \ff (z,w)-d\bar w d_z\p_{\bar w} \ff (z,w)
\\
&=&
d_w d_z \ff (z,w)+{1\over 2}d\bar w \chi (w) d_z \FF (w,z)
\no \\ && 
- d\bar w  dz\, \mu(w)\p_w\p_z\ln E(z,w)
\nonumber
\eea
modulo $\delta$ functions along coincident points $z=w$. This proves the desired relation.

\sm

The relation (\ref{spint}) shows that, just as in the case of super holomorphic differentials,
the super prime form and its super-derivatives reduce to the ordinary prime form
and its derivatives with respect to the complex structure
of $\hat g_{mn}$, modulo exact differentials and $\delta$ functions. There is a subtlety,
however: the function $\ff (z,w)$ has monodromy in both $z$ and $w$, and the
differential $d_w d_z \ff (z,w)$ is not exact in {\it both} variables $z$ and $w$.
We shall see later that the monodromy of $\ff$ is precisely what will be needed
to supply the resulting holomorphic chiral blocks with the correct monodromy
transformation properties.

\newpage

\section{Kinematic Invariants}
\setcounter{equation}{0}

The combinatorial structure of the chiral blocks of the $N$-point amplitudes
is inherently quite complicated, and it will be very useful to decompose
the blocks according to suitable kinematic invariants. In fact, identifying
these kinematic invariants is a fundamental aspect of the holomorphic
chiral block problem.

\subsection{Grassmann parity of the polarization vectors}

We make systematic use of the differential forms
in superspace, and their commutation relations.  Let $z_r$ be
bosonic and $\theta _r$ be fermionic coordinates,
with corresponding differentials $dz_r$ and $d\theta_r$.
Their commutation rules are
\bea
\label{superforms}
z_r d\theta _s &=& + d\theta _s z_r
\qquad \qquad
dz_r \wedge dz_s = - dz_s \wedge dz_r \nonumber \\
dz_r \theta _s &=& + \theta _s dz_r
\qquad \qquad
dz_r \wedge d\theta _s = - d\theta _s \wedge dz_r \nonumber \\
\theta _r d \theta _s &=& - d\theta _s \theta _r
\qquad \qquad
d\theta _r \wedge d\theta _s = + d\theta _s \wedge d \theta _r
\eea
In particular, the {\it chiral volume form} $dz \wedge d \theta$ is even
as a form but odd Grassmann valued.

\sm

With respect to local complex coordinates $z,\bar z$ associated with the
complex structure induced by the metric $\hat g_{mn}$, the polarization
differential forms, introduced in section 2.5 take the form,
\bea
\e _o ^\mu = \ep^\mu \wedge \hat e^z
& \hskip 0.6in  \ep ^\mu  \equiv  \e^\mu d\theta \hskip 0.6in &
\hat e^z  =  dz-\left ( \mu (z) + {1\over 2}\theta \, \chi (z) \right ) d\bar z
\no \\
\bar \e _o ^\mu  = \bar \ep ^\mu  \wedge \hat e^{\bar z}
& \hskip 0.6in \bar \ep ^\mu  \equiv  \bar \e^\mu d\bar \theta  \hskip 0.6in &
\hat e^{\bar z} =
d\bar z-\left ( \bar \mu (z) + {1\over 2}\bar \theta \, \bar \chi (z) \right ) dz
\eea
The polarization vectors $\e^\mu , \bar \e^\mu$ are ordinary complex numbers,
so that $\e_o ^\mu, \bar \e_o^\mu$ and $\ep ^\mu , \bar \ep ^\mu$ are Grassmann
odd; $\e_o ^\mu, \bar \e_o^\mu$ are differential 2-forms, while
$\ep ^\mu , \bar \ep ^\mu$ are differential 1-forms.
In particular, the chiral amplitude $\cF[\delta]$ will be viewed as a differential
form in each of the vertex insertion points $\z_r$. Finally, we shall use the following
notations throughout,
\bea
d_r = \p _r + \bar \p_r
& \hskip 1in &
\p _r \equiv dz_r \p _{z_r}
\no \\
& \hskip 1in &
\bar \p_r \equiv d\bar z_r \p _{\bar z_r}
\eea
where $d_r$ is the total differential in $r$, while $\p_r$ and $\bar \p_r$ are
its $(1,0)$ and $(0,1)$ form components respectively.

\subsection{The quantity $K_r^\mu$}

We introduce the following key quantity
\bea
\label{K}
K^\mu _r  \equiv  \ep ^\mu _r \wedge dz_r  + i k^\mu _r \theta _r
= \e^\mu d\theta_r \wedge dz_r + i k^\mu _r \theta _r
\eea
which combines  the momentum $k_r^\mu $ and
the polarization vector $\ep _r ^\mu = \e_r ^\mu d\theta _r $.
In view of the properties of $\ep_r ^\mu$, the
quantity $K_r ^\mu$ is odd Grassmann valued, as well as a linear combination
of a differential form of degree 0 and a differential form of degree 2.

\sm

The combination $K^\mu_r$ plays a central role in the organization of the
space-time structure of the chiral amplitudes. This
may be seen, for example, by concentrating on those terms occurring in the
chiral amplitude $\cF [\delta ]$ of (\ref{FE}) which involve the Szeg\"o kernel
$S (z_r,z_s)$ only. Using the explicit form of the super prime form
in (\ref{suprime2}) and (\ref{suprime3}),  as well as its derivatives in (\ref{derprime}),
we find that the Szeg\"o kernel $S (z_r,z_s)$ in the argument of the
exponential in (\ref{FE})  has the following multiplicative coefficient,
\bea
 - k_r ^\mu  k_s ^\mu + \e _{or} ^\mu \e _{os} ^\mu
- i k^\mu _r \e ^\mu _{os} \theta _r +  i k^\mu _s \e ^\mu _{or} \theta _s
= \tilde K_r ^\mu \tilde K_s ^\mu
\eea
where $\tilde K_r ^\mu$ is defined in analogy with $K^\mu _r$ by
\bea
\tilde K_r ^\mu \equiv \e _{or} ^\mu + i k_r ^\mu \theta _r
\hskip 1in
\e_{or} ^\mu = \e^\mu _r d \theta _r \wedge \hat e^z _r
\eea
The quantity $\tilde K^\mu _r$ reduces to $K^\mu _r$ up to terms linear in $\mu$
and $\chi$, since $\e_{or}^\mu$ reduces to $\ep ^\mu _r \wedge dz_r$.
Next, we list some of the key properties of $K$ and $\tilde K$,
\bea
&(1)&
\{ K_r ^\mu , \theta _s \} = \{ K_r ^\mu , d\theta _s \}
= \{ K^\mu _r , \ep _s ^\nu \} = [K ^ \mu _r , dz_s ] = [ K ^\mu _r , e^s ] = 0
\no \\
&(2)&
\{ \tilde K_r ^\mu , \theta _s \} = \{ \tilde K_r ^\mu , d\theta _s \}
= \{ \tilde K^\mu _r , \ep _s ^\nu \} =  [\tilde K ^ \mu _r , dz_s ] =
[ \tilde K ^\mu _r , e^s ] = 0
\no \\
&(3)&
K_r^\mu K_r^\nu= i(\ep_r^\mu k_r^\nu-\ep_r^\nu k_r^\mu)dz_r \theta_r
= i(\e_r^\mu k_r^\nu-\e_r^\nu k_r^\mu) d \theta _r \wedge d z_r \theta_r
\no \\
&(4)&
\tilde K_r ^\mu \tilde K_r ^\nu = i (\epsilon ^\mu _{or} k^\nu _r -
\epsilon _{or} ^\nu k_r ^\mu ) \theta _r
=
i ( \epsilon ^\mu _r k^\nu _r - \epsilon _r ^\nu k_r ^\mu ) d \theta _r  \wedge e^r \theta _r
\no \\
&(5)&
k^\mu _r K^\mu _r = k^\mu _r \tilde K^\mu _r =
\ep _ r ^\mu \theta _r K_r ^\nu = \ep _ r ^\mu \theta _r
\tilde K_r ^\nu = 0
\eea
The quadratic character of the worldsheet fermions $\psi _+$, both in the action
and in the vertex operators, requires that the number of Szeg\"o kernels
at any vertex insertion point be either 0 or 2. The presence of two Szeg\"o
kernels at an insertion point $s$, for example, then produces two
factors of $K_r $ as well. Two factors of $K_r$ automatically produce
the {\sl field strength combination}
$f_r ^{\mu \nu} = \e_r ^\mu k_r ^\nu - \e _r ^\nu k_r ^\mu$
for the polarization vector $\e_r ^\mu$ and momentum $k_r ^\mu$.
Specifically, we have
\bea
\Big ( S (r,s) K_r ^\mu K_s ^\mu \Big )
\Big ( S (s,t) K_s ^\nu K_t ^\nu \Big )
=
K_s ^\mu K_s ^\nu \Big ( S (r,s) S (s,t) K_r ^\mu K_t ^\nu \Big )
\eea
with $K_r ^\mu K_r ^\nu$ given in terms of $f_r ^{\mu \nu}$ by (3) above.
The tremendous advantage is that the {\sl field strengths combination}
$f_r ^{\mu \nu}=\e_r ^\mu k_r ^\nu - \e _r ^\nu k_r ^\mu$ are automatically invariant under
space-time gauge transformations $ \e_r ^\mu \to \e _r ^\mu + c k^\mu _r$,
for any boost $c$.

\sm

Remarkably, it is not only the worldsheet fermion contributions, produced by
the Szeg\"o kernel, which will admit an organization in terms of the quantities
$K_r^\mu$, but also the worldsheet boson contributions. The latter is not seen
directly, but will be established starting in the subsequent section.

\sm

We note that the {\sl Bianchi identity} on $f_r ^{\mu \nu}$ can be expressed
simply as the following identity on the quantity $K_r$ and the associated momenta $k_r$,
\bea
\label{bianchi}
k^\mu _r K^\nu _r K^\rho _r + k^\nu _r K^\rho _r K^\mu _r +
 k^\rho _r K^\mu _r K^\nu _r =0
\eea
In fact, all the kinematic invariants which we shall encounter can be expressed
simply in terms of $K_r^\mu$, and we can describe them now.

\subsection{Kinematic factors for linear chains}

We shall denote all kinematic-type invariants by the same generic letter $K$,
and distinguish them one from another only by their indices.
The first type of kinematic factor arises from contracting indices along
a linear chain with open ends. These open ends will be further contracted
with internal loop momenta $p_I ^\mu$, external momenta $k^\mu _r$
or polarization forms $\ep ^\mu _r$. The key ingredient for the linear
chain structure is given by
\bea
\label{Kstring}
K^{\mu \nu} _{[t_1 t_2 \cdots t_{n-1} t_n]} & \equiv &
K^\mu _{t_1} K^{\rho_1} _{t_1} K^{\rho_1} _{t_2} K^{\rho_2} _{t_2} \cdots
K^{\rho_{n-2}} _{t_{n-1}} K^{\rho_{n-1}} _{t_{n-1}} K^{\rho_{n-1}} _{t_n} K^{\nu} _{t_n}
\no \\
& = & K^\mu _{t_1} \left ( \prod _{k=1} ^{n-1} K^{\rho_k} _{t_k} K^{\rho_k} _{t_{k+1}}
\right ) K^{\nu} _{t_n}
\eea
By construction, these quantities involve only the gauge invariant field strengths
$f_{t_i}^{\mu \nu}$. They have the following reflection and cyclic permutation
symmetry properties,
\bea
K^{\nu \mu} _{[t_1 t_2 \cdots t_{n-1} t_n]} & = &
(-)^n K^{\mu \nu} _{[t_n t_{n-1}  \cdots t_2 t_1]}
\no \\
K^{\mu \mu} _{[t_1 t_2 \cdots t_{n-1} t_n]} & = &
(-)^n K^{\mu \mu} _{[t_n t_{n-1}  \cdots t_2 t_1]}
\no \\
K^{\mu \mu} _{[t_1 t_2 \cdots t_{n-1} t_n]} & = &
K^{\mu \mu} _{[t_2 t_3 \cdots t_{n-1} t_n t_1]}
\eea
and satisfy various identities resulting from contractions,
\bea
k^\mu _s K^{\mu \rho} _{[rs]}
& = & \half K_{[rs]} ^{\mu \mu} ~ k_r ^\rho
\nonumber\\
\half K^\nu _s K^\mu _s K^{\mu \nu} _{[ r t_1 \cdots t_n]}
&=&
- i \ep ^\mu _s \theta _s k^\nu _s K^{\mu \rho} _{[ r t_1 \cdots t_{n-1}]}
K^\rho _{t_n} K^\nu _{t_n}
+ i \ep _s ^{\{ \mu} \theta _s k_s ^{\nu \} } K^{\mu \nu} _{[ r t_1 \cdots t_n]}
\eea
These identities may be proven with the help of the Bianchi identity.

\subsection{Kinematic factors for singly linked chains}

The basic structure of a singly linked kinematic factor is obtained
by taking a linear chain and linking one end to one of the midpoints
of the linear chain with an extra factor of momentum evaluated at that midpoint;
concretely, it is of the form,
\bea
K^{\mu \rho } _{[t_1 \cdots t_m r u_1 \cdots u_\ell ]} ~ k^\rho _r
\eea
Such factors arise from a bosonic field contraction at the end of the linear
chain; the extra factor of momentum arises from a bosonic contraction
with the exponential $\exp \{ i k_r ^\rho x_+ ^\rho (r)\}$.
Using the Bianchi identity, the kinematic combination that will occur later in the
bosonic contractions can be written as,
\bea
K^{\mu \rho } _{[t_1 \cdots t_m r u_1 \cdots u_\ell ]} ~ k^\rho _r
& = &
K^{\mu \nu} _{[t_1 \cdots t_m]} k^\rho _r K^\nu _r K^\sigma  _r
K^{\sigma \rho} _{[ u_1 \cdots u_\ell]}
\no \\ &&
\no \\
& = &
- K^{\mu \nu} _{[t_1 \cdots t_m]} k^\nu _r K^\sigma _r K^\rho  _r
K^{\sigma \rho} _{[ u_1 \cdots u_\ell]}
- K^{\mu \nu} _{[t_1 \cdots t_m]} k^\sigma _r K^\rho _r K^\nu  _r
K^{\sigma \rho} _{[ u_1 \cdots u_\ell]}
\no \\ &&
\no \\
& = &
 K^{\mu \nu} _{[t_1 \cdots t_m]} k^\nu _r
K^{\sigma \sigma} _{[ r u_1 \cdots u_\ell]}
- K^{\mu \nu} _{[t_1 \cdots t_m]} k^\rho _r
K^{\rho \nu} _{[ u_1 \cdots u_\ell r ]}
\no \\ &&
\no \\
& = &
 K^{\mu \nu} _{[t_1 \cdots t_m]} k^\nu _r
K^{\sigma \sigma} _{[ r u_1 \cdots u_\ell]}
+(-)^\ell  K^{\mu \rho} _{[t_1 \cdots t_m r u_\ell \cdots u_1 ]} k^\rho _r
\eea
Equivalently, we have the following anti-symmetrization formula,
\bea
K^{\mu \rho } _{[t_1 \cdots t_m r u_1 \cdots u_\ell ]} ~ k^\rho _r
-
(-)^\ell  K^{\mu \rho} _{[t_1 \cdots t_m r u_\ell \cdots u_1 ]} k^\rho _r
=
 K^{\mu \rho} _{[t_1 \cdots t_m]} ~ k^\rho _r ~
K^{\sigma \sigma} _{[ r u_1 \cdots u_\ell]}
\eea
where the right hand side is the same form as we shall find in the fermionic loop contractions.
Thus, it is natural to decompose the kinematic factor as follows,
\bea
\label{Kplusminus}
K^{\mu \rho } _{[t_1 \cdots t_m r u_1 \cdots u_\ell ]} ~ k^\rho _r
& = & \half \bigg (
K^{\mu \rho } _{[t_1 \cdots t_m r u_1 \cdots u_\ell ]} ~ k^\rho _r
-
(-)^\ell  K^{\mu \rho} _{[t_1 \cdots t_m r u_\ell \cdots u_1 ]} k^\rho _r  \bigg )
\no \\ &&
+ \half \bigg (
K^{\mu \rho } _{[t_1 \cdots t_m r u_1 \cdots u_\ell ]} ~ k^\rho _r
+ (-)^\ell  K^{\mu \rho} _{[t_1 \cdots t_m r u_\ell \cdots u_1 ]} k^\rho _r  \bigg )
\eea
or using the above identity,
\bea
K^{\mu \rho } _{[t_1 \cdots t_m r u_1 \cdots u_\ell ]} ~ k^\rho _r
 =
\half  K^{\mu \rho} _{[t_1 \cdots t_m]} ~ k^\rho _r ~
K^{\sigma \sigma} _{[ r u_1 \cdots u_\ell]}
+ K^{\mu \rho } _{[t_1 \cdots t_m r \{ u_1 \cdots u_\ell \} ]} ~ k^\rho _r
\eea
where we have defined the ``symmetrized" combination
$K^{\mu \rho } _{[t_1 \cdots t_m r \{ u_1 \cdots u_\ell \} ]}$
by
\bea
K^{\mu \rho } _{[t_1 \cdots t_m r \{ u_1 \cdots u_\ell \} ]}
\equiv
\half \bigg (
K^{\mu \rho } _{[t_1 \cdots t_m r u_1 \cdots u_\ell ]}
+ (-)^\ell  K^{\mu \rho} _{[t_1 \cdots t_m r u_\ell \cdots u_1 ]}   \bigg )
\eea
We now define the kinematic invariants
$K_{\pm}^{(m+1|\ell+1)}$
by
\bea
\label{Kpm}
K_+^{(m+1|\ell+1)}
&=&
\ep_s^\mu\theta_s\,K_{[t_1\cdots t_m r\{u_1\cdots u_{\ell}]}^{\mu\rho}ik_r^\rho
\nonumber\\
K_-^{(m+1|\ell+1)}
&=&
\ep_s^\mu\theta_s\,K_{[t_1\cdots t_m]}^{\mu\rho}ik_r^\rho K_{[ru_1\cdots u_{\ell}]}^{\sigma\sigma}
\eea

\newpage

\section{Combinatorial Organization}
\setcounter{equation}{0}

In this section, we shall start from the superspace formulation of chiral blocks as
superholomorphic forms in the vertex insertion points, as well as on super moduli
space, and decompose the blocks with two goals in mind.
\begin{itemize}
\item Separate the terms which are a total derivative in at least one vertex insertion point,
and thus contain a non-vanishing $(0,1)$ form in that vertex insertion point;
we generally denote these terms by $\cD [\delta] $. The remainder will be all the
terms which are $(1,0)$ forms in each vertex insertion point; we
generally denote these terms by $\cZ [\delta] $; the full chiral amplitude will then be
given by the sum
\bea
\cF [\delta ] = \cZ [\delta ] + \cD [\delta]
\eea
\item Bring to the fore the kinematic dependence solely in the form of the space-time
gauge invariant combinations $K^\mu _r$, introduced in the preceding section.
\end{itemize}
It should be kept in mind, however, that the $\cD [\delta ]$-terms will not be closed in the
all insertion points, and as such, do not yet qualify to be exact terms in the sense
of hybrid cohomology. A subsequent regrouping of terms producing such hybrid
exact differentials, will be carried out in section \S 9.

\subsection{Decomposition of chiral blocks}

The starting point is the superspace representation for the chiral amplitudes,
which was derived in (\ref{FE}) as a differential form in each of the vertex insertion points,
and which we recall here for convenience,
\bea
\label{FE1}
\cF [\delta]
& = &
\exp \biggl \{
{i \over 4 \pi} p_I ^\mu \hat \Omega _{IJ} p_J ^\mu
+ p_I ^\mu \sum _{r=1} ^N \left ( \e ^\mu _{or} \hat \omega _I (\z _r)
+ ik ^\mu _r \int ^{\z _r} _ {\z _0} \hat \omega _I  \right ) \biggr \} \times
\no \\
&&
\exp \half \sum _{[r,s]} ^N \biggl [
k_r ^\mu k_s ^\mu \ln \cE _\delta (\z _r, \z _s)
+ \epsilon _{or} ^\mu \epsilon _{os} ^\mu \p_+ ^r \p_+ ^s \ln \cE _\delta (\z _r, \z _s)
\no \\ && \hskip 0.7in
- i k _r ^\mu  \epsilon _{os} ^\mu \p_+ ^s \ln \cE _\delta (\z _r, \z _s)
- i k _s ^\mu  \epsilon _{or} ^\mu \p_+ ^r \ln \cE _\delta (\z _s, \z _r)
\biggr ]
\eea
We use (\ref{suprime2}),  (\ref{suprime3}) as well as (\ref{derprime}) to
recast the super Abelian differentials and the superprime form in terms
of $\hat \o_{Io}, \lambda _I, \FF, \ff$, and $\Psi$, as well as the
Szeg\"o kernel and the prime form $E$. With the help of the expressions
derived in Section 3.5, all $\mu(r)$ and $\chi (r)$ dependence
encountered in the polarization forms $\e_{or}$ is recombined
in such a manner that all $(0,1)$ forms that occur are converted into total differentials
applied to $\lambda _I, \FF$, or $\ff$. All dependence of the Szeg\"o kernel and the
prime form evaluated between vertex insertion points is regrouped into an overall
factor which will be denoted by $Z_0$; it is independent
of $\mu (z) $ and $\chi (z)$, and thus independent of odd supermoduli.
The final result is
\bea
\label{FE2}
{\cal F}[\delta]
&=&
Z_0\, \exp \bigg \{ \,p_I^\mu \sum _r \bigg ( \ep_r^\mu \theta_r d_r \lambda_I(r)
+ik_r^\mu \lambda_I(r)+K_r^\mu \hat\o_{I0}(r) \bigg ) \bigg\}
\no \\
&& \hskip 0.1in
\times {\rm exp}\bigg\{\sum_{[rs]}\bigg(-{1\over 2}k_r^\mu k_s^\mu \ff (r,s)+
ik_r^\mu\ep_s^\mu \theta_s d_s \ff (r,s)
+{1\over 2} \ep_r^\mu \ep _s ^\mu \theta_r \theta_s d_r d_s \ff (r,s)
\no \\
&&
\hskip 1in
+ {1\over 2} (\tilde K_r^\mu \tilde K_s^\mu  - K_r^\mu K_s^\mu ) S(r,s)
+ \half K_r^\mu K_s^\mu  \Psi (r,s)
\no \\
&&
\hskip 1in
+ iK_r^\mu k_s^\mu \FF (r,s)+K_r^\mu \ep_s^\mu d_s \FF (r,s)\bigg)\bigg\}
\eea

\subsection{The factor $Z_0$ and its generalizations}

The factor $Z_0$ occurring in the formula for $\cF [\delta]$ is given by
\bea
Z_0 & = & \exp \biggl \{
{i \over 4 \pi} p_I ^\mu \hat \Omega _{IJ} p_J ^\mu
+ p_I ^\mu \sum _{r=1} ^N \left ( \ep ^\mu _r \omega _I (r)
+ ik ^\mu _r \int ^r  \omega _I  \right ) \biggr \}
\no \\ && \times
\exp \half \sum _{[r,s]} ^N \biggl [ K^\mu _r K^\mu _s S (r,s) +
k_r ^\mu k_s ^\mu \ln E  (r, s)
+ \ep _r ^\mu \ep _s ^\mu \theta _r \theta _s \p_r \p_s  \ln E (r, s)
\no \\ && \hskip 0.8in
- i k _r ^\mu  \ep _s ^\mu \theta_s \p_s \ln E  (r, s)
- i k _s ^\mu  \ep _r ^\mu \theta _r \p_r \ln E  (s, r)
\biggr ]
\eea
The rearrangement of the kinematic factor multiplying $S (r,s)$ has
already been explained in section 4.2. It will turn out to be very convenient
to reformulate the bosonic contributions in $Z_0$ in terms of correlators of
the chiral worldsheet boson field $x_+ ^\mu$, but leave Wick contractions
undone until needed. This reformulation will drastically simplify the
combinatorics, and facilitate the extraction of the kinematic combinations
$K^\mu _r K^\nu _r$ also for the terms involving the worldsheet bosons.
Specifically, we shall introduce the field $x^\mu _+(z)$ with two-point function,
\bea
\< x_+ ^\mu (z) x^\nu _+ (w) \> = - \ln E(z,w)
\eea
Furthermore, we shall use the combination below in order to deal efficiently
with the internal momentum dependence,
\bea
\bar x_+ ^\mu (z) = x_+^\mu (z) + p^\mu _I \int ^z \omega _I
\eea
The $Z_0$ factor is then given by\footnote{For further notational simplification,
we shall not exhibit the vacuum  expectation signs $\< ~ \>$ on $Z_0$. }
\bea
\label{Z0}
Z_0 \equiv
\exp  \left \{ {i \over 4 \pi} p_I ^\mu \hat \Omega _{IJ} p_J ^\mu
+ \half \sum  _{[r,s]}  K^\mu _r K^\mu _s S (r,s)
+  \sum_r ik_r^\mu \bar x_+^\mu(r)
+ \sum_r \ep_r^\mu \theta_r\p_r \bar x_+^\mu(r)
\right \}
\eea
In $Z_0$, one could have left also the worldsheet
fermion field uncontracted in the form $K^\mu _r \psi _+ ^\mu (r)$,
instead of incorporating the Wick contracted Szeg\"o kernel in $Z_0$.
It does not appear, however,  that much will be gained by doing so,
and we shall not do so here.

\sm

To facilitate the organization of the combinatorics, it will be extremely
useful to introduce a generalization of $Z^0$, which may be defined
by removing the polarization vector contribution for the worldsheet bosons
on a sequence of points $r_1, \cdots, r_n$. The reason this will be
a helpful tool is that certain -- but not all -- boson vertex contractions in $Z_0$
will need  to be carried out, and combined with certain combinations
of the functions $\hat \o _{I0}, \lambda _I, \FF, \ff$ in order to
exhibit their kinematic dependence entirely in the form of the gauge invariant
combinations $K^\mu _r K^\nu _r$. The required generalization is defined by
\bea
\label{Z0gen}
Z_0^{r_1\cdots r_n}
\equiv
\exp  \left \{ {i \over 4 \pi} p_I ^\mu \hat \Omega _{IJ} p_J ^\mu
+ \half \sum  _{[r,s]}  K^\mu _r K^\mu _s S (r,s)
+ \sum_r ik_r^\mu \bar x_+^\mu(r)
+ \! \! \! \! \sum_{r\not= r_1,\cdots,r_n} \! \! \! \! \ep_r^\mu \theta_r\p_r \bar x_+^\mu(r)
\right \}
\eea
Here it is assumed that all $r_i$, $i=1,\cdots,n$ are distinct.
Each $Z_0^{r_1\cdots r_n}$ is strictly invariant under each of the monodromy
transformations. Finally, we shall also use the following abbreviation,
\bea
X_r^\mu  \equiv  ik_r^\mu +K_r^\nu K_r^\mu (dr)^{-1}\p_r\bar x_+^\nu (r)
\eea
With these conventions and notations at our disposal, we are now ready to
reformulate the chiral block and decompose them according to
$\cF [\delta ] = \cZ [\delta ]+ \cD [\delta]$, as explained earlier on in this section.

\subsection{Summary of blocks of type $\cZ [\delta ]$}

It will be convenient to first present the result of the calculation explained above first,
and then give a detailed derivation of each of the terms that contributes. We shall organize the
contributions according to the number of vertex insertion points at which
the contribution is {\sl not manifestly holomorphic}. Since the sources
of non-holomorphicity are a single occurrence of $\mu (z)$, which couples
quadratically to  the fields $x_+$ and $\psi _+$, and a double
occurrence of $\chi (z) $, each of which couples bilinearly to $x_+$ and $\psi _+$,
there can be at most 4 vertex insertion points at which any given contribution
is not manifestly holomorphic. The result for the blocks $\cZ$ are as follows,
\bea
\cZ [\delta ] & = &
\cZ_{1a} + \cZ_{1b} + \cZ_{2a} + \cZ_{2b} + \cZ_{2c} + \cZ_{2d} + \cZ_{2e} + \cZ_{2f}
\no \\ &&
+ \cZ_{3a} + \cZ_{3b} + \cZ_{3c} + \cZ_{3d} + \cZ_{3e} + \cZ_4
\eea
The first entry in the subscript stands for the number of non-holomorphic insertion
points, while the supplementary index is used to
distinguish different contributions to that order.

\sm

The corresponding individual contributions are given by
\bea
\label{master1}
\cZ_{1a} & = &
+ \sum _r Z_0 ^{r}  p^\mu _I X^\mu _r  \lambda _I(r)
\no \\
\cZ_{1b} & = &
- \sum _r Z_0 ^{r} \half p^\mu _I K^\mu _r K^\nu _r p^\nu _J
\hat \omega _{I0}(r) \hat \omega _{J0}(r)
\no \\
\cZ_{2a} & = &
- \sum _{[rs]} Z_0 ^{rs} \half p^\mu _I K^\mu _r K^\nu _s p^\nu _J \hat
\omega _{I0}(r) \hat \omega _{J0}(s)
\no \\
\cZ_{2b} & = &
+ \sum _{[rs]} Z_0 ^{rs}  \half  \ff (r,s) X^\mu _r X^\mu _s
\no \\
\cZ_{2c} & = & + \sum _{[r,s]} Z_0 ^{rs} \half K^\mu _r K^\mu _s \Psi (r,s)
\no \\
\cZ_{2d} & = &
+ \sum _{[rs]} Z_0 ^{rs} K^\mu _r X^\mu _s  \FF (r,s)
p^\nu _I \bigg ( K^\nu _r \hat \omega _{I0} (r) + K^\nu _s \hat \omega _{I0} (s) \bigg )
\no \\
\cZ_{2e} & = &
+ \sum _{[rs]} Z_0 ^{rs} {1 \over 2} K^\mu _r
K^\nu _r K^\nu _s K^\mu _s (ds)^{-1} \biggl (\Phi _B (s;r,r) - i k^\sigma _s \p _s
\bar x ^\sigma _+ (s) Q_B(s;r,r) \biggr )
\no \\
\cZ_{2f} & = &
- \sum _{[rs]}  Z_0 ^{rs} \half K^\mu _r
k ^\mu _s K^\nu _s k ^\nu _r  \FF (r,s)  \FF (s,r)
\no \\
\cZ_{3a} & = &
+ \sum _{[rst]} Z_0 ^{rst} K^\mu _r X^\mu _s  \FF (r,s)
p^\nu _I K^\nu _t \hat \omega _{I0} (t)
\no \\
\cZ_{3b} & = &
+ \sum _{[rst]} Z_0 ^{rst} (-) {i \over 2} K^\mu _r K^\nu _t
X^\nu _s k^\mu _s  \FF (r,s)  \FF (t,s)
\no \\
\cZ_{3c} & = &
+ \sum _{[rst]} Z_0 ^{rst} {1 \over 2} K^\mu _r
K^\nu _t K^\nu _s K^\mu _s (ds)^{-1} \biggl (\Phi _B (s;r,t) - i k^\sigma _s \p _s
\bar x ^\sigma _+ (s) Q_B (s;r,t) \biggr )
\no \\
\cZ_{3d} & = &
- \sum _{[rst]}  Z_0 ^{rst}  K^\mu _r
X^\mu _s K^\nu _t ik ^\nu _r  \FF (r,s)  \FF (t,r)
\no \\
\cZ_{3e} & = &
- \sum _{[rst]} Z_0 ^{rst} \half K^\mu _r
X^\mu _s K^\nu _r X^\nu _t \FF (r,s)  \FF (r,t)
\no \\
\cZ_4 & = &
- \sum _{[rstu]}  Z_0 ^{rstu} \half K^\mu _r
X^\mu _s K^\nu _t X^\nu _u  \FF (r,s)  \FF (t,u)
\eea
All ingredients have been defined in earlier sections, except
for the {\sl three point functions} $Q_B (s;r,t)$ and $\Phi _B (s;r,t)$,
which will be defined during the construction of these blocks to follow,
where also the presence of the inverse differential $(ds)^{-1}$ will be explained.
The functions $Q_B (s;r,t)$ and $\Phi _B (s;r,t)$ will be constructed explicitly in Section 6.

\subsection{Summary of the blocks of type $\cD$}

Similarly, all the blocks of type $\cD$ with an exact differentials in at least one of the
vertex insertion points are as follows. Using the same
nomenclature as for the $\cZ$ blocks, we have
\bea
\cD [\delta ] & = &
\cD_1 + \cD_{2a} + \cD_{2b} + \cD_{2c} + \cD_{2d} + \cD_{3a} + \cD_{3b}
\no \\ &&
 + \cD_{3c} + \cD_{3d} + \cD_{3e} + \cD_{3f} + \cD_{4a} + \cD _{4b}
\eea
The corresponding individual contributions are given by,
\bea
\label{master2}
\cD_1 & = &
- \sum _r d_r \biggl ( Z_0 ^{r} p^\mu _I \ep _r ^\mu \theta _r
\lambda _I(r) \biggr )
\no \\
\cD_{2a} & = &
+ \sum _{[rs]} d_r d_s \biggl (\half Z_0 ^{rs}
\ep _r ^\mu \ep ^\mu _s \theta _r \theta _s \ff (r,s)
\biggr )
\no \\
\cD_{2b} & = &
- \sum _{[rs]} d_s \biggl ( Z_0 ^{rs}   \ep ^\mu _s
\theta _s X^\mu _r \ff (r,s) \biggr )
\no \\
\cD_{2c} & = &
- \sum _{[rs]} d_s \biggl (
Z_0 ^{rs} p^\nu _I K^\mu _r \ep ^\mu _s \theta _s
\FF (r,s)  K^\nu _r \hat \omega _{I0} (r) \biggr )
\no \\
\cD_{2d} & = &
- \sum _{[rs]} d_s \biggl (Z_0 ^{rs}  {1 \over 2} K^\mu _r K^\nu _r
K^\nu _s K^\mu _s  (ds)^{-1} Q_B (s;r,r) \biggr )
\no \\
\cD_{3a} & = &
- \sum _{[rst]} d_s \biggl (
Z_0 ^{rst} p^\nu _I K^\mu _r \ep ^\mu _s \theta _s
 \FF (r,s)  K^\nu _t \hat \omega _{I0} (t) \biggr )
\nonumber \\
\cD_{3b} & = &
+ \sum _{[rst]} d_s \biggl (
Z_0 ^{rst} \half K^\mu _r K^\nu _t i k^\mu _s \ep ^\nu _s \theta _s
 \FF (r,s)  \FF (t,s) \biggr )
\nonumber \\
\cD_{3c} & = &
- \sum _{[rst]}  d_s \biggl (Z_0 ^{rst} {1 \over 2} K^\mu _r K^\nu _t
K^\nu _s K^\mu _s (ds)^{-1}  Q_B (s;r,t) \biggr )
\no \\
\cD_{3d} & = &
+ \sum _{[rst]}  d_t d_s \bigg (
Z^{rst} _0 \half K^\mu_r \ep ^\mu _s \theta _s \FF (r,s)
K^\nu _r \ep ^\nu _t \theta _t \FF (r,t) \bigg )
\no \\
\cD_{3e} & = &
- \sum _{[rst]}  d_s \bigg (
Z^{rst} _0  K^\mu_r \ep ^\mu _s \theta _s \FF (r,s)
K^\nu _r X ^\nu _t  \FF (r,t) \bigg )
\no \\
\cD_{3f} & = &
- \sum _{[rst]}  d_s \bigg (
Z^{rst} _0  K^\mu_r \ep ^\mu _s \theta _s \FF (r,s)
K^\nu _t ik  ^\nu _r  \FF (t,r) \bigg )
\no \\
\cD_{4a} & = &
+ \sum _{[rstu]}  d_u d_s \bigg (
Z^{rstu} _0 \half K^\mu_r \ep ^\mu _s \theta _s \FF (r,s)
K^\nu _t \ep ^\nu _u \theta _u \FF (t,u) \bigg )
\no \\
\cD_{4b} & = &
- \sum _{[rstu]}  d_s \bigg (
Z^{rstu} _0  K^\mu_r \ep ^\mu _s \theta _s \FF (r,s)
K^\nu _t X ^\nu _u  \FF (t,u) \bigg )
\eea
We stress that each contribution is exact in one or two insertion points only,
but not necessarily  closed in any of the others.
Note that the number of $\ep$'s occurring in each contribution always coincides
with the number of total derivatives in each contribution.

\sm

To derive the above results for $\cZ$ and $\cD$, we proceed by expanding
(\ref{FE2}) in powers of  $\hat \o _{I0}, \lambda _I, \FF, \ff, \Psi, Q_B$ and $\Phi_B$,
inductively on the number of non-holomorphic vertex insertion points.
Since $\l_I, \ff, \Psi , Q_B, \Phi _B$ are of order $\cO (\zeta ^1 \zeta ^2)$,
and $\hat \o _{I0}$ and $\FF$ are of order $\cO(\zeta)$ in the supermoduli,
only terms proportional to $\l_I$, $\ff$, $\Psi $, $Q_B$, $\Phi _B$,
and the bilinears $\hat \o_{I0} \hat \o_{J0}$,
$\hat \o_{I0} \FF$, and $\FF \FF$ must be retained. All others would be at least
trilinear in supermoduli $\zeta$ and must thus vanish.

\subsection{Contributions in $\lambda _I$}

The contributions in $\lambda _I$ can only occur to first order, and  arise
from two sources in (\ref{FE2}): one in $d_r\lambda_I$, and another in $\lambda _I$.
Collecting both contributions yields all the dependence of $\cF [\delta ]$ on $\lambda_I$,
and gives,
\bea
\label{lambda1}
\cF_\lambda = Z_0  \sum _r  p^\mu _I \biggl ( \ep ^\mu _r \theta _r d_r \lambda
_I (r) + i  k^\mu _r \lambda _I (r) \biggr )
\eea
The term in $Z_0 d_r \lambda_I$ is not a total differential in $r$, because
$Z_0$ itself has non-trivial $r$-dependence. We wish to isolate a total
differential which includes also $Z_0$. Furthermore, $Z_0$ contains
a contribution from $\ep ^\mu _r \theta _r \p_r \bar x _+ ^\mu (r) $, which
provides with a further differential form contribution of type $(1,0)$.
It will be advantageous to exhibit this contribution, and then partially
Wick contract it. To this end, we use
\bea
\label{Z01}
Z_0 = Z_0 ^r \bigg  ( 1 + \ep ^\mu _r \theta _r \p_r \bar
x^\mu _+ (r) \biggr )
\eea
This relation is an exact Taylor expansion in view of the fact that
the term $\ep ^\mu _r \theta _r \p_r \bar x _+ ^\mu (r) $ is linear in
$\theta _s$, so that its square vanishes. Using this identity in (\ref{lambda1}),
we obtain,
\bea
\label{Flambda}
\cF _\lambda & = &
\sum _r Z_0 ^r p^\mu _I \biggl ( \ep ^\mu _r \theta _r d_r
\lambda _I (r) + i  k^\mu _r \lambda _I (r)
+ \ep ^\nu _r \theta _r \p_r \bar x^\nu _+ (r) i  k^\mu _r
\lambda _I (r)
\biggr )
\no \\ & = &
- \sum _r d_r \biggl (Z_0 ^r p^\mu _I \ep ^\mu _r \theta _r \lambda _I(r) \biggr )
\no \\ &&
+ \sum _r Z_0 ^r  p_I ^\mu \biggl (
i k^\nu _r  \p_r \bar x^\nu _+(r) \ep ^\mu _r \theta _r
+i k^\mu _r + \ep ^\nu _r \theta _r \p_r \bar x^\nu _+ (r) i k^\mu_r
\biggr ) \lambda _I (r)
\nonumber
\eea
Note that care is required in properly ordering standard differential 1-forms such as
$d_r$ and $\p_r$, with respect to to $\ep _r ^\mu$, which is itself a differential
1-form. This ordering is responsible for the minus sign multiplying the first
term on the right hand side of (\ref{Flambda}). Finally, in differentiating $Z_0$,
we have used the relation
\bea
d_r \biggl (Z_0 ^r  \ep ^\mu _r \theta _r  \biggr )
= Z_0 ^r \biggl ( i k^\nu _r  \p_r \bar x^\nu _+(r) \ep ^\mu _r \theta _r \bigg )
\eea
Note that the contributions from the Szeg\"o kernel in $Z_0$ cancel out in this
differentiation, in view of the presence of the factor $\ep^\mu _r \theta_r$,
and the fact that $(\theta _r)^2=0$, as well as the fact that the polarization vectors
$\ep _r ^\mu$ cannot occur to an order higher than first. Finally, making use of the
definition of $K^\mu _r $, we may rearrange the  terms in $\cF_\lambda$ as follows,
\bea
i k^\nu _r  \p_r \bar x^\nu _+(r) \ep ^\mu _r \theta _r
+i k^\mu _r + \ep ^\nu _r \theta _r \p_r \bar x^\nu _+ (r) i k^\mu_r
& = &
i k^\mu _r - i (   k^\nu _r   \ep ^\mu _r - \ep ^\nu _r   k^\mu_r)
\theta _r \p_r \bar x^\nu _+(r)
\no \\ & = &
i k^\mu _r -  K_r ^\mu K_r ^\nu (dr)^{-1} \p_r \bar x^\nu _+(r)
\no \\
& = & X^\mu _r
\eea
As a result, we have
\bea
\label{Flambda1}
\cF _\lambda =
- \sum _r d_r \biggl (Z_0 ^r p^\mu _I \ep ^\mu _r \theta _r \lambda _I(r) \biggr )
+ \sum _r Z_0 ^r  p_I ^\mu X^\mu _r  \lambda _I (r)
\eea
The first term on the right hand side produces $\cD_1$ of (\ref{master2}),
while the second term produces $\cZ_{1a}$ of (\ref{master1}).

\subsection{Contributions bilinear in $\hat \omega _{I0}$}

By expanding the exponential of the terms in $\hat \o _{I0}$ to
second order, one obtains bilinears in $\hat \o_{I0}$, which may be grouped
into contributions at coincident points $r$, which yield $\cZ_{1b}$,
and contributions at distinct points $r$ and $s$, which
yield $\cZ_{2a}$. No $\cD$-terms appear.

\subsection{Contributions in $ \ff $}

The contributions in $ \ff (r,s)$ can occur to first order only, and
arise in (\ref{FE2}) from 3 sources: one term in $\ff$, one term in
$d_r \ff$ and one term in $d_r d_s \ff$,
\bea
\cF _\ff =  Z_0  \sum _{[r,s]} \biggl (
-\half k^\mu _r k^\mu _s  \ff (r,s)
+ i k^\mu _r \ep ^\mu _s \theta _s d_s  \ff (r,s)
+ \half \ep ^\mu _r \ep ^\mu _s \theta _r \theta _s d_r d_s  \ff (r,s) \biggr )
\eea
We proceed as in  the derivation of the contributions in
$\lambda _I$, except that this time we need to work in both variables
$r$ and $s$. The generalization of (\ref{Z01}) is given by
\bea
\label{Z02}
Z_0 = Z_0 ^{rs}
\bigg  ( 1 + \ep ^\mu _r \theta _r \p_r \bar x^\mu _+ (r) \biggr )
\bigg  ( 1 + \ep ^\nu _s \theta _s \p_s \bar x^\nu _+ (s) \biggr )
\eea
As a result, we have
\bea
\cF _\ff & = &
\sum _{[r,s]} Z_0 ^{rs} \biggl (1 + \ep ^\mu _r \theta _r \p_r \bar
x^\mu _+ (r) \biggr ) \biggl (1 + \ep ^\nu _s \theta _s \p_s \bar
x^\nu _+ (s) \biggr )
\\ && \qquad \times
\biggl (
-\half k^\mu _r k^\mu _s  \ff (r,s) + i k^\mu _r \ep ^\mu _s \theta _s d_s \ff (r,s)
 + \half \ep ^\mu _r \ep ^\mu _s \theta _r \theta _s d_r d_s \ff (r,s) \biggr )
\no \eea
The single derivative term is handled as $\lambda _I$ was,
\bea
Z_0 ^{rs} i k^\mu \ep ^\mu _s \theta _s d_s  \ff (r,s)
& = & - d_s \Big (Z_0 ^{rs} i k^\mu _r \ep ^\mu _s \theta _s \ff (r,s) \Big )
\no \\ &&
+ i k^\nu _s \p_s \bar x ^\nu _+(s) i k^\mu _r \ep ^\mu _s
\theta _s d_s  \ff (r,s) Z_0 ^{rs}
\nonumber
\eea
while the double derivative must be dealt with twice, and produces
\bea
Z_0 ^{rs} \half \ep ^\mu _r \ep ^\mu _s \theta _r \theta _s d_r d_s \ff (r,s)
 & = &
+ d_r \biggl (Z_0 ^{rs} \half \ep ^\mu _r \ep ^\mu _s \theta _r
\theta _s  d_s  \ff (r,s) \biggr )
\nonumber \\ &&
+ d_s \biggl (Z_0 ^{rs} \half i k^\nu _r \p_r \bar x^\nu _+ (r) \ep
^\mu _r \ep ^\mu _s \theta _r \theta _s  \ff (r,s) \biggr )
\nonumber \\ &&
- Z_0 ^{rs} \half i k^\sigma _s \p_s \bar x^\sigma _+ (s) i k^\nu _r \p_r
\bar x^\nu _+ (r) \ep ^\mu _r \ep ^\mu _s \theta _r \theta _s
 \ff (r,s)
\nonumber
\eea
The same technique is now reiterated on the first term on the
right hand side to bring out the differential $d_s$ as well.
Putting together all the {\sl non-derivative terms}, and
re-expressing the results in terms of $K$ factors, we obtain
\bea
\sum _{[rs]} Z_0 ^{rs}  \biggl (
- \half k^\mu _r k^\mu _s - i K^\mu _s K^\nu _s k^\mu _r \p_s \bar x^\nu
_+(s) + \half K^\mu _r K^\sigma _r K^\nu _s K^\sigma _s \p_r \bar x ^\mu
_+(r) \p_s \bar x ^\nu _+(s) \biggr ) \ff (r,s)
\eea
Expressing the terms in parentheses in terms of $X^\mu _r X^\mu _s$,
produces the contribution $\cZ _{2b}$. Collecting the {\sl exact differential
terms}, we get
\bea
\label{eff}
-\sum _{[rs]} d_s \biggl ( Z_0 ^{rs}  \ep _s ^\mu \theta _s X^\mu _r  \ff (r,s) \biggr )
+ \sum _{[rs]} d_r d_s \biggl (\half  Z_0 ^{rs}  \ep ^\mu _r
\ep ^\mu _s \theta _r \theta _s  \ff (r,s) \biggr )
\eea
which produces the terms ${\cal D}_{2b}+{\cal D}_{2a}$.

\subsection{Contributions in $\Psi $}

There is only a single source for this contribution in (\ref{FE2}),
and its expansion to first order readily gives $\cZ_{2c}$.

\subsection{Contributions in  $\hat \o _{I0} \FF$}

Terms in $\FF$ may arise to first order, in which case they must
ultimately be multiplied by one power of $\hat \omega _{I0}$, or to
second order. Here, we work out the first order case, leaving
the case of second order to the next subsection. The double
expansion, to first order in $\hat \o_{I0}$ and to first order
in $\FF$ produces the following contributions,
\bea
\cF _{\hat \o \FF} =
\sum _{[rs]} \sum _t Z_0 K^\mu _r \biggl (i  k^\mu _s  \FF (r,s)
+ \ep ^\mu _s \theta _s d_s \FF (r,s) \biggr ) p^\nu _I K^\nu _t \hat \o _{I0} (t)
\eea
Since $\ep ^\rho _r \theta _r K_r ^\mu =0$, a simplification occurs in
the use of formula (\ref{Z02}), and we have
\bea
Z_0 K^\mu _r = Z_0 ^{rs} K^\mu _r
\bigg  ( 1 + \ep ^\nu _s \theta _s \p_s \bar x^\nu _+ (s) \biggr )
\eea
which yields
\bea
\cF _{\hat \o \FF} =
\sum _{[rs]} \sum _t Z_0^{rs}  K^\mu _r  \biggl (i k^\mu _s  \FF (r,s) +
\ep ^\mu _s \theta _s d_s  \FF (r,s)
+ i k^\mu _s  \FF (r,s) \ep ^\nu _s \theta _s \p_s \bar x^\nu _+(s)   \biggr )
p^\nu _I K^\nu _t \hat \o _{I0} (t)
\nonumber
\eea
Using the following rearrangement of the derivative term,
\bea
Z_0 ^{rs} K^\mu _r \ep ^\mu _s \theta _s d_s  \FF (r,s)
=
-d_s \biggl ( Z_0 ^{rs} K^\mu _r \ep ^\mu _s \theta _s  \FF (r,s)
\biggr ) + i k^\nu _s \p _s \bar x^\nu _+ (s) Z_0 ^{rs} K^\mu _r \ep
^\mu _s \theta _s \FF (r,s)
\nonumber
\eea
we isolate an exact differential as well as the non-derivative term as follows,
\bea
\label{lineareff}
\cF _{\hat \o \FF} ^{(1)} & = &
- \sum _{[r,s]} \sum _t d_s \biggl ( Z_0 ^{rs}
p^\nu _I K^\mu _r \ep ^\mu _s \theta _s  \FF (r,s)
 K^\nu _t \hat \omega _{I0}(t) \biggr )
\nonumber \\
\cF _{\hat \o \FF} ^{(2)} & = &
+ \sum _{[r,s]} \sum _t Z_0 ^{rs}
K^\mu _r X^\mu _s  \FF (r,s) p^\nu _I K^\nu _t \hat \omega _{I0}(t)
\eea
In the sums defining $ \cF _{\hat \o \FF} ^{(1)}$, the point $t$ can never coincide with
the point $s$, because the corresponding kinematic factor $\ep ^\mu _s \theta _s
K^\nu _t$ vanishes. But $t$ is allowed to coincide with the point $r$, and this
contribution yields the term $\cD_{2c}$. The remaining contributions are when
$t$ is different from $r$ and $s$, and they produce the term $\cD_{3a}$ of
(\ref{master2}).

\sm

In the sums defining $ \cF _{\hat \o \FF} ^{(2)}$, the point $t$ may coincide with
the point $r$ or with the point $s$, yielding the two different terms in $\cZ_{2d}$.
The remaining contributions are when $t$ is different from both $r$ and $s$,
and they produce the term $\cZ_{3a}$ of (\ref{master1}).

\subsection{Contributions in  $\FF \FF$ and $\tilde K \tilde K$}

The contributions bilinear in $\FF$ are by far the most involved kinematically,
and they will also force us to introduce new basic functions $Q_B$ and $\Phi_B$
in addition to the building blocks $\hat \o_{I0}$, $\lambda _I$, $\FF, \ff$, and
$\Psi $. A specific part of the contributions bilinear in $\FF$ will naturally
combine with the contribution from the $(\tilde K \tilde K - KK)$ term in (\ref{FE2}).
The total contribution of the terms bilinear in $\FF$ is given by,
\bea
\cF _{FF} & = & \half Z_0 \sum _{[rs]}
\biggl (i K^\mu _r k^\mu _s  \FF (r,s) + K^\mu _r \ep ^\mu _s
\theta _s d_s  \FF (r,s) \biggr )
\no \\ && \hskip 0.4in \times
\sum _{[tu]}
\biggl (i K^\nu _t k^\nu _u  \FF (t,u) + K^\nu _t \ep ^\nu _u
\theta _u d_u  \FF (t,u) \biggr )
\nonumber
\eea
We split this quadruple sum into three parts,
\bea
\cF _{FF} = \cF _{FF} ^{(1)} + \cF _{FF} ^{(2S)} + \cF _{FF} ^{(2A)}
\eea
Here, $\cF _{FF} ^{(1)}$ is the part of $\cF _{FF} $ for which $s\not= u$,
while $\cF _{FF} ^{(2S)} + \cF _{FF} ^{(2A)}$ is the part of $\cF _{FF}$ for which $s=u$.
The last contribution is further separated into $\cF _{FF} ^{(2S)} $ which is
symmetric under interchange of $r$ and $t$, and $\cF _{FF} ^{(2A)} $
which is anti-symmetric under interchange of $r$ and $t$.
The reason for separating the $u \not= s$ from the $u=s$ parts is that in
$\cF _{FF} ^{(1)}$,
the differentials in $s$ and in $u$ can be pulled out without hitting
$t$ or $r$ respectively, while they cannot in $\cF _{FF} ^{(2S)}$ or
$\cF _{FF} ^{(2A )}$. The reason for separating into symmetric and anti-symmetric
parts under $r \leftrightarrow t$ will be explained later.

\subsubsection{Calculation of $\cF _{FF} ^{(1)}$}

The contributions in $\cF _{FF} ^{(1)}$ are naturally gotten by iterating the splitting
off the exact differential twice for the result gotten for the terms linear in $\FF $,
\bea
\label{quadFI}
\cF _{FF} ^{(1)} & = & + \half \sum _{[rs]} \ \sum _{[tu], s\not= u} Z_0 ^{rt}
K^\mu _r X^\mu _s \FF (r,s) K^\nu _t X^\nu _u  \FF (t,u)
\no \\ &&
+ \half \sum _{[rs]} \ \sum _{[tu], s\not= u}
d_u d_s \biggl ( Z_0 ^{rstu} K^\mu _r \ep _s ^\mu \theta _s
 \FF (r,s) K^\nu _t \ep _u ^\nu \theta _u  \FF (t,u) \biggr )
\no \\ &&
- \sum _{[rs]} \ \sum _{[tu], s\not= u} d_s \biggl (
Z_0 ^{rst} K^\mu _r \ep _s ^\mu \theta _s  \FF (r,s) K^\nu _t
X^\nu _u  \FF (t,u) \biggr )
\eea
In the first term of (\ref{quadFI}), the summation instructions for the points $r,s,t,u$
leave the following options.

\begin{itemize}
\item Either $u=r$, for which there are two possibilities.
Either $t=s$ as well, which produces the term $\cZ_{2f}$ of (\ref{master2});
or $t\not= s$, which produces instead the term
\bea
-\sum_{[rst]}Z_0^{rst}{1\over 2}K_r^\mu X_s^\mu K_t^\nu X_r ^\nu \FF (r,s) \FF (t,r)
\eea
After a further simplification of this term, in view of the fact that $K_r^\mu X_r ^\nu
= K_r^\mu i k_r ^\nu$, {\sl half of the contribution} $\cZ_{3d}$ of (\ref{master1})
is produced.
\item Or $u\not= r$, for which there are three possibilities. The first is $t=s$,
which produces the other half of the contribution $\cZ _{3d} $ of (\ref{master1}).
The second is $t=r$, which produces $\cZ_{3e}$. The third is $t\not=r,s$,
which produces $\cZ_4$.
\end{itemize}

In the second term in (\ref{quadFI}), the summation instructions for the points $r,s,t,u$
leave only two possibilities. In view of the structure of the kinematic factors,
we are forced to have $u \not= r$ and $t \not= s$. Thus we can have that either all
4 points are pairwise distinct, yielding $\cD_{4a}$, or $t=r$, yielding $\cD_{3d}$.

\sm

In the third term in (\ref{quadFI}), the summation instructions for the points $r,s,t,u$
leave only three possibilities. In view of the kinematic factors, we are forced to have
$s \not= t$. The first possibility is to have all  4 points are pairwise distinct,
yielding $\cD_{4b}$; the second possibility is $t=r$, yielding $\cD_{3e}$;
and the third possibility is to have $u=r$ yielding $\cD_{3f}$.

\subsubsection{Calculation of $\cF _{FF} ^{(2A)}$}

Next, we calculate $\cF _{FF} ^{(2A)}$, which may be rewritten as
\bea
\cF _{FF} ^{(2A)} =
\half  \sum _{[rst]} Z_0 ^{rt} K^{\{ \nu} _t K^{\mu \} } _r
\biggl (i  k^\mu _s  \FF (r,s) +  \ep ^\mu _s \theta _s d_s  \FF (r,s) \biggr )
\biggl (i  k^\nu _s  \FF (t,s) +  \ep ^\nu _s \theta _s d_s \FF (t,s) \biggr )
\quad
\eea
Splitting off a single exact differential term, we obtain,
\bea
\label{quadFIIA}
\cF _{FF} ^{(2A)}
& = &
- {i \over 2} \sum _{[rst]} Z_0 ^{rst} K^\mu _r K^\nu _t
k^\mu _s X^\nu _s  \FF (r,s)  \FF (t,s)
\no \\
&&
+ \sum _{[rst]} d_s \biggl ( \half Z_0 ^{rst} K^\mu _r K^\nu _t
i k^\mu _s \ep _s ^\nu \theta _s  \FF (r,s)  \FF (t,s)
\biggr )
\eea
The first and second terms produce $\cZ_{3b}$ and $\cD_{3b}$ respectively.
In both terms, anti-symmetry under exchange of $r$ and $t$ is guaranteed
by the anti-symmetry of $\FF (r,s) \FF (t,s)$; it renders the tensor
structure symmetric under interchange of $\mu$ and $\nu$.

\subsubsection{Contributions from $ (\tilde K \tilde K - KK)$}

Before turning to the contributions of $\cF _{FF} ^{(2S)}$, we work out the
contribution $\cF _{\tilde K \tilde K}$ arising from the differences between
$(\tilde K_r ^\mu  \tilde K ^\mu _s  - K_r^\mu K ^\mu _s) S (r,s)$.
This is because exact differentials can only be extracted after these terms are
recombined with the terms from $\cF _{FF} ^{(2S)}$.
This contribution is given by,
\be
\cF _{\tilde K \tilde K} =
- \sum _{[rs]} Z_0 ^{rs} K^\mu _r \ep ^\mu _s \mu (s) d\bar s S (r,s)
\ee
At the point $s$, there must be brought down an additional $K_s$
factor from a fermion propagator term. (One cannot bring down a factor of
$\hat \omega _{I0}$ because this would be third order in $\chi$ and thus
vanish.) Doing this in all possible ways gives rise to
\be
\cF _{\tilde K \tilde K} =
- \sum _{[rs]} \sum _{t \not= s} Z_0 ^{rs} K^\mu _r \ep ^\mu _s \mu (s)
d\bar s S (r,s) K^\nu _t K^\nu _s S (t,s)
\ee
This term is a $(0,1)$ form in $s$, and may be rearranged using
$K^\nu _s \ep ^\mu _s = - i \ep ^\mu _s \theta _s k_s ^\nu$.
Upon symmetrization of $r$ and $t$, $\mu$ and $\nu$ are anti-symmetrized
and we get
\bea
\label{fermionicdifference}
\cF _{\tilde K \tilde K} & = &
\sum _{[rs]} \sum _{t \not= s} \half Z_0 ^{rs} K^\mu _r K^\mu _s K^\nu _s K^\nu _t
\mu (s) {d\bar s \over ds} S (r,s)  S (t,s)
\no \\ & = &
\sum _{[rs]}  \half Z_0 ^{rs} K^\mu _r K^\mu _s K^\nu _s K^\nu _r
\mu (s) {d\bar s \over ds} S (r,s)  S (r,s)
\no \\ &&  +
\sum _{[rst]}  \half Z_0 ^{rst} K^\mu _r K^\mu _s K^\nu _s K^\nu _t
\mu (s) {d\bar s \over ds} S (r,s)  S (t,s)
\eea
where we have separated the contributions from two and three points.
Both contributions will be combined with those of $\cF _{FF} ^{(2S)}$
in the next subsection.

\subsubsection{Contributions from $\cF _{FF} ^{(2S)}$}

We can return now to the term $\cF _{FF} ^{(2S)}$. We shall extract an
exact differential by recombining it with the preceding contributions from
the differences  $(\tilde K_r^\mu\tilde K_s^\mu - K^\mu _r K^\mu _s) S(r,s)$.
\bea
\cF _{FF} ^{(2S)} & = &
\half  \sum _{[rs]} \sum _t Z_0 ^{rs} K^{[ \nu} _t K^{\mu ] } _r
\biggl (i  k^\mu _s  \FF (r,s) +  \ep ^\mu _s \theta _s d_s  \FF (r,s) \biggr )
\no \\ && \hskip 1in \times
\biggl (i  k^\nu _s  \FF (t,s) +  \ep ^\nu _s \theta _s d_s \FF (t,s) \biggr )
\quad
\eea
In the product, the contribution in $\FF (r,s) \FF (t,s)$ cancels out since
$K^{[ \nu} _t K^{\mu ] } _r  k^\mu _s k^\nu _s=0$, while the contribution
in  $d_s \FF (r,s) d_s\FF (t,s)$ cancels in view of $(\theta _s)^2=0$.
As a result, we are left with the  following contributions,
\bea
\label{quadFIIS}
\cF _{FF} ^{(2S)} =
{1 \over 4} \sum _{[rs]} \sum _t Z_0 ^{rst} K^\mu _r K^\nu _t K^\mu _s K^\nu _s (ds)^{-1}
\biggl (  \FF (r,s) d_s  \FF (t,s)  - d_s  \FF (r,s) \FF (t,s)  \biggr )
\eea
The terms in parentheses cannot be recast as a total differential
in terms of the basic building blocks $\FF$ and $\ff$ only. To extract
an exact differential requires introducing an intrinsic three-point function.

\sm

In the next section, it will be shown that there exists an intrinsic three
point function $Q_B(s;r,t)$ which satisfies the following $\bar \p$ equation,
\bea
\label{Qequation1}
- 2 \p_{\bar s} Q_B (s;r,t)
=  \FF (r,s) \p_{\bar s} \FF (t,s) - \p_{\bar s} \FF(r,s) \FF (t,s)
+ 2 \mu (s) S(r,s) S(t,s)
\eea
Thanks to the presence of the last term in $\mu (s)$, it will turn out
that the function $Q_B(s;r,t)$ is well-defined in $s,r$ and $t$.
This will be shown in the next section, where an explicit form for
$Q_B(s;r,t)$ will also be calculated. Note that the $\mu(s)$ term
needed in the definition of $Q_B$ is precisely provided by the terms
$\cF _{\tilde K \tilde K}$ computed in the preceding subsection.

\sm

In view of the definition of $Q_B(s;r,t)$, the term (\ref{fermionicdifference})
and the $d\bar s$ terms of $\cF _{FF} ^{(2S)}$ combine into a single $\bar \p_s$
differential of type $(0,1)$ in $s$,
\be
-\half \bar \p_s \sum_{[rst]}\bigg(Z_0^{rst}K_r^\mu K_t^\nu K_s^\nu K_s^\mu
(ds)^{-1}\, Q_B(s;r,t)\bigg)
\ee
We can express $\bar \p_s$ as $\bar \p_s=d_s-\p_s$, and use the $d_s$
part to isolate a total differential in $s$. The remaining terms are pure $(1,0)$ in $s$.
They may be regrouped by the use of another intrinsic three-point functions
$\Phi _B(s;r,t)$, defined by,
\bea
\Phi _B(s;r,t) \equiv \half \p_s \FF (t,s) \FF (r,s) + \half \p_s \FF (r,s) \FF (t,s)
- \p _s Q_B(s;r,t)
\eea
Combining all contributions of $\cF _{FF} ^{(2S)}$ and $\cF _{\tilde K \tilde K}$,
using $Q_B$ and $\Phi_B$, we find
\bea
\cF _{FF} ^{(2S)} + \cF _{\tilde K \tilde K}
&=&
\cZ_{2e}+ \cZ_{3c}+ \cD _{2d}+ \cD_{3c}
\eea
where we have separated out the contributions with two and three points.
This completes the proof of the decomposition of chiral blocks $\cF [\delta]$
into the blocks $\cZ[\delta]$ and $\cD [\delta]$, listed in (\ref{master1}) and
(\ref{master2}).

\newpage

\section{Solving $\bar\partial$ Equations with Monodromy}
\setcounter{equation}{0}

In the derivation of a holomorphic representative for the cohomology
class of the chiral blocks of the $N$-point function, we shall need several
key quantities which are solutions of $\bar\p$ equations where the
right hand side has monodromy. A basic example is the quantity
$Q_B(s;r,t)$ defined by the following equation
\bea
\label{Qequation}
\p_{\bar s} Q_B (s;r,t)
= \half \p_{\bar s} \FF(r,s) \FF (t,s) + \half \p_{\bar s} \FF (t,s) \FF (r,s)
- \mu (s) S(r,s) S(t,s)
\eea
subject to the requirement that $Q_B(s;r,t)$ have no monodromy in $r$ and in $t$.
The main difficulty stems from the fact that the right-hand-side of this equation
has non-trivial monodromy in $s$ since $\FF(r,s)$ and $\FF(t,s) $ do.
The existence and properties of quantities such as $Q_B(s;r,t)$ are novel features
in the function theory of Riemann surfaces, and have to be treated with some care.

\sm

In fact, the monodromy on the right-hand-side may be traced back to the monodromy
of the scalar Green function in the definition of $\FF$. This allows the problem to be
reduced to the study of a more basic quantity $Q_0(s;x,y)$, which is defined to obey,
\bea
\label{defL}
\p _{\bar s} Q_0(s;x,y) = \half \p_{\bar s} G(x,s) G(y,s) - \half \p_{\bar s} G(y,s) G(x,s)
\eea
The solution $Q_0(s;x,y)$ to this equation is a scalar function in $s$ with monodromy,
which remains to be specified. In the variables $x$ and $y$, $Q_0(s;x,y)$ must be a
$(1,0)_x \otimes (1,0)_y$ form with vanishing monodromy (so that the integrals
over $\Sigma$, needed to obtain $\FF$ in (\ref{Qequation}), will be well-defined).
Naturally, we also require that it be odd under interchange of $x$ and $y$,
namely $Q_0(s;y,x)= - Q_0(s;x,y)$.
The solution of this equation  then immediately allows us to solve for $Q_B$ in
(\ref{Qequation}), and
we have
\bea
Q_B(s;r,t) & = & {1 \over 16 \pi^2} \int _x \int _y S(r,x) S(t,y) \chix \chiy Q_0(s;x,y)
\no \\ &&
+ {1 \over 2 \pi} \int _x G(x,s) \mu (x) S(r,x) S(t,x).
\eea
All other cases needed here of $\bar \p$ equations whose right-hand-side has
monodromy can be
reduced and solved with the help of the function $Q_0(s;x,y)$. Before setting out to solve
for $Q_0(s;x,y)$, some simple considerations on monodromy will be reviewed first.

\subsection{The simplest $\bar\p$ equation}

Let $\varpi =\varpi _{\bar s}d\bar s$ be a single-valued  $(0,1)$-form on the
compact Riemann surface $\Sigma$. A basic fact of function theory on Riemann
surfaces is that the equation
\be
\label{peq}
\bar\p_s f = \varpi
\ee
admits a single-valued smooth solution $f$ on $\Sigma$ (unique up to an additive constant)
if and only if $\varpi $ satisfies the orthogonality conditions
\be
\label{ort}
\int_\Sigma \o_I \wedge \varpi =0
\ee
for any holomorphic Abelian differential $\o_I$ on $\Sigma$. The equation may then
be solved explicitly in terms of the scalar Green function $G$,
\bea
\label{fsol}
f(s) = - {1 \over 2 \pi} \int _\Sigma d^2 x\,  G(x,s) \varpi _{\bar x}
\eea
When $\varpi$ is single-valued, but the orthogonality condition (\ref{ort}) is
violated, the function $f$ defined in (\ref{fsol}) still solves (\ref{peq}), but $f$ now has
non-trivial monodromy in $s$,
\bea
\label{monf}
f(s+A_K)&=& f(s)\nonumber\\
f(s+B_K)&=& f(s)+ \int_\Sigma \o_K\wedge \varpi
\eea
As a result, the solution $f$ should be viewed as defined on the universal covering
space of $\Sigma$ (in fact, on its quotient $\hat\Sigma$ by the subgroup
of $\pi_1(\Sigma)$ generated by commutators).

\sm

It should be stressed that, if functions $f$ with monodromy are allowed,
the solution $f$ is no longer unique. Since one can add to any solution $f$ of
(\ref{peq}) an arbitrary holomorphic Abelian integral on $\hat\Sigma$ and
obtain a new solution, we cannot have uniqueness without some
a priori constraints on the monodromy of $f$. Clearly, solutions $f$
will be unique (up to constants) if their monodromies on both $A_K$ and $B_K$
cycles are fixed. A useful observation is that, under the assumption that
the monodromies of $f$ are all {\sl constant},  the $A$-monodromies
determine the $B$-monodromies. Indeed, by subtracting a suitable linear
combination of Abelian integrals, we may assume that the $A$-monodromies
of $f$ are all $0$. Then the $B$-monodromies of $f$ are determined by
\bea
\int_\Sigma\o_L\wedge \varpi =\int_\Sigma\o_L\wedge df
=
\sum_K\oint_{A_K} \! \o_L \Big ( f(z+B_K)-f(z) \Big )
= f(z+B_L)-f(z),
\eea
which are indeed the monodromies exhibited by the solution (\ref{monf}).
Thus, once a canonical homology basis $(A_K,B_K)$ has been chosen, we may view
the solution with vanishing $A$-monodromies as the canonical choice among solutions
with constant monodromies.

\sm

Finally, the case of interest here is when $\varpi$ itself has monodromy, so that
it becomes more appropriate to view it as a differential form on $\hat \Sigma$.
The solution (\ref{fsol}) given above in terms of {\sl surface integrals} does not
extend to the case where $\varpi$ has monodromy, since the resulting $f$
would then depend upon the fundamental domain chosen to represent
$\Sigma$ in its covering space. But it is possible to solve (\ref{peq}) in terms
of {\sl line integrals} on the cover space $\hat \Sigma$.
One proceeds by completing the $\bar \p$ equation of (\ref{peq})
into a total differential equation for $f$,
\bea
\label{deq}
d_s f = \varpi + \rho
\eea
where $\rho  = \rho _s ds$ is required to be a pure $(1,0)$ form. Clearly, the $(0,1)$
component of (\ref{deq}) coincides with $\bar \p_s f = \varpi$, and the $(1,0)$
component simply determines $\rho$. Since $\varpi$ has monodromy on $\Sigma$,
and is properly defined only on $\hat \Sigma$, the entire equation (\ref{deq})
is viewed on $\hat \Sigma$. Given $\varpi$ and a differential form $\rho$
which satisfies the integrability condition,
\bea
0 = d_s \varpi + d_s \rho \hskip 0.5in \Leftrightarrow \hskip 0.5in
0 = \p_s \varpi _{\bar s} - \p_{\bar s} \rho _s
\eea
the unique solution $f$ on $\hat \Sigma$ is obtained  in terms
of a line integral,
\bea
f(s)  =  f(s_0) + \int ^s _{s_0} (\varpi  + \rho)
\eea
The form $\rho$ is, of course, not unique, since the addition to $\rho$ of
any holomorphic differential on $\hat \Sigma$ will produce again a solution
to the integrability condition.

\sm

The  non-trivial monodromy of $\varpi$ results in non-trivial $s$-dependence of the
monodromy of $f$. To see this, we compute the monodromy of $f$ around any
homology cycle $C$,
\bea
\label{genmon}
f(s+C) - f(s) & = & \int _{s_0} ^{s+C} (\varpi + \rho)(t) - \int _{s_0} ^{s} (\varpi + \rho)(t)
\\ &=&
\int _{s_0+C} ^{s+C} (\varpi + \rho)(t) - \int _{s_0} ^{s} (\varpi + \rho)(t)
+ \int _{s_0} ^{s_0+C} (\varpi + \rho)(t)
\no \\ & = &
 \int _{s_0} ^{s} \bigg ( (\varpi + \rho)(t+C)  - (\varpi + \rho)(t) \bigg )
+ \int _{s_0} ^{s_0+C} (\varpi + \rho)(t) \no
\eea
Thus, the $s$-dependence of the monodromy of $f$ is given by the line
integral of the monodromy of the form $\varpi + \rho$. The $s$-independent
part of the monodromy, given by the last integral, is generally more complicated
to compute explicitly, but it will fortunately not be needed here. In the case of the
function $Q_0(s;x,y)$, the form $\rho$ will in fact be monodromy-free.

\subsection{Solving the integrability equation for $Q_0(s;x,y)$}

The methods outlined in the preceding subsection will now be applied to the calculation
of the function $Q_0(s;x,y)$, which is required to be single-valued on $\Sigma$
in $x$ and $y$. The defining equation (\ref{defL}) may be simplified
by carrying out the $\bar \p$ derivatives on the Green function, and we obtain,
\bea
\label{deltaL}
\p _{\bar s} Q_0(s;x,y) = - \pi \delta (s,x) G(y,s) + \pi \delta (s,y) G(x,s)
\eea
The presence of only $\delta$-functions on the right hand side reveals that
$Q_0(s;x,y)$ is holomorphic in $s$ away from $x$ and $y$, making it natural to
seek solutions  for $Q_0$ by meromorphic line integrals.
The first step is to complete the $\bar \p_s$
equation (\ref{defL}) into a $d_s$ equation,
\bea
\label{dsL}
d_s Q_0(s;x,y) = \half d_s G(x,s) G(y,s) - \half d_s G(y,s) G(x,s) + \rho (s;x,y)
\eea
where $\rho(s;x,y)$ is required to be a pure $(1,0)$ form in $s$, and a single-valued
$(1,0)$ form in $x$ and $y$. The integrability condition on $Q_0$  is,
\bea
0 = d_s G(x,s) \wedge d_s G(y,s) + ds \wedge d\bar s \, \p_{\bar s} \rho _s (s;x,y)
\eea
It is a special property of the problem at hand that the first term has vanishing
monodromy and vanishing $s$-intgeral over $\Sigma$. As a result, the
$\bar \p _s \rho _s$ equation admits a solution $\rho_s$ with vanishing monodromy
in $s$,
\bea
\rho _s (s;x,y) & = &
- {1 \over 2 \pi} \int _w G(s,w) \bigg (
\p_w G(x,w) \p_{\bar w} G(y,w) - \p_w G(y,w) \p_w G(x,w) \bigg )
\no \\
&& + \sum _I c_K(x,y) \o _I (s)
\eea
where $c_I(x,y)$ are two as yet arbitrary functions of $x$ and $y$ only,
subject to the condition $c_I(y,x)= - c_I(x,y)$.
The $w$-integral may be carried out, and we obtain,
\bea
\rho _s (s;x,y) =  \bigg ( G(s,y) - G(s,x) \bigg )\p_x \p_y \ln E (x,y) + \sum _I c_I(x,y) \o _I (s)
\eea
where we have used the identity $\p_y G(x,y) = \p_x G(y,x) = \p_x \p_y \ln E(x,y)$.
To summarize, we have the following set of two integrable equations for $Q_0$,
\bea
\label{pLeq}
\p_s Q_0(s;x,y) & = & \half \p_s G(x,s) G(y,s) - \half \p_s G(y,s) G(x,s)
\no \\ && +
\bigg ( G(s,y) - G(s,x) \bigg )\p_x \p_y \ln E (x,y) + \sum _I c_I (x,y) \o _I (s)
\no \\
\p _{\bar s} Q_0(s;x,y) & = & - \pi \delta (s,x) G(y,s) + \pi \delta (s,y) G(x,s)
\eea
An essential property of $\p_s Q_0$ is its behavior as $s\rightarrow x$,
which may be read off from the $\p_s Q_0$ equation above, and is given by,
\be
\p_s  Q_0(s;x,y) = +  {1/2 \over (s-x)^2} G(y,x) + \cO ( 1 )
\ee
(and similarly as $s\rightarrow y$). The fact that $\p_s Q_0(s;x,y)$ has only double poles
(no residues) at $s=x$ and $s=y$ is what allows $\p_s Q_0$ to be integrated
unambiguously along any path which avoids these singularities,
\bea
Q_0(s;x,y)
&=&
\int^s\bigg[{1\over 2}d_sG(x,s)G(y,s)-{1\over 2}d_s G(y,s)G(x,s)\\
&&
\qquad -\bigg(G(s,x)-G(s,y)\bigg)\p_x\p_y\ln \,E(x,y)\bigg]
+
\sum_I c_I (x,y)\int^s\o_I(s)
\nonumber
\eea
In particular,
we have
\bea
Q_0 (s;x,y)=-  {G(y,x) \over 2(s-x)} + \cO \left ( 1 \right )
\eea
as $s\rightarrow x$ (and similarly as $s\rightarrow y$). The resulting  $Q_0(s;x,y)$ is
meromorphic in $s$ with only poles and no logarithmic  branch cuts.

\subsection{Solving for $Q_0(s;x,y)$}

Single-valuedness on $\Sigma$ of $Q_0(s;x,y)$ in $x$ and $y$ requires that the right
hand side of (\ref{pLeq}) be single-valued in $x$ and $y$. This is manifest for the
first and third lines in (\ref{pLeq}), but single-valuedness of the second line imposes
conditions on $c_I(x,y)$: it must have vanishing monodromy around  $A$-cycles,
while around a cycle $B_K$, its monodromy must be,
\bea
c_I (x + B_K, y) = c_I(x,y) + 2 \pi i \delta _{IK} \p_x \p_y \ln E(x,y)
\eea
The monodromy equation for $c_I$, combined with its symmetry $c_I (y,x)=-c_K(x,y)$,
allow for the following general solution in terms  the scalar function $\varphi_I(w)$.

\sm

The function $\f _I (w)$ is meromorphic and multiple-valued on $\Sigma$.
It may be defined by its pairing relation with holomorphic  1-forms $\o_J$,
\bea
\int _w \o_J (w) \p_{\bar w} \f _I (w) = \delta _{IJ}
\eea
the requirement $\f _I (w_0)=0$, and its  monodromy
\bea
\f _I (w+ A_J) & = & \f _I(w)
\no \\
\f _I (w+ B_K) & = & \f _I (w) - 2 \pi i \delta _{IJ}
\eea
The associated Green's function $G^\f (z,w)$, obtained by setting
\bea
\label{GG1}
G^\f (z,w) = G(z,w) + \sum _I \o _I(z) \f _I(w)
\eea
is single-valued in both $z$ and $w$, and satisfies the following equations
\bea
\label{g1}
\p _{\bar z} G^\f (z,w) & = & + 2 \pi \bigg ( \delta (z,w) - \delta (z,w_0) \bigg )
\no \\
\p _{\bar w} G^\f (z,w) & = & - 2 \pi  \delta (z,w) + \sum _I \o_I(z) \p_{\bar w} \f _I (w)
\eea
In terms of the function $\f _I (w)$, we may now easily write down the solution for
$c_I(x,y)$,
\bea
c_I (x,y) = c^{(0)} _I(x,y) - \bigg ( \varphi _I(x) - \varphi _I(y) \bigg ) \p_x \p_y \ln E(x,y)
\eea
where $c_I^{(0)} (x,y) = - c_I^{(0)} (y,x)$ is an arbitrary $(1,0)_x \otimes (1,0)_y$ form
which is single-valued in $x$ and $y$ on $\Sigma$. In terms of the single-valued Green
function $G^\varphi (s,x)$, the result may be expressed equivalently as,
\bea
\p_s Q_0(s;x,y) & = & \half \p_s G(x,s) G(y,s) - \half \p_s G(y,s) G(x,s)
\no \\ && +
\bigg ( G^\varphi (s,y) - G^\varphi (s,x) \bigg )\p_x \p_y \ln E (x,y) +
\sum _I c_I ^{(0)} (x,y) \o _I (s)
\eea
Although $\varphi_I$ appears to introduce new data, the dependence on $\varphi_I$
will be absorbed by the functions $c_I^{(0)}$.
To see this, we perform an infinitesimal variation $\delta \varphi _I$ on $\varphi_I$,
under which we have
\bea
\p_s \delta Q_0 (s;x,y) = \sum _I \bigg ( \Big (\delta \varphi_I(y) - \delta \varphi _I(x) \Big )
\p_x \p_y \ln E(x,y) + \delta c_I ^{(0)} (x,y) \bigg ) \o _I(s)
\eea
All dependence on $\varphi_I$ may be eliminated by requiring that the
differential $\p_s Q_0$ have vanishing $A$-periods in $s$. Since the differential
$\p_s Q_0$ has vanishing monodromy on $A$-cycles, this condition is equivalent to
\bea
c_I ^{(0)} (x,y) & = & - \half \oint _{A_I} ds \bigg (
\p_s G(x,s) G(y,s) - \p_s G(y,s) G(x,s)
\no \\ && \hskip 0.7in + 2 \Big (
G^\varphi (s,y) - G^\varphi (s,x) \Big )\p_x \p_y \ln E (x,y) \bigg )
\eea
Since the integrand on the right hand side is manifestly single-valued
on $\Sigma$ in $x$ and $y$,
and all the poles in $s$ have vanishing residue, the form $c_I ^{(0)} (x,y)$
defined this way is automatically single-valued in $x$ and $y$ as well.
It is easy to check that $Q_0$ defined this way is invariant under changes of $\varphi_I$,
and is thus intrinsically defined.

\sm

In practice, it will often be convenient to calculate with the help of the function
$Q_0^\varphi (s;x,y)$ which is defined to obey
\bea
\label{pLeq1}
\p_s Q_0^\varphi (s;x,y) & = & \half \p_s G(x,s) G(y,s) - \half \p_s G(y,s) G(x,s)
\no \\ && +
\bigg ( G^\varphi (s,y) - G^\varphi (s,x) \bigg )\p_x \p_y \ln E (x,y)
\no \\
\p _{\bar s} Q_0^\varphi (s;x,y) & = & - \pi \delta (s,x) G(y,s) + \pi \delta (s,y) G(x,s)
\eea
and then obtain the function $Q_0(s;x,y)$ by subtracting the $A$-periods in $s$ of $Q_0^\f$,
\bea
Q_0(s;x,y) = Q_0^\varphi (s;x,y) - \sum _I \left ( \int ^s _{w_0} \o_I \right )
\left ( \oint _{A_I} dt \, \p_t Q_0^\varphi (t;x,y) \right )
\eea
The two constructions of $Q_0(s;x,y)$ are equivalent, as may be checked by
calculation.

\subsection{Definition of $Q_B$, $Q_F$, and $Q_\ff $}

During the course of our construction of holomorphic chiral blocks and
analysis of their cohomology, we shall need solutions to the following $\bar \p$
equations which have right hand sides with non-zero monodromy in $s$,
\bea
\p_{\bar s} Q_B (s;r,t)
& = & \half \p_{\bar s} \FF(r,s) \FF (t,s) + \half \p_{\bar s} \FF (t,s) \FF (r,s)
- \mu (s) S(r,s) S(t,s)
\no \\
\p_{\bar s} Q_F (s;r,t) & = & \half \p_{\bar s} \FF (t,s) G(r,s) -
\half \p_{\bar s} G(r,s) \FF (t,s)
\no \\
\p _{\bar s} Q_\ff (s;r,t)  & = &  \half \p _{\bar s} \ff (r,s) G(t,s)
- \half \ff (r,s) \p_{\bar s} G(t,s)
\eea
Each function is a scalar in $s$.  $Q_B$ is a spinor in $r,t$, and has vanishing
monodromy  in these variables; by construction, we have $Q_B(s;r,t) = Q_B(s;t,r)$.
 $Q_F$ is a spinor in $t$, and a $(1,0)$ form in $r$,
and has vanishing monodromy in these variables. $Q_\ff$ is a 1-form in $t$
in which it has no monodromy, and a scalar in $r$ in which it does have monodromy.
All three equations are readily solved using the function $Q_0(s;x,y)$
constructed earlier;
\bea
Q_B(s;r,t) & = & {1 \over (4 \pi)^2} \int _x \int _y S(r,x) S(t,y) \chix \chiy Q_0(s;x,y)
\no \\ && \hskip 0.5in + {1 \over 2 \pi} \int _x G(x,s) \mu (x) S(r,x) S(t,x)
\no \\
Q_F (s;r,t) & = & - {1 \over 4 \pi} \int _x S(t,x) \chix Q_0(s;r,x)
\no \\
Q_\ff (s;r,t)
& = & { 1 \over 16 \pi^2} \int _x \int _y G(x,r) \chi (x) S(x,y) \chi (y) Q_0(s;y,t)
\no \\ && \hskip 0.5in
- {1 \over 2 \pi } \int _x G(x,r) \mu (x) Q_0(s;x,t).
\eea
All integrals in $x$ and $y$ are well-defined on $\Sigma$ since $Q_0(s;x,y)$
was constructed to be single-valued in $x$ and $y$.

\subsection{Monodromies of $Q_0$, $Q_B$, $Q_F$, and $Q_\ff$}

We shall now evaluate the monodromies in $s$ of the quantities $Q_0(s;x,y)$,
$Q_B(s;x,y)$, $Q_F(s;x,y)$, and $Q_\ff (s;x,y)$.
By construction, the $A$-cycle monodromies in $s$ of $Q_0(s;x,y)$, and hence of
$Q_B(s;x,y)$, $Q_F(s;x,y)$, and $Q_\ff (s;x,y)$ all vanish. Thus, we need to evaluate
only the $B$-cycle monodromies. Actually, we shall require only
the $B$-cycle monodromies {\sl up to additive terms which are independent of $s$},
and the formulas given below will hold only up to such terms.

\sm

From the form of $d_s Q_0(s;x,y)$ in  (\ref{pLeq}), we can read off the monodromy
of $d_s Q_0$ under $s\to s'=s + B_K$. Similarly,  the monodromies of $d_s Q_B$,
$d_sQ_F$, and $d_sQ_\ff $ follow from the definitions of $Q_B$, $Q_F$, and $Q_\ff $,
\bea
d_s Q_0(s';x,y) -
d_s Q_0 (s;x,y) & = & - i \pi d_s G(y,s) \o_K(x) + i \pi d_s G(x,s) \o_K (y)
\no \\
d_s Q_B(s';r,t)-
d_s Q_B(s;r,t) &=&- i \pi d_s \FF (t,s) \hat \o _{K0} (r)
- i \pi d_s \FF (r,s) \hat \o _{K0} (t)
\no \\
d_s Q_F(s';r,t)-
d_s Q_F(s;r,t) &=&+ i \pi d_s \FF (t,s) \o_K(r) - i \pi d_s G(r,s) \hat \o_{K0}(t)
\no \\
d_s Q_\ff (s';r,t)
-
d_s Q_\ff (s;r,t)
&=&  + i \pi d_s G(t,s) \lambda _K(r) + i \pi d_s \ff (r,s) \o_K (t)
\eea
Applying now the general formula (\ref{genmon}), we find the monodromy of $Q_0(s;x,y)$.
The monodromies of $Q_0(s;x,y)$, $Q_B(s;r,t)$, $Q_F(s;r,t)$, and $Q_\ff (s;r,t)$ follow at once,
\bea
Q_0 (s';x,y) -  Q_0 (s;x,y)
& = &
- i \pi G(y,s) \o_K(x) + i \pi G(x,s) \o_K (y)
\no \\
Q_B (s';r,t) - Q_B (s;r,t)
& = &
- i \pi  \FF (t,s) \hat \o _{K0} (r) - i \pi  \FF (r,s) \hat \o _{K0} (t)
\no \\
Q_F(s';r,t) - Q_F(s;r,t)
& = & + i \pi \FF (t,s) \o_K(r) - i \pi  G(r,s) \hat \o_{K0}(t)
\no \\
Q_\ff (s';r,t)- Q_\ff (s;r,t)
&=& + i \pi G(t,s) \lambda _K(r) + i \pi \ff (r,s) \o_K (t)
\eea
up to an additive contribution which is independent of $s$.
The $s$-independent integral term arising in the monodromy of $Q_0$,
for example, is given by,
\bea
 \int _{w_0} ^{w_0 + B_K} \! \! d_t Q_0(t;x,y)
\eea
Its  form is akin to the integral occurring in the construction of the
Riemann vector $\Delta _I$, but its explicit expressions will not be needed here.

\subsection{Derivatives of $Q_0$, $Q_B$, $Q_F$, and $Q_\ff $}

It will be convenient to group together in this section all the derivatives
of the quantities $Q_B$, $Q_F$, and $Q_\ff $ that we will need.
The $\p_{\bar s}$ derivatives are given by their defining equations,
while the other derivatives are obtained by differentiating
their integral formulas, as well as the following result,
\bea
\p_{\bar y} Q_0(s;x,y) = \pi G(x,s) \bigg ( 2 \delta (y,x) - \delta (y,s) - \delta (y,w_0) \bigg )
\eea
We find for $Q_B(r,s;t)$
\bea
\p_{\bar s} Q_B(s;r,t) =
- {1 \over 4} \chi(s) \bigg  ( S(t,s) \FF (r,s) + S(r,s) \FF (t,s) \bigg )
- \mu (s) S(r,s) S(t,s)
\eea
as well as the derivatives,
\bea
\p_{\bar r} Q_B(s;r,t) & = & - \half \chi(r) Q_F(s;r,t) + G(r,s) \mu(r) S(t,r)
\no \\
\p_{\bar r} Q_B(s;r,r) & = & - \chi(r) Q_F(s;r,r) - \p_r \bigg ( \mu(r) G(r,s) \bigg )
\eea
For $Q_F$, we find,
\bea
\p_{\bar s} Q_F(s;r;t) & = & - {1 \over 4} \chi(s) S(t,s) G(r,s) + \pi \delta (r,s) \FF (t,r)
\no \\
\p_{\bar r} Q_F (s;r;t) & = & \half \chi(r) S(t,r) G(r,s)
- \pi \left (\delta (r,s) + \delta (r,w_0) \right  ) \FF (t,s)
\no \\
\p_{\bar t} Q_F (s;r;t) & = & - \half \chi (t) Q_0(s;r,t)
\eea
For $Q_\ff $, we find,
\bea
\p_{\bar s} Q_\ff (s;r,t) & = & - {1 \over 4} \chi (s) \FF (s,r) G(t,s) + \half \mu(s) G(s,r) G(t,s)
+ \pi \ff (r,s) \delta (t,s)
\no \\
\p_{\bar r} Q_\ff  (s;r,t) & = &
- \half \chi (r) Q_F(s;t,r) + \mu (r) Q_0(s;r,t)
 \\
\p_{\bar t} Q_\ff (s;r,t) & = & \half \chi (t) \FF (t,r) G(t,s) - \mu (t) G(t,r) G(t,s)
- \pi \bigg ( \delta (t,s) + \delta (t,w_0) \bigg ) \ff (r,s)
\no \eea

\subsection{Transformations under changes of slice}

The changes under slice $\mu$ are given by (up to contributions
due to $v(w_0)$ and $\xi (w_0)$.)
\bea
\delta _v Q_B(s;r,t) & = &
- v(s) S(r,s) S(t,s) + v(r) G(r,s) S(t,r)
\no \\ &&  
+ v(t) G(t,s) S(r,t)
\no \\
\delta _v Q_F(s;r,t) & = & 0
\no \\
\delta _v Q_\ff (s;r,t) & = & 
v(r) Q_0 (s;r,t) - v(t) G(t,s) G(t,r) 
\no \\ &&
+ \half v(s) G(t,s) G(s,r)
\eea
The changes under slice $\chi$ are given by
\bea
\delta _\xi Q_B (s;r,t) & = &  \half \xi  (s) \FF (t,s) S(r,s) + \half \xi (s) \FF (r,s) S(t,s)
\no \\ &&
+ \xi (r) Q_F(s;r,t) + \xi (t) Q_F (s;t,r)
\no \\
\delta _\xi Q_F (s;r,t) & = & \half \xi (s) G(r,s) S(t,s) - \xi (r) G(r,s) S(t,r)
\no \\ &&
+ \xi (t) Q_0(s;r,t)
\no \\
\delta _\xi Q_\ff (s;r,t) & = &
\xi (r) Q_F (s;t,r) - \xi (t) G(t,s) F(t,r) 
\no \\ && 
+ \half \xi (s) G(t,s) F(s,r)
\eea

\subsection{Definition and properties of  $\Phi _B$, $\Phi_F$, and $\Phi _\ff $}

We define the $(1,0)$ forms in $s$, denoted by
$\Phi _B (s;r,t)$, $\Phi_F(s;r,t)$, and $\Phi _\ff (s;r,t)$ as follows,
\bea
\Phi _B(s;r,t) & \equiv & \half \p_s \FF (t,s) \FF (r,s) + \half \p_s \FF (r,s) \FF (t,s)
- \p _s Q_B(s;r,t)
\no \\
\Phi _F (s;r;t) & \equiv & \half \p_s \FF (t,s) G(r,s) - \half \p_s G(r,s) \FF (t,s)
- \p _s Q_F(s;r,t)
\no \\
\Phi _\ff (s;r,t) & \equiv & \half \p_s \ff (r,s) G(t,s) - \half \p_s \ff (t,s) \p_s G(t,s)
- \p _s Q_\ff (s;r,t)
\eea
Each term separately is monodromy-free in $r$ and $t$, but has monodromy in $s$.
The quantities $\Phi_B(s;r,t)$ and $\Phi _\ff (s;r,t)$ satisfy the symmetry properties,
$\Phi _B (s;r,t)= \Phi _B (s;t,r)$, and $\Phi _\ff (s;r,t) = - \Phi _\ff (s;t,r)$. Using the
explicit expressions for $Q_B, Q_F$, and $Q_\ff$, one shows that they are
monodromy-free in $s$. In fact, these quantities are essentially the analogs
of the $\rho$ differentials which entered into our general considerations
in section 4.2, which also confirms why they should be monodromy-free in $s$.

\sm

The derivatives of these quantities are readily obtained from their definitions,
\bea
\p_{\bar s} \Phi_B (s;r,t) & = &
+ \half \chi (s) \bigg ( \p_s \FF (t,s)  S(r,s) + \p_s \FF (r,s)  S(t,s) \bigg )
\no \\ &&
+ \p_s \bigg ( \mu(s) S(r,s) S(t,s) \bigg )
\no \\
\p_{\bar r} \Phi _B (s;r,t) & = & - \half \chi (r) \Phi _F (s;r;t) - \mu (r) \p_s G(r,s) S(t,r)
\eea
as well as,
\bea
\p_{\bar s} \Phi _F (s;r;t) & = &
\half \chi(s) S(t,s) \p_s G(r,s) - 2 \pi \delta (r,s) \p_s \FF (t,s)
\no \\
\p_{\bar r} \Phi _F (s;r;t) & = &
- \half \chi(r) S(t,r) \p_s G(r,s) + 2 \pi \delta (r,s) \p_s \FF (t,s)
\no \\
\p_{\bar t} \Phi _F (s;r;t) & = & - \half \chi (t) \bigg ( G(s,r) - G(s,t) \bigg ) \p_r \p_t \ln E(r,t)
\eea

\subsection{Definition and derivatives of $\Psi (r,s)$}

Finally, we also require the following composites, whose definitions
are straightforward, but which we also include in this section
for easy reference:
\bea
\label{Tdelta}
\Psi (r,s) & = &
- {1 \over 16 \pi^2} \int _x \int _y  S(r,x) \chix \p_x \p_y \ln E(x,y) \chiy S(y,s)
\no \\&&
+ {1 \over 4 \pi} \int _x \mu(x) \bigg ( S(x,r) \p_x S(x,s) - S(x,s) \p_x S(x,r) \bigg )
\eea
From this definition, we have $\Psi  (s,r) = - \Psi (r,s)$. The derivatives are
\bea
\p_{\bar r} \Psi (r,s)
=
- \half \p_r \FF (s,r) \chi(r) - \half \mu(r) \p_r S(r,s) - \half \p_r \bigg (\mu(r) S(r,s) \bigg )
\eea

\newpage

\section{Linear Chain Blocks}
\setcounter{equation}{0}

In this section and the next, we construct the blocks which will ultimately be extracted
from the decomposition of chiral blocks given in the previous section.
The extraction process itself will be given in subsequent sections.
Since it is combinatorially involved, it is helpful
to know in advance what the blocks are. The blocks are of several types:
holomorphic in all insertion points, holomorphic in all but one insertion
point, holomorphic in all but two insertion points
(all away from the diagonal $z_r=z_s$; henceforth, this is to be understood
and we shall not mention it explicitly.)
The corresponding differentials
will always include a full exterior differential $d_r$ with respect to
each insertion point $z_r$ if the block is not holomorphic in $z_r$.
Furthermore, the blocks can be classified by a corresponding kinematic invariant.

This section is devoted to the definition and construction of the holomorphic
blocks which have the connectivity of a  {\sl linear chain}. The most basic
such block is the object
$\Pi ^{(n+2)}$ which plays a key role in lifting the obstruction to holomorphicity
caused by the presence of $(0,1)\otimes (0,1)$ forms in the chiral amplitudes.
It is in this role that they were already identified in the Introduction. The monodromy
and derivative operations on the blocks $\Pi ^{(n+2)}$ produce new blocks
$\Pi _I ^{(n+2)}$, $\Pi _{IJ} ^{(n+2)}$, and $\Pi _F^{(n+2)}$ whose role in the
cohomology construction of chiral amplitudes will be discussed in later sections.

\subsection{Notation}

It is convenient to introduce the following notation
\bea
\label{notation}
G^{n}(t_1,\cdots ,t_{n+1})
& \equiv & G(t_1,t_2)G(t_2,t_3)\cdots G(t_{n},t_{n+1}), \qquad \qquad G^{(0)}(t,t)=1
\nonumber\\
S^{n}(t_1,\cdots ,t_{n+1})
& \equiv & S(t_1,t_2)S(t_2,t_3)\cdots S(t_n,t_{n+1}), \qquad \qquad S^{(0)}(t,t)=1
\eea
As with kinematic invariants, we will denote all blocks by the same generic letter,
in this case $\Pi$, and distinguish them from one another only by their indices.

\subsection{Motivation and definition of the blocks $\Pi$}

A natural starting point for the definition and construction of the blocks
$\Pi ^{(n+2)}$ is found in their role in lifting the obstruction to holomorphicity
caused by the presence of $(0,1)_r\otimes (0,1)_s$ forms in the chiral amplitude.
The contribution of such a  form arises from the third line
in the
component form of the chiral amplitude given in (\ref{Ys}), and is given by
\bea
\left \< Q(p_I) \, \cV_r ^{(1)} \cV _s ^{(1)} \prod _{l \not= r,s} ^N \cV _l ^{(0)} \right \>
\eea
The proportionality of $\cV_r ^{(1)}$ to $\chi (r) \psi _+ (r)$ forces this correlator
to always contain a linear strand of RNS fermion $\psi _+$ contractions.
Therefore, the above correlator will always contain a linear strand of
Szeg\"o kernels, arranged as follows,
\bea
d\bar r \, \chi (r) S(r,t_1) S(t_1,t_2) \cdots S(t_{n-1} , t_n) S(t_n, s)  \chi (s) d\bar s
\eea
Here, the points $t_1,t_2, \cdots, t_{n-1}, t_n$ are all distinct from one another
and from  $r$ and $s$, as is guaranteed by the structure of the Wick contractions
for the free field $\psi _+$. The proposition to be proven here is that all such
terms in the chiral amplitudes must in fact be exact differentials of well-defined
functions. Lifting the holomorphicity obstruction just for this term thus requires
that there exist a function $\Pi ^{(n+2)} (r;t_1,t_2, \cdots, t_{n-1}, t_n;s)$, which is
a $(0,0)$ form in $r$ and $s$, and a $(1,0)$ form in $t_1,\cdots, t_n$, and is
such that\footnote{The factor $1/4$ is introduced as a
convenient normalization; henceforth we use the notation of (\ref{notation}).}
\bea
\p_{\bar r} \p_{\bar s} \Pi ^{(n+2)} (r;t_1,t_2, \cdots, t_{n-1}, t_n;s)
= {1 \over 4} \chi (r) S(r,t_1,t_2, \cdots, t_{n-1} , t_n, s)  \chi (s)
\eea
The blocks $\Pi ^{(n+2)} (r;t_1, \cdots,  t_n;s)$ must be holomorphic
in each of the points $t_1, \cdots, t_n$, away from coincident points $t_i= t_j$
for $i \not= j$, and away from the points $r$ and $s$. Finally, the blocks
must have the same mirror symmetry as the correlator, so that we require
\bea
\label{symmPi1}
\Pi ^{(n+2)}  (r;t_1, \cdots , t_n;s)
= (-)^n \Pi ^{(n+2)}  (s;t_n, \cdots , t_1;r)
\eea
Finally, we require that its monodromy in $r$ is independent of $r$,
and the monodromy in $s$ is independent of $s$. Although one might
initially have hoped that blocks $\Pi ^{(n+2)}$ would exist without monodromy
in $r,s$, it turns out to be impossible to achieve without introducing new
singularities at extraneous points. Instead of being a problem, the non-trivial
monodromies of $\Pi ^{(n+2)}$ will  combine precisely with the
monodromies of the $(0,1)$ form obstructions and help lift those as well.

\subsection{Construction of the blocks $\Pi $}

The simplest case arises for $n=0$, where the block must satisfy
\bea
\p_{\bar r} \p_{\bar s} \Pi ^{(2)} (r;s) = {1 \over 4} \chi (r) S(r, s)  \chi (s)
\eea
The solution to this equation may be gathered from the derivative fromulas
of the basic functions $\ff$ and $\FF$ given in (\ref{differentialrelations}), and we find that
$\Pi ^{(2)}(r,s) = \ff (r,s)$. The next few lowest order blocks may be
obtained by combining $\ff$, $\FF$ and the bosonic Green function $G$
and Szeg\"o kernel $S$, and we find for $0\leq n \leq 2$,
\bea
\label{Pilow}
\Pi ^{(2)} (r,s) & = & \ff (r,s)
\\
\Pi ^{(3)} (r;t;s) & = &
\ff (r,t) G(t,s) - G(t,r) \ff (t,s) - \FF (t,r) \FF (t,s)
\no \\
\Pi ^{(4)} (r;t_1,t_2;s) & = &
+ \ff (r,t_1) G(t_1,t_2) G(t_2,s)
- G(t_1,r) \ff (t_1,t_2) G(t_2,s)
\no \\ &&
+ G(t_1,r) G(t_2,t_1) \ff (t_2,s)
- \FF (t_1,r) \FF (t_1,t_2) G(t_2,s)
\no \\ &&
- \FF (t_1,r) S(t_1,t_2) \FF (t_2,s)
+ G(t_1,r) \FF (t_2,t_1) \FF (t_2,s)
\no \eea
A clear pattern for all values of $n$ is recognized, and takes the following form,
\bea
\Pi ^{(n+2)} = \sum _{\a=0} ^n (-)^\a G^\a \ff G^{n-\a}
- \sum _{\a=0} ^{n-1} ~ \sum _{\b =0} ^{n-1-\a} (-)^\a G^\a \FF S^\b \FF G^{n-1-a-\b}
\eea
It is useful to write out the dependence on the points of $\Pi ^{(n+2)}(r;t_1, \cdots, t_n;s)$,
\bea
\Pi^{(n+2)} & = &
\sum _{\alpha = 0} ^n (-)^\alpha G^\alpha(t_\alpha,\cdots ,t_1,r)\ff (t_\alpha,t_{\alpha+1})
 G ^{n-\alpha}(t_{\alpha+1},\cdots ,t_n,s)
\no \\
&&
- \sum _{\alpha =0}^{n-1}\sum_{\beta =0}^{n-1-\alpha}
    (-)^\alpha G ^\alpha(t_\alpha,\cdots ,t_1,r) \FF (t_{\alpha+1},t_\alpha)
    S^\beta(t_{\alpha+1},\cdots ,t_{\alpha+\beta+1})
\nonumber\\
&&
\qquad\qquad
\times\, \FF (t_{\alpha+\beta+1},t_{\alpha+\beta+2})
    G ^{n-1 -\alpha -\beta }(t_{\alpha+\beta+2},\cdots ,t_n,s)
\eea
By construction, the blocks $\Pi^{(n+2)}(r;t_1,\cdots ;t_n;s)$ are $(0,0)$ forms
in $r$ and $s$, and $(1,0)$ forms in  $t_1,\cdots,t_n$, and obey the symmetry
relation (\ref{symmPi}). The expressions for $n=0,1,2$ manifestly reproduce
those found in (\ref{Pilow}). We shall later verify that $\Pi ^{(n+2)}(r;t_1, \cdots, t_n;s)$
is indeed holomorphic in $t_1, \cdots, t_n$, away from coincident points $t_i=t_j$
for $i \not= j$ and $t_i$ away from the points $r$ and $s$.

\subsection{Monodromy of $\Pi $ produces new blocks $\Pi _I $}

Monodromy of $\Pi ^{(n+2)}(r;t_1, \cdots, t_n;s)$ produces new blocks.
The $A$-cycle monodromy is absent in any of the arguments of $\Pi ^{(n+2)}$,
but the blocks in general have $B$-cycle monodromy. We shall define the
following $B$-cycle monodromy operator
\bea
\Delta _K ^{(z)} f(z) \equiv {1 \over 2 \pi i} \bigg ( f(z+B_K) - f(z) \bigg )
\eea
while leaving any possible other independent variables in $f$ unchanged.
By using the monodromy properties of $\ff$, $\FF$, and $G$, we find the
monodromy of $\Pi ^{(n+2)}$. The monodromies at the end points are given by,
\bea
\Delta _K ^{(r)} \Pi ^{(n+2)}  (r;t_1, \cdots , t_n;s)
& = & -    \Pi ^{(n+1)} _K (t_1;t_2, \cdots , t_n;s)
\no \\
\Delta _K ^{(s)} \Pi  ^{(n+2)} (r;t_1 ,\cdots , t_n;s)
& = &
-   (-)^n \Pi ^{(n+1)} _{K} (t_n; t_{n-1}, \cdots , t_1;r)
\eea
where the blocks $\Pi_K^{(n+1)}$ are defined below.
The monodromies at the midpoints $t_k, \, k=1,\cdots, n$, are given by,
\bea
\Delta _K ^{(t_k)}  \Pi ^{(n+2)} (r;t_1 ,\cdots , t_n;s)
& = &
-   (-)^k  G^k \Pi _K ^{(n+1-k)} (t_{k+1};t_{k+2} ,\cdots , t_n;s)
\no \\ &&
+   \Pi ^{(k)} _{K} (r; t_1, \cdots , t_{k-2};t_{k-1}) G^{n+1-k}
\eea
where we use the abbreviations $G^k = G^k (t_k, \cdots, t_1,r)$ and
$G^{n+1-k} = G^{n+1-k} (t_k, \cdots, t_n,s)$.

\sm

The new blocks $\Pi _I ^{(n+2)}(r;t_1, \cdots, t_n;s)$, defined for $n\geq -1$,
may be computed for the  first few lowest orders,
\bea
\label{PiIlow}
\Pi _I ^{(1)} (s) & = & \lambda _I (s)
\no \\
\Pi _I ^{(2)} (r,s) & = &
\lambda _I (r) G(r,s)
+ \o _I (r) \ff (r,s) - \hat \o_{I0} (r) \FF (r,s)
\no \\
\Pi _I ^{(3)} (r;t;s) & = &
\lambda _I (r) G(r,t) G(t,s)
+ \o _I (r) \ff (r,t) G(t,s)
\no \\ &&
- \hat \o_{I0} (r) \FF (r,t)G(t,s)
+ \hat \o_{I0}(r) S(r,t) \FF (t,s)
\no \\ &&
+ \o_I(r) \FF (t,r) \FF (t,s)
+ \o_I(r) G(t,r) \ff (s,t)
\eea
The expressions for the blocks for general $n$ are given as follows,
\bea
\label{PiI}
\Pi ^{(n+2)} _I (r;t_1, \cdots , t_n;s)
&=&
+ \l_I(r) G ^{n+1}(r,t_1,\cdots,t_n,s)
\\
&& +
\o _I(r) \Pi ^{(n+2)} (r;t_1,\cdots, t_n;s)
 \nonumber\\
 &&- \hat \o _{I0}(r) \Pi _F ^{(n+2)} (r;t_1,\cdots, t_n;s)
\no \eea
Here, we have introduced yet another block $\Pi _F ^{(n+2)} (r;t_1,\cdots, t_n;s)$
which is Grassmann odd. Its explicit form will be presented in the next subsection.

\subsection{The fermionic blocks $\Pi _F$}

The monodromy calculation of the blocks $\Pi$ yields the blocks $\Pi_I$, which
in turn are directly expressed in terms of the fermionic blocks $\Pi _F$.
The latter take the following explicit form in terms of the building blocks $\ff$,
$\FF$, $S$ and $G$,
\bea
\label{PiF}
\Pi _F ^{(n+2)} (r;t_1,\cdots, t_n;s) =
\sum _{\alpha =0 } ^n  S^\alpha(r,t_1,\cdots, t_\alpha) \FF (t_\alpha,
    t_{\alpha+1})G ^{n-\alpha}(t_{\alpha+1},\cdots,t_n,s)
\eea
The lowest order expressions for these blocks are
\bea
\Pi ^{(2)} _F (r;s) & = & \FF (r,s)
\no \\
\Pi ^{(3)} _F (r;t;s) & = & \FF (r,t) G(t,s) + S(r,t) \FF (t,s)
\no \\
\Pi ^{(4)} _F (r;t_1,t_2;s) & = & \FF (r,t_1) G(t_1,t_2) G(t_2,s) +
S(r,t_1) \FF (t_1,t_2) G(t_2,s)\no \\ &&  + S(r,t_1) S(t_1,t_2) \FF (t_2,s)
\eea
The blocks $\Pi _F ^{(n+2)}$ satisfy a generalized recursion relation,
which allows one to split this block at any midpoint. This relation takes the form,
\bea
\label{PiFrecursion}
&&
\Pi _F ^{(m+\ell+3)} (r;t_1,\cdots, t_m, w, u_1\cdots, u_\ell;s)
\no \\ && \hskip 0.7in =
\Pi _F ^{(m+2)} (r;t_1,\cdots, t_m, w) G^{\ell +1} (w,u_1,\cdots, u_\ell,s)
\no \\ && \hskip 0.9in
+ S(r,t_1, \cdots ,t_m,w) \Pi _F ^{(\ell+2)} ( w, u_1\cdots, u_\ell;s)
\eea
The use of $\Pi _F$ also allows one to obtain an analogous generalized
recursion relation for the block $\Pi$ itself,
\bea
\label{recurPi}
&&
\Pi  ^{(m+\ell+3)} (r;t_1,\cdots, t_m, w, u_1\cdots, u_\ell;s)
\no \\ && \hskip 0.5in  =
\Pi  ^{(m+2)} (r;t_1,\cdots, t_m, w) \, G^{\ell +1} (w,u_1,\cdots, u_\ell,s)
\no \\ &&  \hskip 0.7in
- (-)^m G^{m+1} (w,t_m, \cdots, t_1,r) \, \Pi  ^{(\ell+2)} ( w, u_1\cdots, u_\ell;s)
\no \\ &&  \hskip 0.7in
- (-)^m \Pi _F ^{(m+2)}  (w;t_m, \cdots , t_1;r) \, \Pi _F ^{(\ell+2)} ( w, u_1\cdots, u_\ell;s)
\eea
The monodromies of the blocks $\Pi _F ^{(n+2)} $ do not produce any new blocks,
but it will nonetheless be useful to record their form,
\bea
\Delta _K ^{(r)} \Pi _F ^{(n+2)}  (r;t_1, \cdots , t_n;s)
& = & 0
\\
\Delta _K ^{(s)} \Pi _F ^{(n+2)} (r;t_1 ,\cdots , t_n;s)
& = &
+   \Pi _F ^{(n+1)} (r;t_1, \cdots, t_{n-1};t_n) \o _K (t_n)
\no \\ &&
-   S^n(r,t_1,\cdots, t_n) \hat \o_{I0}(t_n)
\no \\
\Delta _K ^{(t_k)}  \Pi _F ^{(n+2)} (r;t_1 ,\cdots , t_n;s)
& = &
+   \Pi _F ^{(k)} (r;t_1, \cdots, t_{k-2};t_{k-1}) \o _K (t_{k-1}) G^{n+1-k}
\no \\ &&
-   S^{k-1}(r,t_1,\cdots, t_{k-1}) \hat \o_{I0}(t_{k-1}) G^{n+1-k}
\no \eea
where we use again the abbreviation
$G^{n+1-k} = G^{n+1-k} (t_k, \cdots, t_n,s)$.

\subsection{Monodromy of $\Pi _I$ produces new blocks $\Pi _{IJ}$}

The monodromy in the end points is given by
\bea
\label{monPiI}
\Delta _K ^{(r)} \Pi ^{(n+2)} _I (r;t_1, \cdots , t_n;s)
& = &
-   \o _I(r) \Pi ^{(n+1)} _K (t_1;t_2, \cdots , t_n;s)
\no \\
\Delta _K ^{(s)} \Pi _I ^{(n+2)} (r;t_1 ,\cdots , t_n;s)
& = &
+   \Pi ^{(n+1)} _{IK} (r; t_1, \cdots , t_{n-1};t_n)
\eea
while the monodromies at the midpoints are given as follows,
\bea
\Delta _K ^{(t_k)} \Pi _I ^{(n+2)} (r;t_1 ,\cdots , t_n;s)
& = &
-   (-)^k \o_I(r) G^k \Pi _K ^{(n+1-k)} (t_{k+1};t_{k+2} ,\cdots , t_n;s)
\no \\ &&
+    \Pi ^{(k)} _{IK} (r; t_1, \cdots , t_{k-2};t_{k-1}) G^{n+1-k}
\eea
where we use again $G^k = G^k (t_k, \cdots, t_1,r)$ and
$G^{n+1-k} = G^{n+1-k} (t_k, \cdots, t_n,s)$.

\sm

The blocks $\Pi ^{(n+2)} _{IJ} (r;t_1,\cdots, t_n;s)$ are defined for all $n \geq -1$,
and are new. They may be evaluated from the second relation in (\ref{monPiI}),
and the monodromies of $G$, $\Pi$ and $\Pi_F$. This monodromy variation will
produce one term involving $\Pi_F$ and one term involving $\l_I$. Both may
be eliminated using (\ref{PiI}), and we obtain the expression,
\bea
\label{PiIJ}
\Pi ^{(n+2)} _{IJ} (r;t_1, \cdots, t_n;s)
& = &
+  \Pi _I ^{(n+2)} (r;t_1, \cdots, t_n;s) \o_J(s)
\no \\ &&
+ (-)^n \o_I (r) \Pi _J ^{(n+2)} (s;t_n, \cdots, t_1;r)
\no \\ &&
-  \o _I(r) \Pi  ^{(n+2)} (r;t_1, \cdots, t_n;s) \o_J(s)
\no \\ &&
+ \hat \o _{I0}(r) S ^{n+1} (r,t_1, \cdots, t_n,s) \hat \o_{J0}(s)
\eea
which satisfies the following symmetry relation,
\bea
\label{reflection}
\Pi ^{(n+2 )} _{IJ} (r; t_1, \cdots , t_n;s)
= (-)^n \Pi ^{(n+2 )} _{JI} (s; t_n, \cdots , t_1;r)
\eea
The lowest order blocks are given by
\bea
\label{Pi-low}
\Pi_{IJ}^{(1)}(r)
&=& +\hat\o_{I0}(r)\hat\o_{J0}(r)-\o_I(r)\l_J(r)+\l_I(r)\o_J(r)
\nonumber\\
\Pi_{IJ}^{(2)}(r,s)
&=& +\hat\o_{I0}(r)S(r,s)\hat\o_{J0}(s) +
\o_I(r)F(s,r)\hat\o_{J0} - \hat\o_{I0}(r)F(r,s)\o_J(s)
\nonumber\\
&&
+\o_I(r)G(s,r)\l_J(s)+\l_I(r)G(r,s)\o_J(s)
+
\o_I(r)\ff (r,s)\o_J(s)
\nonumber\\
\Pi_{IJ}^{(3)}(r;t;s)
&=& +\hat\o_{I0}(r)S(r,t)S(t,s)\hat\o_{J0}(s)
+ \o_I(r)F(t,r)S(t,s)\hat\o_{J0}(s)
\nonumber\\
&&
- \hat\o_{I0}(r)S(r,t)F(t,s)\o_J(s)
-\o_I(r)G(t,r)F(t,s)\hat\o_{J0}(s)
\nonumber\\
&&
- \hat\o_{I0}(r)F(r,t)G(t,s)\o_J(s) - \o_I(r)G(s,t)G(t,r)\l_J(s)
\nonumber\\
&&
+\l_I(r)G(r,t)G(t,s)\o_J(s)
+
\o_I(r)\ff (r,t)G(t,s)\o_J(s)
\nonumber\\
&&
- \o_I(r)G(t,r)\ff (r,s)\o_J(s) - \o_I(r)F(t,r)F(t,s)\o_J(s)
\eea
The monodromies of the blocks $\Pi _{IJ} ^{(n+2)} (r;t_1, \cdots,
t_n;s)$ are readily evaluated from their expression in (\ref{PiIJ}),
and we find, \bea \Delta _K ^{(r)} \Pi ^{(n+2 )} _{IJ} (r; t_1,
\cdots , t_n;s) & = & -  \omega _I(r) \Pi ^{(n+1 )} _{KJ} (t_1; t_2,
\cdots , t_n;s)
\no \\
\Delta _K ^{(s)} \Pi ^{(n+2 )} _{IJ} (r; t_1, \cdots , t_n;s)
& = & +   \Pi ^{(n+1 )} _{IK} (r; t_1, \cdots ; t_n) \omega _J(s)
\eea
and for the midpoints, for $1 \leq k \leq n$,
\bea
\Delta _K ^{(t_k)} \Pi ^{(n+2 )} _{IJ} (r; t_1, \cdots , t_n;s)
& = & -  (-)^k \,\o_I(r) G^k
\Pi ^{(n+1-k)}_{KJ} (t_{k+1}; t_{k+2}, \cdots, t_n;s)
\no \\ &&
+ \Pi ^{(k)} _{IK} (r; t_1, \cdots , t_{k-2};t_{k-1})
\, G^{n+1-k} \o_J(s) \qquad
\eea
where we use again $G^k = G^k (t_k, \cdots, t_1,r)$ and
$G^{n+1-k} = G^{n+1-k} (t_k, \cdots, t_n,s)$.
These relations show that no further new blocks are being produced
by these final monodromies.

\subsection{Derivatives of the blocks $\Pi, \Pi_I, \Pi_{IJ}$ and $\Pi_F$}

We make use of a series of very convenient differentiation formulas,
whose role is to simplify the combinatorics of the derivative calculations.
It is easy to prove the following identities,
\bea
\label{midpoints}
\p_{\bar t} \Big (  S(u,t) \FF(t,v) + \FF (u,t) G(t,v) \Big )
& = &
2 \pi \Big ( \delta (t,v) - \delta (t,u) \Big ) \FF (u,v)
\no \\
\p_{\bar t} \Big (  \FF (t,u) S(t,v) - G(t,u) \FF(v,t) \Big )
& = &
2 \pi \Big ( \delta (t,v) - \delta (t,u) \Big ) \FF (v,u)
\no \\
\p_{\bar t} \Big (  \ff (u,t) G(t,v) - G(t,u) \ff(t,v)
 -\FF (t,u) \FF (t,v) \Big )
& = &
2 \pi \Big ( \delta (t,v) - \delta (t,u) \Big ) \ff (u,v)
\no \\
\p_{\bar r} \Big (
\l _I (r) G(r,t)  + \o _I (r) \ff (r,t) - \hat \o _{I0} (r) \FF (r,t) \Big )
& = & 2 \pi \delta (r,t) \lambda _I (t)
\no \\
\p_{\bar r} \Big ( \hat \o _{I0} (r) S(r,t)  + \o_I (r) \FF (t,r) \Big )
& = & 2 \pi \delta (r,t) \hat \o _{I0}(t)
\eea
Effectively, the differentiations are governed just by
the poles in the propagators $G$ and $S$.

\sm

$\bullet$ The derivatives of the blocks $\Pi^{(n+2)}(r;t_1, \cdots , t_n;s) $
at the midpoints  are given by
\bea
&&
\p_{\bar t_k} \Pi ^{(n+2)}  (r;t_1, \cdots , t_n;s)
\no \\ && \qquad
=
 2 \pi \bigg ( \delta (t_k, t_{k+1}) - \delta (t_k, t_{k-1}) \bigg )
 \Pi ^{(n+1)}  (r; t_1, \cdots , \widehat{\, t_k \, }, \cdots  , t_n;s)
\eea
where the caret \, $\widehat{}$ \,  denotes that this variable
is to be omitted. In particular, this formula shows that
$\Pi^{(n+2)}(r;t_1,\cdots,t_n;s)$ is holomorphic in the vertex
insertion points $t_1,\cdots,t_n$, with simple poles. The
derivatives at the end points are readily computed and we find,
\bea
\p_{\bar r} \Pi ^{(n+2)}  (r;t_1, \cdots , t_n;s) & = & 2 \pi \delta
(r,t_1) \Pi ^{(n+1)}  (t_1;t_2, \cdots , t_n;s) \no \\ && - \half
\chi (r) \Pi _F ^{(n+2)} (r;t_1, \cdots , t_n;s) \no \\ && + \mu (r)
G^{n+1} (r,t_1, \cdots , t_n, s)
\eea
where $\Pi_F$ is the fermionic
block given by (\ref{PiF}).

\sm

$\bullet$ The derivatives of the blocks $\Pi_F^{(n+2)}(r;t_1, \cdots , t_n;s)$,
may be obtained from the definition and generalized recursion relation for $\Pi_F$,
and we find,
\bea
\p_{\bar r} \Pi _F ^{(n+2)}  (r;t_1, \cdots , t_n;s) & = &
+ \half \chi (r)  G^{n+1} (r,t_1, \cdots , t_n,s)
\no \\
\p_{\bar s} \Pi _F ^{(n+2)}  (r;t_1, \cdots , t_n;s) & = &
- \half S(r,t_1, \cdots ,t_n,s) \chi (s)
\no \\
\p_{\bar t_k} \Pi _F ^{(n+2)}  (r;t_1, \cdots , t_n;s) & = & 0
\eea

\sm

$\bullet$ The derivatives of the blocks $\Pi_I^{(n+2)}(r;t_1, \cdots , t_n;s)$
may be obtained from the expression for $\Pi_I$ in (\ref{PiI}), and the
derivatives of $\Pi$ and $\Pi_F$, and we find,
\bea
\p_{\bar r} \Pi ^{(n+2)} _I (r;t_1, \cdots , t_n;s)
& = & + 2 \pi \delta (r, t_1) \Pi ^{(n+1)} _I (t_1; t_2, \cdots , t_n;s)
\\
\p_{\bar s} \Pi ^{(n+2)} _I (r;t_1, \cdots , t_n;s)
& = & \half \hat \o_{I0}(r) S^{n+1} \chi (s) + \o_I (r) \p_{\bar s} \Pi ^{(n+2)}
(r;t_1, \cdots , t_n;s)
\no \eea
while the derivatives at the midpoints are given by
\bea
&&
\p_{\bar t_k} \Pi ^{(n+2)} _I (r;t_1, \cdots , t_n;s)
\no \\ && \qquad
=
+ 2 \pi \bigg ( \delta (t_k, t_{k+1}) - \delta (t_k, t_{k-1}) \bigg )
 \Pi ^{(n+1)} _I (r; t_1, \cdots , \widehat{\, t_k \, }, \cdots  , t_{n-1};s)
\eea
Note that $\Pi_I^{(n+2)}(r;t_1,\cdots,t_n;s)$ is holomorphic in
$r$ and in $t_1,\cdots,t_n$, but not in $s$.

\sm

$\bullet$ The derivatives of the blocks $\Pi_{IJ}^{(n+2)}(r;t_1, \cdots , t_n;s)$
are obtained from the expression for $\Pi_{IJ}$ in (\ref{PiIJ}), and the
derivatives of $\Pi$, $\Pi_I$, and $\Pi_F$, already computed earlier.
Derivatives at the end points yield,
\bea
\label{rder}
\p _{\bar r}  \Pi ^{(n+2)} _{IJ} (r; t_1, t_2, \cdots , t_n;s)
& = &
+ 2 \pi   \delta (r,t_1) \Pi ^{(n+1)} _{IJ} (t_1; t_2, \cdots  , t_n;s)
\no \\
\p _{\bar s}  \Pi ^{(n+2 )} _{IJ} (r; t_1,  \cdots , t_{n-1} , t_n;s)
& = &
- 2 \pi   \delta (s,t_n)
\Pi ^{(n+1 )} _{IJ} (r; t_1, \cdots  , t_{n-1};t_n)
\eea
Derivatives at the midpoints gives,
\bea
\label{tder}
&&
\p _{\bar t_k}  \Pi ^{(n+2 )} _{IJ} (r; t_1, \cdots , t_n;s)
\\ && \qquad
=
+ 2 \pi \bigg ( \delta (t_k, t_{k+1}) - \delta (t_k, t_{k-1}) \bigg )
 \Pi ^{(n+1)} _{IJ} (r; t_1, \cdots , \widehat{\, t_k \, }, \cdots  , t_{n-1};s)
\no
\eea
where the caret \, $\widehat{}$ \,  denotes that this variable is to be omitted.
These identities show in particular that $\Pi_{IJ}^{(n+2)}(r;t_1,\cdots,t_n;s)$
is holomorphic in all insertion points $r$, $t_1,\cdots,t_n$, $s$.

\newpage

\section{Singly Linked Chain Blocks}
\setcounter{equation}{0}

In addition to the linear chain blocks, constructed and studied in the preceding section,
further blocks occur in the chiral amplitudes corresponding to a chain with one
link or with two links. The linking of the chain arises when a bosonic field $\p x_+(r)$ in the
supercurrent at the end of a linear chain is being contracted onto the
$\exp \{ ik_{t_i} \cdot x_+(t_i) \}$ operator of a vertex operator occurring in the linear chain.
Since there are only two supercurrents, the linking can occur either once or twice.
The basic building blocks of the linked chains are again the blocks $\Pi $, with either
one or two end points linked to a midpoint,
\bea
{\rm singly ~ linked \, } & \hskip 0.6in & \Pi ^{(n+2)} (t_i;t_1, \cdots , t_n; s) \delta _{r, t_i}
\no \\
{\rm doubly ~ linked} & \hskip 0.6in & \Pi ^{(n+2)} (t_i;t_1, \cdots , t_n; t_j) \delta _{r, t_i}
\delta _{s, t_j}
\eea
The doubly linked chains will occur only in the construction of the holomorphic
chiral blocks $\cZ$, but not for the differential chiral blocks $\cD$. Since the
focus in this paper is on the differential chiral blocks, we shall postpone the
study of the doubly linked chain blocks to a sequel paper in which the holomorphic
chiral blocks $\cZ$ will be explicitly constructed, and deal here only with the
simply linked chain blocks.

\sm

The single  linking of a linear chain creates a closed loop. The linked chain
block function is multiplied in the chiral amplitude by a kinematic factor.
The kinematic factor, and thus the associated block function, is naturally
decomposed into parts with definite symmetry properties under reflection
of the linked chain. As  a  result, we define two singly linked block functions,
$\Pi _+$ and $\Pi _-$, whose basic building blocks will be
\bea
\label{linkblocks}
\Pi _+ ^{(m+1|\ell+1)}
& \sim & \Pi ^{(m+\ell +3)} (s;t_1,\cdots, t_m,r,u_1,\cdots, u_\ell;r)
\no \\ && \qquad
+ (-)^\ell \Pi ^{(m+\ell +3)} (s;t_1,\cdots, t_m,r,u_\ell,\cdots, u_1;r)
\no \\
\Pi _- ^{(m+1|\ell+1)}
& \sim & \Pi ^{(m+\ell +3)} (s;t_1,\cdots, t_m,r,u_1,\cdots, u_\ell;r)
\no \\ && \qquad
- (-)^\ell \Pi ^{(m+\ell +3)} (s;t_1,\cdots, t_m,r,u_\ell,\cdots, u_1;r)
\eea
In both blocks, the loop in the chain contains $\ell +1$ points, $u_1, \cdots, u_\ell$
and $r$. This loop is connected at the point $r$ to a linear chain containing
$m+2$ points $s, t_1, \cdots, t_m,r$.
By construction, the right hand sides of (\ref{linkblocks}) is holomorphic
in $t_1, \cdots, t_m$, and in $u_1, \cdots, u_\ell$ with simple poles
occurring only between neighboring points, and between $t_1$ and $s$.
The blocks are not holomorphic at the end point $s$ of the linear chain,
just as the original block $\Pi$ is not holomorphic in its endpoints.

\sm

The key difficulty in completing the construction of the blocks $\Pi _\pm$
that are holomorphic in all variables, but the end point $s$, is to complete
the blocks so that holomorphicity  in $r$ is achieved. This will now be done
separately for the blocks $\Pi _+$ and $\Pi_-$. The completion of the blocks
$\Pi _+$ is straightforward and is achieved by further use of the basic
block $\Pi$. The completion of the block $\Pi_-$ is considerably more
involved and will necessitate the use of the functions $Q_0, Q_B, Q_F$, and $Q_\ff$.

\subsection{The  blocks $\Pi_+ ^{(m+1|\ell+1)}$}

The completion of the singly linked block $\Pi _+ ^{(m+1|\ell+1)}$ from
(\ref{linkblocks}) is straightforward, and is achieved through the addition
of a closed loop $\Pi ^{(\ell +2)} (r; u_1, \cdots, u_\ell;r)$ multiplied by
a linear chain of bosonic Green functions $G$, as follows,
\bea
\label{Piplus}
&&
2 \Pi_+ ^{(m+1|\ell+1)} (s;t_1,\cdots, t_m, r, u_1, \cdots, u_\ell)
\\ && \hskip 1in =
\Pi ^{(m+\ell+3)} (s; t_1 , \cdots , t_m, r, u_1 , \cdots , u_\ell, r)
\no \\ && \hskip 1.2in
+  (-)^\ell \Pi ^{(m + \ell +3)} (s; t_1 , \cdots , t_m, r, u_\ell , \cdots , u_1, r)
\no \\ && \hskip 1.2in
+ (-)^{m} G^{m+1} (r,t_m, \cdots, t_1,s) \Pi ^{(\ell +2)} (r;u_1, \cdots, u_\ell;r)
\no \eea
The construction holds for  $\ell \geq 1$, and $m \geq 0$; these singly linked
chain blocks $\Pi_+$ correspond to a loop with $\ell +1$ points, attached
at the point $r$ to a linear chain with $m+2$ points, including the point $r$.
When $m=0$, the points $t_k$ are absent, and the linear chain has only the
points $r$ and $s$.
These blocks are $(1,0)$-forms in the insertion points $t_1,\cdots$, $u_1,\cdots, u_{\ell}$,
and $r$, and scalars in $s$, and clearly holomorphic in $t_1,\cdots, t_m$ and
$u_1,\cdots, u_{\ell}$. They satisfy the following reflection symmetry property,
\bea
\Pi_+^{(m+1|\ell+1)}(s;t_1,\cdots,t_m,r,u_1,\cdots,u_\ell)
= (-)^\ell\Pi_+^{(m+1|\ell+1)}(s;t_1,\cdots,t_m,r,u_\ell,\cdots,u_1)
\eea
We now show that they are also holomorphic in $r$,
thanks to the addition of the last term.

\sm

To carry out this check, we  consider  the sum of the first and
second terms on the right hand side of (\ref{Piplus}), and denote this sum by
$2 \tilde \Pi_+^{(m+1|\ell+1)}(s;t_1,\cdots, t_m, r, u_1, \cdots, u_\ell)$.
We begin by computing its $\p_{\bar r}$ derivative, using the formulas for
the derivatives of $\Pi^{(n+2)}$ evaluated earlier for the case $n=m+\ell +1$.
These derivatives are readily expressed in terms of the fermionic blocks
$\Pi_F^{(n+2)}$, and we shall omit the $\delta$-function contributions
at coincident vertex operator points. We find,
\bea
\label{derPiplus}
&&
2 \p _{\bar r} \tilde \Pi _+ ^{(m+|\ell+1)} (s;t_1,\cdots, t_m, r, u_1, \cdots, u_\ell)
\no \\ && \hskip 1in =
(-)^{m+\ell +1} \mu(r) G^{m+\ell +2} (r,u_\ell, \cdots, , u_1,r,t_m, \cdots, t_1,s)
\no \\ && \hskip 1.2in
+ (-)^{m+1} \mu(r) G^{m+\ell +2} (r,u_1, \cdots, , u_\ell,r,t_m, \cdots, t_1,s)
\no \\ && \hskip 1.2in
+ \half  (-)^{m+\ell} \chi (r) \Pi _F ^{(m+\ell +3)} (r; u_\ell, \cdots, , u_1,r,t_m, \cdots, t_1; s)
\no \\ && \hskip 1.2in
+ \half  (-)^{m+1} \chi (r) \Pi _F ^{(m+\ell +3)} (r; u_1, \cdots, , u_\ell,r,t_m, \cdots, t_1; s)
\eea
We use the recursion relation (\ref{PiFrecursion}) for $\Pi _F$, with $w=r$,
in the third and fourth terms on the right hand side of (\ref{derPiplus}). The resulting
expression consists of two terms involving $\Pi _F ^{(m+2)}$ and two terms
involving $\Pi _F ^{(\ell +2)}$.
The sum of the terms involving $\Pi _F ^{(m+2)}$ is given by,
\bea
&&
+ \half (-)^{m+\ell} \chi (r) S(r,u_\ell) \cdots S(u_1,r) \Pi _F ^{(m+2)} (r;t_m,\cdots, t_1; s)
\no \\
&&
+ \half (-)^{m} \chi (r) S(r,u_1) \cdots S(u_\ell,r) \Pi _F ^{(m+2)} (r;t_m,\cdots, t_1; s)
\eea
Using the reflection symmetry  of the product of $\ell +1$ Szeg\"o kernels,
\bea
S(r,u_\ell) \cdots S(u_1,r) = (-)^{\ell +1} S(r,u_1) \cdots S(u_\ell,r)
\eea
it is clear that these terms cancel in $\p_{\bar r} \Pi_+ ^{(m +1| \ell +1)}$.
The remaining terms are
\bea
&&
2 \p _{\bar r} \tilde \Pi_+ ^{(m+1|\ell+1)} (s;t_1,\cdots, t_m, r, u_1, \cdots, u_\ell)
\no \\ && \hskip 0.1in =
(-)^{m+\ell +1} G^{m+1}
\bigg [ \mu (r) G^{\ell +1} (r,u_\ell , \cdots, u_1,r)
- \half \chi (r) \Pi _F ^{(\ell +2)} (r; u_\ell, \cdots, u_1;r) \bigg ]
\no \\ && \hskip 0.3in
+ (-)^{m +1} G^{m+1}
\bigg [ \mu (r) G^{\ell +1} (r,u_1 , \cdots, u_\ell,r)
- \half \chi (r) \Pi _F ^{(\ell +2)} (r; u_1, \cdots, u_\ell;r) \bigg ] \hskip 0.3in
\eea
where we  use the abbreviation $G^{m+1} = G^{m+1} (r,t_m, \cdots, t_1,s)$.
We recognize the coefficient of $(-)^{m+1} G^{m+1}$ in the above expression
as the $\p_{\bar r}$ derivative of
$\Pi ^{(\ell +2)} (r;u_1,\cdots, u_\ell;r) = (-)^\ell \Pi ^{(\ell +2)} (r;u_\ell,\cdots, u_1;r)$.
As  a  result, we have shown that,
\bea
 \p _{\bar r}  \Pi_+ ^{(m+1|\ell+1)} (s;t_1,\cdots, t_m, r, u_1, \cdots, u_\ell) =0
\eea
and the desired assertion follows.

\sm

An alternative formula for the blocks $\Pi _+$ may be obtained using the
recursion relation (\ref{recurPi}) for $\Pi$ itself. One finds,
\bea
\label{Piplus1}
&&
2 \Pi_+ ^{(m+1|\ell+1)} (s;t_1,\cdots, t_m, r, u_1, \cdots, u_\ell;r)
\\ && \hskip 0.1in =
\Pi ^{(m+2)} (s; t_1 , \cdots , t_m;r) \bigg [ G^{\ell +1}( r, u_1 , \cdots , u_\ell, r)
+(-)^\ell G^{\ell +1}( r, u_\ell , \cdots , u_1, r) \bigg ]
\no \\ && \hskip 0.3in
- (-)^m \Pi ^{(m+2)}_F  (r; t_m , \cdots , t_1;s)
\bigg [ \Pi ^{(\ell +2)}_F ( r; u_1 , \cdots , u_\ell; r)
+(-)^\ell \Pi ^{(\ell +2)}_F ( r; u_\ell , \cdots , u_1; r)  \bigg ]
\no \\ && \hskip 0.3in
- (-)^{m} G^{m+1} (r,t_m, \cdots, t_1,s) \Pi ^{(\ell +2)} (r;u_1, \cdots, u_\ell;r)
\no \eea
Notice the sign change in the last term compared to the last term in (\ref{Piplus}).

\subsection{The blocks $\Pi ^{(\ell+1)}_+$}

There is a natural extension of the connectivity of the blocks $\Pi_+^{(m+1|\ell+1)}$
to the case where formally $m=-1$; we shall denote these blocks by $\Pi ^{(\ell+1)}_+$.
They may be simply defined by
\bea
\Pi ^{(\ell+1)}_+ (s;u_1, \cdots , u _\ell) = \Pi ^{(\ell+2)} (s; u_1, \dots, u_\ell; s)
\eea
They are $(0,0)$ forms in $s$, and holomorphic $(1,0)$ forms in $u_1, \cdots, u_\ell$.
Because of the symmetry of the block $\Pi$ itself, $\Pi ^{(\ell+2)} (s; u_1, \dots, u_\ell; s)
= (-)^\ell \Pi ^{(\ell+2)} (s; u_\ell, \dots, u_1; s)$, it is manifest that
$\Pi ^{(\ell+1)} _+ (s;u_1, \cdots, u_\ell) =0$ for $\ell$ odd.

\subsection{The blocks $\Pi _-^{(m+1|\ell+1)}$}

The construction of the blocks  $\Pi _-^{(m+1|\ell+1)}$ originates  with
the symmetrization of the block $\Pi ^{(m+\ell +3)}$ given in (\ref{linkblocks}).
The blocks $\Pi _-^{(m+1|\ell+1)} $ have the following  reflection symmetry,
\bea
\label{symmPiminus}
\Pi_-^{(m+1|\ell+1)}(s;t_1,\cdots,t_m,r,u_1,\cdots,u_\ell)
= - (-)^\ell\Pi_-^{(m+1|\ell+1)}(s;t_1,\cdots,t_m,r,u_\ell,\cdots,u_1)
\eea
Using the generalized recursion relation (\ref{recurPi}), the symmetrized
combination on the right hand side of $\Pi _- ^{(m+1|\ell+1)}$ in (\ref{linkblocks})
may alternatively be expressed as
\bea
\label{Piminus1}
&&
- \Pi ^{(m+2)} (s; t_1 , \cdots , t_m;r) \bigg [ G^{\ell +1}( r, u_1 , \cdots , u_\ell, r)
+ (-)^\ell G^{\ell +1}( r, u_\ell , \cdots , u_1, r) \bigg ]
\\ &&
+ (-)^m \Pi ^{(m+2)}_F  (r; t_m , \cdots , t_1;s)
\bigg [ \Pi ^{(\ell +2)}_F ( r; u_1 , \cdots , u_\ell; r)
- (-)^\ell \Pi ^{(\ell +2)}_F ( r; u_\ell , \cdots , u_1; r)  \bigg ]
\no \eea
Notice that the terms of the type $G^{m+1} \Pi ^{(\ell +2)} (r;u_1, \cdots, u_\ell;r)$,
which occurred for $\Pi _+$ in (\ref{Piplus}), do not survive the symmetrization
needed for $\Pi _-$.
There is, however, another term, of the form $\Pi ^{(m+2)} S^{\ell +1}$, which naturally
arises in the chiral blocks and
which does respect the symmetry of $\Pi_-$. Including this term, we shall use the
following intermediate expression $\tilde \Pi _-$ to complete the symmetrized form of
(\ref{Piminus1}) into a holomorphic form in $r$,
\bea
\label{Piminustilde}
&&
2 \tilde \Pi_- ^{(m+1|\ell+1)} (s;t_1,\cdots, t_m, r, u_1, \cdots, u_\ell)
\\ && \hskip 0.1in =
- \Pi ^{(m+2)} (s; t_1 , \cdots , t_m;r) \bigg [  G^{\ell +1}( r, u_1 , \cdots , u_\ell, r)
- (-)^\ell G^{\ell +1}( r, u_\ell , \cdots , u_1, r) \bigg ]
\no \\ && \hskip 0.2in
+ (-)^m \Pi ^{(m+2)}_F  (r; t_m , \cdots , t_1;s)
\bigg [ \Pi ^{(\ell +2)}_F ( r; u_1 , \cdots , u_\ell; r)
- (-)^\ell \Pi ^{(\ell +2)}_F ( r; u_\ell , \cdots , u_1; r)  \bigg ]
\no \\ && \hskip 0.2in
+ 2 \Pi ^{(m+2)} (s;t_1, \cdots, t_m;r) S(r,u_1) S(u_1,u_2) \cdots S(u_{\ell-1}, u_\ell) S(u_\ell , r)
\no \eea
This expression is defined for $\ell \geq 1$ and $m \geq 0$; in the case $m=0$,
this singly linked chain consists of a loop containing the $\ell +1$ points
$u_1, \cdots, u_\ell$ and $r$, attached to a linear chain between $r$ and $s$.
$\tilde \Pi_-^{(m+1|\ell+1)}$ satisfies the same symmetry relation as $\Pi ^{(m+1|\ell+1)}$ does
in (\ref{symmPiminus}). It is holomorphic in $t_1,\cdots, t_m$ and $u_1, \cdots, u _\ell$,
but  fails to be holomorphic in $r$. Its $\p _{\bar r}$ derivative is readily calculated
from the derivatives of $\Pi$ and $\Pi_F$. Neglecting contact terms between
coincident points, we have
\bea
\p_{\bar r} \tilde \Pi_- ^{(m+1 | \ell+1)} (s;t_1, \cdots , t_m, r, u_1, \cdots ,u_\ell)
 =
(-)^m G^{m+1} B _- ^{(\ell+1)} (r;u_1,\cdots, u_\ell)
\eea
where we have used the abbreviation $G^{m+1} = G^{m+1}(r,t_m,\cdots, t_1,s)$.
The function $B _-^{(\ell+1)}$ may be calculated explicitly, and takes the form,
\bea
B _-^{(\ell+1)} (r;u_1,\cdots, u_\ell)
& = &
 \mu(r)  S(r,u_1) \cdots S(u_\ell, r)
 \\ &&
- \half \mu(r) \bigg [ G^{\ell +1} (r,u_1,\cdots, u_\ell, r)
- (-)^\ell G^{\ell +1} (r,u_\ell ,\cdots, u_1, r)
\bigg ]
\no \\ &&
+ {1 \over 4} \chi (r)
\bigg [ \Pi _F ^{(\ell +2)} (r;u_1,\cdots , u_\ell;r)
- (-)^\ell \Pi _F ^{(\ell +2)} (r;u_\ell ,\cdots , u_1;r) \bigg ] \quad
\no
\eea
Remarkably, the function $B _- ^{(\ell +1)}$ is independent of $m$.
In part, this is thanks to the addition of the term $\Pi ^{(m+2)} S^{\ell+1}$
in (\ref{Piminustilde}). In the subsequent subsection, we shall solve the
equation that formally corresponds to the case $m=-1$,
in terms of a new block $\Pi _- ^{(\ell +1)}$, defined by
\bea
\label{defPiminus}
\p_{\bar r} \Pi_- ^{( \ell+1)} (r; u_1, \cdots ,u_\ell)
& = & B _- ^{(\ell+1)} (r;u_1,\cdots, u_\ell)
\no \\
\Pi_- ^{( \ell+1)} (r; u_\ell, \cdots ,u_1) & = &
(-)^{\ell +1}\Pi_- ^{( \ell+1)} (r; u_1, \cdots ,u_\ell)
\eea
The block $\Pi _- ^{(\ell +1)} (r;u_1 , \cdots, u_\ell)$ will be holomorphic in
$u_1, \cdots, u_\ell$ with simple poles only at coincident vertex insertion points.
Given this block, it is now straightforward to construct the desired block
$\Pi _- ^{(m+1|\ell +1)}$, and it is given by
\bea
&&
\Pi_- ^{(m+1|\ell+1)} (s;t_1,\cdots, t_m, r, u_1, \cdots, u_\ell)
\no \\ && \hskip 0.6in =
\tilde \Pi_- ^{(m+1|\ell+1)} (s;t_1,\cdots, t_m, r, u_1, \cdots, u_\ell)
\no \\ && \hskip 0.8in
- (-)^m G^{m+1}(r,t_m,\cdots, t_1,s) \Pi _- ^{(\ell +1)} (r; u_1, \cdots ,u_\ell)
\eea

\subsection{The blocks $\Pi_-^{(\ell +1)}$}

To construct the blocks $\Pi _- ^{(\ell +1)} (r;u_1,\cdots, u_\ell)$, we start from its
defining equation (\ref{defPiminus}), and render all $r$-dependence explicit.
To so so, we use the generalized recursion relation (\ref{PiFrecursion}) for
$m=0$ and $\ell \to \ell -1$, to obtain,
\bea
\Pi _F ^{(\ell +2)} (r;u_1, \cdots, u_\ell ;r) & = &
S(r,u_1) G(u_\ell, r) \Pi _F ^{(\ell)} (u_1; u_2, \cdots, u_{\ell-1};  u_\ell)
\no \\ &&
+ \FF (r,u_1) G(u_\ell, r) G^{\ell -1} (u_1, \cdots, u_\ell)
\no \\ &&
+ S(r,u_1) \FF (u_\ell, r) S^{\ell-1} (u_1, \cdots, u_\ell)
\eea
Next, we substitute this expressions (and its mirror with $u_i \to u_{\ell -i +1}$)
for $\Pi _F ^{(\ell+1)}$ on the right hand side of (\ref{defPiminus}).
The result for $\p_{\bar r} \Pi _- ^{(\ell +1)} (r;u_1,\cdots, u_\ell)$
is then given by,
\bea
\label{derPiminus}
\p_{\bar r} \Pi _- ^{(\ell +1)} (r;u_1,\cdots, u_\ell)
& = &
+ \mu (r)  S^{\ell+1} (r,u_1, \cdots, u_\ell,r)
\no \\ &&
- \half \mu(r)  G^{\ell+1}(r, u_1,\cdots, u_\ell,r)
\no \\ &&
+ \half (-)^\ell \mu(r)  G^{\ell+1}(r, u_\ell,\cdots, u_1,r)
\no \\ &&
+{1 \over 4} \chi (r) S(r,u_1) G(u_\ell, r) \Pi _F ^{(\ell)} (u_1; u_2, \cdots, u_{\ell-1}; u_\ell)
\no \\ &&
+{1 \over 4} \chi (r) \FF (r,u_1)  G ^\ell (u_1,\cdots, u_\ell,r)
\no \\ &&
+{1 \over 4} \chi (r)  \FF (u_\ell, r) S ^\ell (r, u_1,\cdots, u_\ell)
\no \\ &&
-{1 \over 4} (-)^\ell \chi (r) S (r,u_\ell) G (u_1, r) \Pi _F ^{(\ell)} (u_\ell; u_{\ell-1}, \cdots,u_2; u_1)
\no \\ &&
-{1 \over 4} (-)^\ell \chi (r) \FF (r,u_\ell)  G ^\ell (u_\ell,\cdots, u_1,r)
\no \\ &&
-{1 \over 4} (-)^\ell \chi (r)  \FF (u_1, r) S ^\ell (r,u_\ell,\cdots, u_1)
\eea
The dependence on $r$ on the right hand side of (\ref{derPiminus})
is typical of that of the derivatives of the three-point functions $Q_0, Q_B,Q_F$ and $Q_\ff$.
We recall their expressions, neglecting any contact terms at coincident vertex
insertion points,
\bea
\label{derLQ}
\p_{\bar r} Q_0 (r;u_1,u_\ell) & = & 0
\no \\
\p_{\bar r} Q_B (r;u_1,u_\ell) & = &
+ \mu (r) S(r,u_1) S(u_\ell, r)
\no \\ &&
- { 1 \over 4} \chi (r) S(u_\ell, r) \FF (u_1,r)
- { 1 \over 4} \chi (r) S(u_1, r) \FF (u_\ell,r)
\no \\
\p_{\bar r} Q_F (r;u_1,u_\ell) & = &
- { 1 \over 4} \chi (r) S(u_\ell , r) G (u_1,r)
\no \\
\p_{\bar r} Q_\ff (r;u_1,u_\ell) & = &
+ \half \mu (r) G (r,u_1) G (u_\ell, r)
- { 1 \over 4} \chi (r) \FF (r,u_1) G(u_\ell,r)
\eea
Comparison of the right hand sides of (\ref{derPiminus}) and (\ref{derLQ})
allows us to construct a particular solution $\Pi _- ^{(\ell +1)}$ to (\ref{derPiminus}),
given by
\bea
\label{bb5}
\tilde \Pi_- ^{(\ell+1)} (r;u_1, \cdots, u_\ell)
& = &
+ Q_B(r;u_1,u_\ell) S^{\ell-1} (u_1,\cdots, u_\ell)
\\ &&
- Q_\ff (r;u_1,u_\ell) G^{\ell-1} (u_1, \cdots, u_\ell)
\no \\ &&
+ (-)^\ell Q_\ff (r;u_\ell, u_1)  G^{\ell-1}  (u_\ell, \cdots, u_1)
\no \\ &&
+ Q_F(r;u_\ell,u_1) \Pi _F ^{(\ell)} (u_1; u_2, \cdots, u_{\ell-1}; u_\ell)
\no \\ &&
- (-)^\ell Q_F (r;u_1,u_\ell) \Pi ^{(\ell)} _F (u_\ell; u_{\ell-1},  \cdots, u_2; u_1)
\no
\eea
As a result, the general solution for (\ref{derPiminus}) is of the form,
\bea
\Pi_- ^{(\ell+1)} (r;u_1, \cdots, u_\ell)
=
\tilde \Pi_- ^{(\ell+1)} (r;u_1, \cdots, u_\ell) + \hat \Pi _- ^{(\ell+1)} (r;u_1, \cdots, u_\ell)
\eea
where $\hat \Pi _- ^{(\ell+1)} (r;u_1, \cdots, u_\ell)$ is holomorphic in $s$, apart from
contact terms at coincident vertex insertion points, and satisfies the symmetry
property of (\ref{derPiminus}).

\subsubsection{Holomorphicity in $u_1$ and $u_\ell$}

The homogeneous solution $\hat \Pi _- ^{(\ell+1)} (r;u_1, \cdots, u_\ell)$ is not
arbitrary, however, but is further constrained by the requirement that the full blocks
$\Pi _- ^{(\ell+1)} (r;u_1, \cdots, u_\ell)$ be holomorphic in all the points
$u_1, \cdots, u_\ell$. This requirement is automatic for the points
$u_2, u_3, \cdots, u_{\ell-2}, u_{\ell-1}$, but not for the points $u_1,u_\ell$.
Thus,  it will suffice to seek a term $\hat \Pi _- ^{(\ell+1)} (r;u_1, \cdots, u_\ell)$
which is holomorphic in $s$, and $u_2, u_3, \cdots, u_{\ell-2}, u_{\ell-1}$, has
the symmetry properties of (\ref{derPiminus}), and compensates for the
non-holomorphicity of $\tilde \Pi _- ^{(\ell+1)} (r;u_1, \cdots, u_\ell)$ in $u_1$.
Given the symmetry property (\ref{derPiminus}), the resulting
$\Pi _- ^{(\ell+1)} (r;u_1, \cdots, u_\ell)$ will then be automatically
holomorphic in $u_\ell$ as well. We begin by computing the $u_1$-derivative
of $\tilde \Pi_- ^{(\ell+1)} (r;u_1, \cdots, u_\ell)$,
\bea
\label{uderPi}
\p_{\bar u_1} \tilde \Pi_- ^{(\ell+1)}
& = &
Q_0 (r;u_1,u_\ell) \bigg [ \half \chi (u_1) \Pi _F ^{(\ell)} (u_1; \cdots ; u_\ell)
- \mu (u_1) G^{\ell-1} (u_1, \cdots, u_\ell) \bigg ]
\no \\ &&
+ G(u_1,r) \bigg [ \mu (u_1) S^\ell (u_1,  u_2, \cdots, u_\ell , u_1)
- (-)^\ell \mu (u_1) G^\ell (u_1, u_2, \cdots, u_\ell, u_1)
\no \\ && \hskip 0.8in
+ \half (-)^\ell \chi (u_1)  \Pi _F ^{(\ell+1)} (u_1;u_\ell, u_{\ell-1}, \cdots, u_2;u_1)
\bigg ]
\eea
where we have used the recursion relation (\ref{PiFrecursion}) in the form
\bea
&&
\FF (u_1, u_\ell) G^{\ell-1} (u_\ell, \cdots, u_1) + S(u_1,u_\ell)
\Pi _F ^{(\ell) } (u_\ell; \cdots ; u_1)
\no \\ && \hskip 0.5in =
\Pi _F ^{(\ell+1) } (u_1;u_\ell \cdots ; u_1)
\eea
to combine two of the terms.

\sm

The observation that the $r$-dependence of the right hand side of (\ref{uderPi})
is holomorphic makes it possible to complete the block $\Pi _- ^{(\ell+1)}$
into a block with the desired holomorphicity and symmetry properties.

\sm

The term proportional to $Q_0 (r;u_1,u_\ell)$ in (\ref{uderPi}) may be recast
as a $\p _{\bar r}$-derivative,
\bea
\label{Lterm}
&&
Q_0 (r;u_1,u_\ell) \bigg [ \half \chi (u_1) \Pi _F ^{(\ell)} (u_1; \cdots ; u_\ell)
- \mu (u_1) G^{\ell-1} (u_1, \cdots, u_\ell) \bigg ]
\no \\ && \hskip 0.5in =
\p_{\bar u_1} \bigg ( - Q_0 (r;u_1,u_\ell) \Pi ^{(\ell)} (u_1; u_2, \cdots, u_{\ell-1};u_\ell) \bigg )
\eea
Note that $Q_0$ is not holomorphic in $u_1$, even up to contact terms in coincident
vertex insertion points, because there is a residual $\delta$-function at $w_0$,
\bea
\p_{\bar u_1}  Q_0(r;u_1,u_\ell) = - G(u_\ell, r) \bigg ( 2 \delta (u_1,u_\ell) - \delta (u_1,r)
- \delta (u_1,w_0) \bigg )
\eea
Here, $w_0$ is the point on $\Sigma$ where $G(z,w_0)=0$ for all $z$,
so that also $\ff(z,w_0)= \FF (z,w_0)=0$ and thus also $\Pi ^{(\ell)}
(w_0;u_2,\cdots , u_{\ell-1};u_\ell)=0$. In view of the last equality,
the pole in $u_1$ at $w_0$ which occurs in $Q_0(r;u_1,u_\ell)$ is cancelled
by a corresponding zero in $\Pi ^{(\ell)} (u_1;u_2,\cdots , u_{\ell-1};u_\ell)$,
thereby yielding (\ref{Lterm}) through differentiation of $\Pi ^{(\ell)}$ only.

\sm

To identify the term proportional to $G(u_1,r)$ in (\ref{uderPi}), we relate it
to the block $\tilde \Pi _- ^{(\ell)} (u_1;u_2, \cdots, u_\ell)$. Specifically, the bracket on the
second and third lines of the right hand side of (\ref{uderPi}) equals,
\bea
\bigg [ \cdots \bigg ] =
B _- ^{(\ell)} (u_1;u_2, \cdots, u_\ell)
+ \half \p_{\bar u_1} \Pi ^{(n+1)} (u_1; u_2, \cdots, u_\ell;u_1)
\eea
as may be verified by explicit calculation.

\subsubsection{General recursive formula for $\Pi ^{(\ell+1)}_-$}

Putting all together, using the fact that $B _- ^{(\ell)} (u_1;u_2, \cdots, u_\ell)
= \p_{\bar u_1} \tilde \Pi ^{(\ell)} _- (u_1;u_2, \cdots, u_\ell)$, and the
symmetry of (\ref{uderPi}), we find the following candidate expression for
$\Pi _- ^{(\ell+1)} (r;u_1, \cdots, u_\ell) $,
\bea
&&
 \tilde \Pi _- ^{(\ell+1)} (r;u_1, \cdots, u_\ell)
+ Q_0 (r;u_1,u_\ell) \Pi ^{(\ell)} (u_1; u_2, \cdots, u_{\ell-1};u_\ell)
\\ &&
- G(u_1,r) \tilde \Pi _- ^{(\ell)} (u_1;u_2, \cdots, u_\ell)
+ (-)^\ell G(u_\ell,r) \tilde \Pi _- ^{(\ell)} (u_\ell;u_{\ell-1}, \cdots, u_1)
\no \\ &&
- \half G(u_1,r) \Pi ^{(\ell +1)} (u_1;u_2, \cdots, u_\ell;u_1)
+ \half (-)^\ell G(u_\ell,r) \Pi ^{(\ell+1)} (u_\ell;u_{\ell-1} , \cdots, u_1;u_\ell)
\no \eea
This result is now holomorphic in $r$, $u_1$ and $u_\ell$ by construction.

\sm

The presence of the terms in $\tilde \Pi ^{(\ell)} _-$ spoils the
holomorphicity  in the points $u_2$ and $u_{\ell-1}$ so that the above
formula cannot yet be the complete expression for
$\Pi _- ^{(\ell+1)} (r;u_1, \cdots, u_\ell) $.

\sm

But it is now manifest how the full
expression should be obtained : one should recursively treat the blocks
$\tilde \Pi ^{(\ell)} _-$, just as one did with the block $\Pi ^{(\ell+1)}_-$,
resulting in a linear recursion relation directly for the blocks,
\bea
\label{Piminusrec}
\Pi _- ^{(\ell+1)} (r;u_1, \cdots, u_\ell)
\! \! & = &  \! \tilde \Pi _- ^{(\ell+1)} (r;u_1, \cdots, u_\ell)
+ Q_0 (r;u_1,u_\ell) \Pi ^{(\ell)} (u_1; u_2, \cdots, u_{\ell-1};u_\ell)
\no \\ &&
- G(u_1,r)  \Pi _- ^{(\ell)} (u_1;u_2, \cdots, u_\ell)
+ (-)^\ell G(u_\ell,r)  \Pi _- ^{(\ell)} (u_\ell;u_{\ell-1}, \cdots, u_1)
\no \\ &&
- \half G(u_1,r) \Pi ^{(\ell +1)} (u_1;u_2, \cdots, u_\ell;u_1)
\no \\ &&
+ \half (-)^\ell G(u_\ell,r) \Pi ^{(\ell+1)} (u_\ell;u_{\ell-1} , \cdots, u_1;u_\ell)
\eea
The recursive argument proceeds as follows. Assuming that
$\Pi _- ^{(\ell)} (r;u_1, \cdots, u_{\ell-1})$ is holomorphic in $u_1,\cdots , u_{\ell-1}$,
and that
\bea
\label{defPiminus1}
\p_{\bar r} \Pi_- ^{( \ell)} (r; u_1, \cdots ,u_{\ell-1})
& = & B_- ^{(\ell)} (r;u_1,\cdots, u_{\ell-1})
\no \\
\Pi_- ^{( \ell)} (r; u_{\ell-1}, \cdots ,u_1) & = &
(-)^\ell \Pi_- ^{( \ell)} (r; u_1, \cdots ,u_{\ell-1})
\eea
then the block $\Pi _- ^{(\ell+1)} (r;u_1, \cdots, u_\ell) $ constructed by
(\ref{Piminusrec}) obeys (\ref{defPiminus}), and is holomorphic in
all points $u_1, \cdots, u_\ell$, up to contact terms between coincident
vertex insertion points. It remains to verify that these properties also
hold on the lowest block $\Pi ^{(2)} _- (r;u)$, which we do next.

\subsubsection{The blocks $\Pi ^{(2)} _- $ and $\Pi ^{(3)} _- $}

Following the above prescriptions for the blocks $\Pi _- ^{(\ell+1)}$,
we compute the two simplest cases. By their very definition, the first
block vanishes,
\bea
 \Pi _- ^{(1)} (r) & = &0
\no \\
\tilde \Pi _- ^{(2)} (r;u) & = & Q_B(r;u,u) - 2 Q_\ff (r;u,u)
\no \\
\tilde \Pi _- ^{(3)} (r;u,v) & = & Q_B(r;u,v) S(u,v)
\no \\ &&
-  Q_\ff (r;u,v) G(u,v) + Q_\ff (r;v,u) G(v,u)
\no \\ &&
+ Q_F (r;v,u) \FF (u,v) - Q_F (r;u,v) \FF( v,u)
\eea
Use of the recursion relation (\ref{Piminusrec}) then gives us the expressions
for the full blocks,
\bea
\Pi _- ^{(2)} (r;u) & = & Q_B(r;u,u) - 2 Q_\ff (r;u,u) - G(u,r) \ff (u,u)
\no \\
\Pi _- ^{(3)} (r;u,v) & = & Q_B(r;u,v) S(u,v) + Q_0 (r;u,v) \ff (u,v)
\no \\ &&
-  Q_\ff (r;u,v) G(u,v) + Q_\ff (r;v,u) G(v,u)
\no \\ &&
+ Q_F (r;v,u) \FF (u,v) - Q_F (r;u,v) \FF( v,u)
\no \\ &&
- G(u,r) \Pi _- ^{(2)} (u;v) + G(v,r) \Pi _- ^{(2)} (v;u)
\eea
Here, we have used the fact that $\Pi ^{(3)}(u;v;u)=0$ by symmetry.

\sm

Holomorphicity of $\Pi _- ^{(2)} (r;u)$ is slightly complicated by the fact
that functions occur which are being evaluated at coincident points. Their
derivatives are readily worked out and we have $\p _{\bar u} \Pi _- ^{(2)} (r;u)=0$,
in view of,
\bea
\p_{\bar u} \ff (u,u)
& = &
- \chi (u) \FF (u,u) + \p_u \mu(u) - 2 \mu(u) \p_u \ln E(u,w_0)
\\
\p_{\bar u} Q _B (r;u,u)
& = &
- \chi (u) Q _F (r;u,u) - \p_u \bigg ( \mu(u) G(u,r) \bigg )
\no \\
\p_{\bar u} Q _\ff (r;u,u)
& = &
- \half \chi (u) Q _F (r;u,u)
+ \half \chi (u) \FF (u,u) G(u,r)
\no \\ &&
- \p_u  \mu(u) G(u,r)
- \half \mu(u) \p_u G(u,r)
+ \mu(u) G(u,r) \p_u \ln E(u,w_0)
\no
\eea
Holomorphicity in $u$ and $v$ of $\Pi _- ^{(3)}(r;u,v)$ may be verified directly.

\newpage

\section{Extraction of  Exact Differentials  \\
{\it Part I ~ Linear Chain blocks} }
\setcounter{equation}{0}

We return now to the task of showing that the cohomology class
of the chiral block contains a pure, holomorphic, $\otimes_{r=1}^N (1,0)_r$
representative, modulo chirally exact terms, that is,
terms which are de Rham exact in at least one and possibly two insertion
points, and are pure $(1,0)$ and holomorphic type in all the others.
Now, from \cite{V}, we know that the chiral block ${\cal F}[\delta]$
is always a closed form in each $z_r$. Thus, if we can show that there exist
chirally exact forms $d_r{\cal S}_r[\delta]$ and $d_rd_s{\cal S}_{rs}[\delta]$
so that
\bea
{\cal F}[\delta]-\bigg(\sum_r d_r{\cal S}_r[\delta]+\sum_{[rs]}
d_rd_s{\cal S}_{rs}[\delta]\bigg)
~ \in ~ \bigotimes_{r=1}^N (1,0) _r
\eea
it will automatically follow that the left-hand side
is holomorphic in all $z_r$, and is the holomorphic representative
that we seek. This is the strategy which we shall adopt.

\sm

The discussion of the extractions of the chirally exact differentials is
divided into two parts. In the first part, the necessary Wick contractions
of both bosonic and fermionic fields will be carried out. Some simple
cases with small number of vertex insertion points will be derived in
detail, and the general contribution of linear chain differential
blocks will be proven. In the second part, the remaining blocks will
be used to extract the final differential blocks involving $\Pi _\pm$.

\subsection{A lemma for bosonic Wick contractions}

Throughout, we shall need to perform partial contractions of the bosonic field
$x_+$, through the composite $X^\mu _s$, which is defined by
\bea
X^\mu _s \equiv i k^\mu _s + K^\nu _s K^\mu _s p^\nu _I \o_I(s)
+ K^\nu _s K^\mu _s \p_s x^\nu _+ (s)
\eea
To organize the combinatorics of these contractions, we first prove
the following,

\begin{lemma}
\label{bosoniclemma}
The contractions of the composite $X^\mu _s$ are given by
\bea
\label{lemma1}
Z_0 ^{r_1 \cdots r_n s} X_s ^\mu & = & Z_0 ^{r_1 \cdots r_n s} \bar X_s ^\mu
+ \sum _{t \not \in \{ r_1\cdots r_n s \} } Z_0 ^{r_1 \cdots r_n st }
K^\mu _s K^\nu _s X^\nu _t G(s,t)
\no \\ && \hskip 0.78in
+ \sum _{t \not \in \{ r_1\cdots r_n s \} } \p_t \bigg ( Z_0 ^{r_1 \cdots r_n st } \ep ^\nu _t \theta _t
K^\nu _s K^\mu _s G(s,t) \bigg )
\eea
where $\bar X^\mu _s$ stands for the instruction that its $x_+$ field is to be
contracted only with the exponentials at $r_1\cdots r_n$.
\end{lemma}

To prove Lemma 1, it suffices to contract the $x_+$ field in $X^\mu _s$
with $Z_0 ^{r_1 \cdots r_n s}$ at all insertion points $t$ which do not belong
to the set $\{ r_1 \cdots r_n s\}$. This contribution is given by
\bea
Z_0 ^{r_1 \cdots r_n s} \left ( X_s ^\mu - \bar X_s ^\mu \right )
 =  Z_0 ^{r_1 \cdots r_n s} K^\nu _s K^\mu _s \p_s x_+^\nu (s)
\eea
with the above instruction on $\p_s x_+^\nu (s)$. It is readily evaluated,
and we get
\bea
Z_0 ^{r_1 \cdots r_n s} \left ( X_s ^\mu - \bar X_s ^\mu \right )
\! &  \! = \!   & \!
 \sum _{t \not \in \{ r_1\cdots r_n s \} } Z_0 ^{r_1 \cdots r_n st } K^\nu _s K^\mu _s
 \bigg [
 -ik^\nu _t G(s,t) \left ( 1 + \ep ^\sigma_t \theta _t \p_t \bar x^\sigma _+ (t) \right )
 \no \\ && \hskip 2in
- \ep ^\nu _t \theta _t \p_t G(s,t) \bigg ]
\eea
Here, the first term in the brackets $[~]$ arises from the contractions with
the exponential, and from the expansion of the un-contracted derivative contribution
in $Z_0 ^{r_1 \cdots r_n s}$ at $t$, while the second term arises from the
contraction of the derivative contribution
in $Z_0 ^{r_1 \cdots r_n s}$ at $t$. Pulling out the differential in $t$, we have
\bea
Z_0 ^{r_1 \cdots r_n s} \left ( X_s ^\mu - \bar X_s ^\mu \right )
\! & \!   = \!   & \!
\sum _{t \not \in \{ r_1\cdots r_n s \} }
\p_t \bigg ( Z_0 ^{r_1 \cdots r_n st } K^\nu _s K^\mu _s \ep ^\nu _t \theta _t G(s,t) \bigg )
\\ &&
-  \sum _{t \not \in \{ r_1\cdots r_n s \} } \! \!
Z_0 ^{r_1 \cdots r_n st } K^\nu _s K^\mu _s
\bigg [ ik^\nu _t +i (k^\nu _t \ep ^\sigma _t \theta _t - k^\sigma _t \ep ^\nu _t \theta _t)
\p_t \bar x_+^\sigma (t) \bigg ] G(s,t)
\no
\eea
The object inside the brackets $[~]$ equals $X^\nu _t$, whence
the result of Lemma~1.

\smallskip

The remaining contractions resulting from $\bar X^\mu_s$ are
simple to express, and we have,
\bea
Z_0 ^{r_1 \cdots r_n s} \bar X_s ^\mu =
Z_0 ^{r_1 \cdots r_n s} \bigg ( i k^\mu _s + K^\nu _s K^\mu _s p^\nu _I \o_I(s)
 -i  K^\nu _s K^\mu _s \sum _{j=1}^n   k_{r_j}^\nu  G(s,r_j) \bigg )
\eea
It will be convenient to introduce a further notation $\hat X_s ^\mu$, defined by
\bea
\bar X_s ^\mu = \hat X^\mu _s + K^\nu _s K^\mu _s p^\nu _I \o_I(s)
\eea
It will allow us to deal directly with the non-trivial $p$-dependence
of the blocks.

\subsection{Extracting the simplest blocks}

Before embarking on the extraction of the general blocks $\Pi$, $\Pi_I$,
and $\Pi _\pm$ from the differential terms $\cD$ in (\ref{master2}), it will
be helpful to understand the extraction mechanism on the simplest blocks first.
This will also help in clarifying the ultimate fate of the total differential
terms that arise in the second line of Lemma 1 formula (\ref{lemma1}).

\sm

$(1)$ The simplest block is $\cD_1$. As it stands in (\ref{master2}),
this block is already manifestly an exact differential in the point $r$,
and holomorphic in all other insertion points. It readily produces a first
differential block,
\bea
\cD _1 & = & \sum _r d_r \cS_r ^{(1)}
\no \\
\cS_r ^{(1)} & = & - Z_0 ^r p_I ^\mu \ep ^\mu _r \t_r \Pi ^{(1)} _I (r)
\eea
where we have used the result $\Pi _I ^{(1)} (r) = \lambda _I (r)$
of (\ref{PiIlow}).

\sm

$(2)$ Another simple block is $\cD_{2a}$, which is manifestly an exact differential
in both the points $r$ and $s$, and holomorphic in all other insertion
points. It readily produces a second differential block,
\bea
\cD_{2a} & = & \sum _{[rs]} d_r d_s S_{rs} ^{(2)}
\no \\
S_{rs}^{(2)} & = & \half Z_0 ^{rs} \ep ^\mu _r \t_r \ep ^\mu _s \t_s \Pi ^{(2)} (r,s)
\eea
where we have used the result $\Pi ^{(2)} (r,s) = \ff (r,s)$ of (\ref{Pilow}).
The remaining blocks in (\ref{master2}) are more complicated and will
require some degree of recombination
of various contributions in (\ref{master2}) to produce further suitable blocks.

\smallskip

$(3)$ A simple recombination of two blocks is obtained by regrouping the
differential block $\cD_{2c}$ with that part of block $\cD_{2b}$ which is
produced by the $p$-dependence of the composite $X^\mu_r$ in $\cD_{2b}$;
we shall denote this contribution by $\cD_{2bp}$. (The remainder  $\cD_{2b}
- \cD_{2bp}$ will be dealt with later.) Combining $\cD _{2bp}$ with $\cD_{2c}$
gives,
\bea
\cD_{2bp} + \cD_{2c} =
- \sum _{[rs]} d_s \bigg ( Z_0 ^{rs} p^\nu _I \ep ^\mu _s \theta _s K^\nu _r K^\mu_r
\Big [ - \hat \o _{I0}(r) \FF (r,s) + \o_I(r) \ff (r,s) \Big ] \bigg )
\eea
Using the definition of the block in (\ref{PiIlow}),
\bea
\Pi _I ^{(2)} (r,s) =
\lambda _I (r) G(r,s)  + \o _I (r) \ff (r,s) - \hat \o_{I0} (r) \FF (r,s)
\eea
the above combination of blocks may be recast as follows,
\bea
\label{D2}
\cD_{2bp} + \cD_{2c} & = & \sum _s d_s \cS_s ^{(2)} +
\sum _{[rs]} \p_s \bigg ( Z_0 ^{rs} p^\nu _I \ep ^\mu _s \theta _s K^\nu _r K^\mu_r
\lambda _I (r) G(r,s) \bigg )
\no \\
\cS^{(2)}_s & = &
 - \sum _{r \not= s} \bigg ( Z_0 ^{rs} p^\nu _I \ep ^\mu _s \theta _s
K^\nu _r K^\mu_r \Pi ^{(2)} _I (r,s) \bigg )
\eea
The block $d_s \cS^{(2)}_s$ is exact in $s$ and holomorphic in $r$ and is of the type
we seek. The additional term with a $\p_s$ derivative which arises on the
first line of (\ref{D2}) is holomorphic in $s$, but not holomorphic in $r$.
At first sight, the appearance of this term looks worrisome.
Actually, it is a term which is of type $(1,0)$ in every insertion point,
and it must be recombined with the $\cZ$ terms, which are all of this type.
It will turn out that this purely $(1,0)$ derivative term, and others that will arise
from later recombinations, cancel similar terms that arise from the
contractions of terms in $\cZ$. This mechanism will be illustrated in
section 10 for these simple cases, and will be instrumental in obtained general
systematic  formulas for holomorphic blocks, to be worked out in the next paper.

\sm

$(4)$ The remainder of $\cD_{2b}- \cD_{2bp}$ is given by partial
contraction using the Lemma 1,
\bea
\label{D2b}
\cD_{2b} - \cD_{2bp} & = &
- \sum _{[rs]} d_s \biggl ( Z_0 ^{rs}  \ff (r,s) \ep ^\mu _s
\theta _s \hat X^\mu _r \biggr )
\no \\ &&
- \sum _{[rst]}  d_s \bigg ( Z_0 ^{rst} \ep ^\mu _s \theta _s \ff (r,s) K^\mu _r K^\nu _r
X^\nu _t G(r,t) \bigg )
\no \\ &&
- \sum _{[rst]}  d_s \p_r \bigg ( Z_0 ^{rst} \ep ^\nu _r \theta _r  \ep_s ^\mu \theta _s
\ff (t,s) K^\nu _t K^\mu _t  G(t,r) \bigg )
\eea
In the last line, we have interchanged $r$ and $t$.
The derivative term on the last line combines with a contribution from $\cD_{3d}$.
To see this, we use the structure of the block $\Pi ^{(3)} (s;t;r)$,
\bea
\label{Pi3}
\Pi ^{(3)} (s;t;r) = - \FF (t,s) \FF (t,r) + \ff (t,s) G(t,r) - \ff (r,t) G(t,s)
\eea
to recast $\cD_{3d}$ in the following form,
\bea
\cD _{3d} & = & \sum _{[rs]} d_r d_s \cS _{rs}^{(3)} + \cD _{3d}'
\no \\
\cS _{rs}^{(3)} & = & \sum _{t \not= r,s} Z_0 ^{rst} \half \ep_s ^\mu \theta _s
 \ep ^\nu _r \theta _r  K^\mu _t K^\nu_t \Pi ^{(3)} (s;t;r)
\no \\
\cD _{3d}' & = &
- \sum _{[rst]}  d_r d_s \bigg ( Z_0 ^{rst} \ep_s ^\mu \theta _s
 \ep ^\nu _r \theta _r  K^\mu _t K^\nu_t \ff (t,s) G(t,r) \bigg )
\eea
We see that the block $d_r d_s \cS _{rs}^{(3)}$ is exact in $r$ and $s$,
and holomorphic in $t$ because $\Pi ^{(3)} (s;t;r)$ is holomorphic in $t$.
In the $\cD_{3d}'$ term, the term inside the parentheses is actually holomorphic in $r$,
so that the total differential $d_r$  acts as a holomorphic $\p_r$ differential.
Using the relation $d_s \p_r = - \p_r d_s$ on these differential forms, we see
that the last term of $\cD_{2b} - \cD_{2bp}$ is cancelled precisely by $\cD_{3d}'$.
As  a result, we have
\bea
\label{Pi4}
\cD_{2b} - \cD_{2bp} + \cD _{3d} & = &
\sum _{[rs]} d_r d_s \cS _{rs}^{(3)}
- \sum _{[rs]} d_s \biggl ( Z_0 ^{rs}  \ff (r,s) \ep ^\mu _s \theta _s \hat X^\mu _r \biggr )
\no \\ &&
- \sum _{[rst]}  d_s \bigg ( Z_0 ^{rst} \ep ^\mu _s \theta _s \ff (r,s) K^\mu _r K^\nu _r
X^\nu _t G(r,t) \bigg )
\eea

\smallskip

$(5)$ Next, the terms $\cD_{2b} - \cD_{2bp} + \cD_{3d}'$ combine with $\cD_{3e}$
as follows. In $\cD_{3e}$, we use again equation (\ref{Pi3}) to eliminate the $\FF \FF$
term, so that
\bea
\cD_{3e} & = & \cD_{3e}'
+ \sum _{[rst]}  d_s \bigg (
Z^{rst} _0  K^\mu_r K^\nu _r \ep ^\mu _s \theta _s \Big [ \ff (r,s) G(r,t) - \ff (r,t) G(r,s) \Big ]
 X ^\nu _t  \bigg )
\no \\
\cD_{3e} ' & = &
- \sum _{[rst]}  d_s \bigg (
Z^{rst} _0  K^\mu_r K^\nu _r \ep ^\mu _s \theta _s \Pi ^{(3)} (s;r;t)
 X ^\nu _t  \bigg )
\eea
The block $\cD_{3e}'$ is holomorphic in $r$, but not in $t$. It will need
to be further expanded and combined with other blocks to be rendered
holomorphic also in $t$.
As a result, the terms in $\ff (r,s)$ cancel between the second line of
(\ref{Pi4}) and $\cD_{3e} - \cD_{3e}' $, yielding
\bea
\cD_{2b} - \cD_{2bp} + \cD_{3d} + \cD_{3e}
& = &
\sum _{[rs]} d_r d_s \cS _{rs}^{(3)}  + \cD_{3e}'
- \sum _{[rs]} d_s \biggl ( Z_0 ^{rs}  \ff (r,s) \epsilon ^\mu _s
\theta _s \hat X^\mu _r \biggr )
\no \\ &&
- \sum _{[rst]}  \p_s \bigg (
Z^{rst} _0  K^\mu_r K^\nu _r \ep ^\mu _s \theta _s  \ff (r,t) G(r,s)
 X ^\nu _t  \bigg )
\eea
The $\p_s$-derivative term on the last line is a purely $(1,0)$ form in
all insertion points, and it must be recombined with the $\cZ$ terms,
which are all of this type.

\sm

$(6)$ We shall now show how the block involving
$\Pi ^{(3)}_I$ arises by combining some of these results. It is the block
involving one internal momentum. One contribution arises from the
fermionic completion of $\cD_{3a}$ with a single
Szeg\"o kernel $K_r ^\sigma K_t ^\sigma S(r,t)$, and yields
\bea
\cD_{3a}'  & = &
- \sum _{[rst]} d_s \biggl (
Z_0 ^{rst} p^\nu _I K^{\mu \nu} _{[rt]}   \ep ^\mu _s \theta _s
 \FF (r,s) S(r,t)  \hat \omega _{I0} (t) \biggr )
\eea
Combining this with the $p$-dependent term of $\cD_{3e}'$, we obtain
\bea
\cD_{3a}'  + \cD_{3ep}' =
\sum _{[rst]} d_s \biggl (
Z_0 ^{rst} p^\nu _I K^{\mu \nu} _{[rt]} \ep ^\mu _s \theta _s
\Big [ \FF (r,s) S(r,t)  \hat \omega _{I0} (t)
+ \Pi ^{(3)} (s;r;t) \o_I(t) \Big ] \biggr ) \quad
\eea
Using the definitions of $\Pi ^{(3)}$ and $\Pi ^{(3)}_I$, we deduce the
following expression for the term in the square brackets,
\bea
&&
\FF (r,s) S(r,t)  \hat \omega _{I0} (t) + \Pi ^{(3)} (s;r;t) \o_I(t)
\no \\ && \hskip 0.5in =
- \Pi ^{(3)}_I (t;r;s) + \lambda _I(t) G(t,r) G(r,s)  - \hat \o_{I0}(t) \FF (t,r) G(r,s)
\quad
\eea
As a result, we have
\bea
\cD_{3a}'  + \cD_{3ep}' & = & \sum _s d_s \cS_s ^{(3)} +
\sum _{[rst]} \p_s \biggl (
Z_0 ^{rst} p^\nu _I K^{\mu \nu} _{[rt]} \ep ^\mu _s \theta _s
\Big [  \lambda _I(t) G(t,r)
- \hat \o_{I0}(t) \FF (t,r)   \Big ] G(r,s) \biggr )
\no \\
\cS^{(3)}_s  & \equiv &
- \sum _{r \not= s} ~ \sum _{t \not= r,s} d_s \biggl (
Z_0 ^{rst} p^\nu _I K^{\mu \nu} _{[rt]} \ep ^\mu _s \theta _s \Pi ^{(3)}_I (t;r;s) \biggr )
\eea
The block $d_s \cS^{(3)}_s$ is exact in $s$. It is holomorphic in $r$ and $t$,
since $\Pi ^{(3)}_I(t;r;s)$ is. The first term inside the bracket of the first line
produces a $\p_s$ derivative term which is a pure $(1,0)$ form in all insertion points.
It will combine with terms in $\cZ$.

\subsection{General pattern for linear chain blocks}

The results obtained from the few simple cases above clearly suggest a pattern
for the structure of the linear chain blocks in $\cD$. The contributions $\cD_{dd}$
to $\cD$ with two exact differentials are given by the sum over all $n\geq 0$ of
\bea
\label{S}
\cD _{dd} & = & \sum _{n\geq0} ~ \sum _{[rs]} d_r d_s S^{(n+2)} _{rs}
\no \\
S^{(n+2)} _{rs} & = &
- \sum _{r,s \not \in [t_1 \cdots t_n ]}  \! \!
Z_0 ^{rst_1 \cdots t_n} \half \ep_r ^\mu \theta _r
 \ep ^\nu _s \theta _s  K^{\mu \nu}  _{[t_1 \cdots  t_n]}
 \Pi ^{(n+2)} (r;t_1,\cdots , t_n;s)
\eea
The contributions $\cD _{dp}$ to $\cD$ with one exact differential and one
factor linear in the internal momenta $p$ is given by,
\bea
\label{Sp}
\cD _{dp} & = & \sum _{n\geq0} ~ \sum _{[rs]} d_s S^{(n+1)} _s
\no \\
\cS ^{(n+1)} _s & = &
- \sum _{s \not \in [t_1 \cdots t_n ]}
Z_0 ^{st_1 \cdots t_n} p^\mu _I  K^{\mu \nu} _{[t_1 \cdots t_n]}
 \ep ^\nu _s \theta _s   \Pi ^{(n+1)} _I (t_1;t_2,\cdots , t_n;s)
\eea
We shall prove these formulas systematically in the next subsection,
and complete them to include all differential blocks $\cD$.

\subsection{Partial bosonic Wick contractions using Lemma \ref{bosoniclemma}}

To deal with the structure of the blocks and the various rearrangements
into holomorphic blocks in a systematic way, we need to deal with the
contractions of the $x_+$ field inside the object $X^\mu _s$. Partial
contractions may be effected with the help of Lemma 1. We shall organize the
contractions according to the number of iterations using Lemma 1 that
have been performed. The differential blocks that involve these contractions
are $\cD_{2b}$, $\cD_{3e}$ and $\cD_{4b}$, and we decompose the
result of their contractions as follows,
\bea
\cD_{2b} & = &
\sum _{n=0} ^{N-2} \bar \cD _{2b}^{(n+2)}
+ \sum _{n=1}  ^{N-2} \cD _{2b\p}^{(n+2)}
\no \\
\cD_{3e} & = &
\sum _{n=0} ^{N-3} \bar \cD _{3e}^{(n+3)}
+ \sum _{n=1} ^{N-3} \cD _{3e\p}^{(n+3)}
\no \\
\cD_{4b} & = &
\sum _{n=0} ^{N-4} \bar \cD _{4b}^{(n+4)}
+ \sum _{n=1} ^{N-4} \cD _{4b\p}^{(n+4)}
\eea
where we define the contributions with a bar as those arising from $\bar X$,
and the contributions labeled by $\p$ as those arising from the derivative
terms in Lemma 1.

\sm

In the notation (\ref{Kstring}) for kinematic invariants and (\ref{notation}) for iterated
products of Green's functions, the $\bar \cD$ contributions are given by
\bea
\bar \cD_{2b} ^{(n+2)} & = &
- \! \sum _{[rst_1 \cdots t_n]} \!
d_s \bigg ( Z_0 ^{rs t_1 \cdots t_n}
\ep ^\mu _s \theta _s K^{\mu \rho}_{[rt_1\cdots t_{n-1}]}
\ff (r,s) G^n(r,t_1, \cdots, t_n)  \bar X^\rho _{t_n}  \bigg )
\no \\
\bar \cD_{3e} ^{(n+3)}& = &
 \sum _{[rsut_1 \cdots t_n]} \!
d_s \bigg ( Z_0 ^{rsu t_1 \cdots t_n}
\ep ^\mu _s \theta _s K^{\mu \rho} _{[rut_1 \cdots t_{n-1}]}
\FF (r,s) 
\no \\ && \hskip 1.5in \times 
\FF (r,u)  G^n(u,t_1,\cdots,t_n)  \bar X^\rho _{t_n}  \bigg )
\no \\
\bar \cD_{4b} ^{(n+4)} \! & = &
\! \!  \sum _{[rsuvt_1 \cdots t_n]} \!
d_s \bigg ( Z_0 ^{rsuv t_1 \cdots t_n}
\ep ^\mu _s \theta _s K^\mu _r K^\nu _u K^{\nu \rho} _{[v t_1 \cdots t_{n-1}]}
\FF (r,s)  
\no \\ && \hskip 1.5in \times  
\FF (u,v) G^n(v,t_1,\cdots , t_n)   \bar X^\rho _{t_n}  \bigg )
\eea
The differential contributions are given by
\bea
 \cD_{2b\p} ^{(n+2)}& = & - \sum _{[rst_1 \cdots t_n]}
 d_s \p_{t_n} \bigg ( Z_0 ^{rs t_1 \cdots t_n}
\ep ^\mu _s \theta _s K^{\mu \rho}_{[rt_1\cdots t_{n-1}]}
\ep ^\rho _{t_n} \theta _{t_n}
\ff (r,s) G^n(r,t_1,\cdots, t_n) \bigg )
\no \\
 \cD_{3e\p} ^{(n+3)}& = &  \sum _{[rsut_1 \cdots t_n]}
 d_s \p_{t_n} \bigg ( Z_0 ^{rsu t_1 \cdots t_n}
\ep ^\mu _s \theta _s K^{\mu \rho} _{[rut_1 \cdots t_{n-1}]}
\ep ^\rho _{t_n} \theta _{t_n}
\FF (r,s) 
\no \\ && \hskip 1.5in \times
\FF (r,u)  G^n(u,t_1,\cdots ,t_n)  \bigg )
\no \\
 \cD_{4b\p} ^{(n+4)}& = &  \sum _{[rsuvt_1 \cdots t_n]}
 d_s \p_{t_n} \bigg ( Z_0 ^{rsuv t_1 \cdots t_n}
\ep ^\mu _s \theta _s K^\mu _r K^\nu _u K^{\nu \rho} _{[vt_1\cdots t_{n-1}]}
 \ep ^\rho _{t_n} \theta _{t_n}
\no \\ && \hskip 1.5in \times
\FF (r,s) \FF (u,v)  G^n(v,t_1, \cdots , t_n)  \bigg )
\eea
Here we have systematically used the notations of (\ref{notation}) for linear chains of
bosonic and fermionic Green functions, as well as the notations
of (\ref{Kstring}) for the associated kinematic factors.

\subsection{Partial fermionic Wick contractions in $\cD _{4b}$}

It will be useful to render the fermionic contractions partially explicit with a string
of Szeg\"o kernels between the points $r$ and $u$ in the blocks $\cD _{4b}$.
By an abuse of notation, we shall now denote by $\cD_{4b} ^{(n+4)}$
the derivative block with a total of $n+4$ points, including the fermionic
contraction points. 

\sm

The derivative block is straightforward, and we find,
\bea
\cD_{4b\p} ^{(n+4)}& = &  \sum _{\beta =1} ^n \sum _{[rsuvt_1 \cdots t_n]}
d_s \p_{t_n} \bigg ( Z_0 ^{rsuv t_1 \cdots t_n}
\ep ^\mu _s \theta _s K^{\mu \rho} _{[rt_1 \cdots t_{\beta -1} uv t_\beta \cdots t_{n-1}]}
 \ep ^\rho _{t_n} \theta _{t_n} \FF (r,s)
\no \\ && \hskip 1in \times
S ^\beta (r,t_1, \cdots ,t_{\beta -1}, u) \FF (u,v)  G^{n-\beta +1}(v,t_\beta, \cdots t_n)  \bigg )
\quad
\eea
The blocks $\bar \cD_{4b} ^{(n+4)}$ are given by,
\bea
\bar \cD_{4b} ^{(n+4)} \! & \! = \! & \!
\sum _{\beta =1} ^n \sum _{[rsuvt_1 \cdots t_n]}
d_s \bigg ( Z_0 ^{rsuv t_1 \cdots t_n}
\ep ^\mu _s \theta _s K^{\mu \rho} _{[rt_1\cdots t_{\beta-1}uv t_\beta \cdots t_{n-1}]}
 \bar X^\rho _{t_n} \FF (r,s)
\\ && \hskip 1.1in \times
S^\beta  (r,t_1, \cdots,  t_{\beta-1},u)  \FF (u,v)
G^{n-\beta+1}(v,t_\beta, \cdots t_n)  \bigg )
\no \\ && +
\sum _{[rsuvt_1 \cdots t_n]}
d_s \bigg ( Z_0 ^{rsuv t_1 \cdots t_n}
\ep ^\mu _s \theta _s K^{\mu \rho} _{[rt_1\cdots t_n u v ]}
 \bar X^\rho _v
\FF (r,s) S^\beta (r,t_1, \cdots , t_n ,u)  \FF (u,v)    \bigg )
\no
\eea
It is important to recall the correct instructions that are to be imposed on the
operator $\bar X^\rho _{t_n}$ in the first $n$ terms of $\bar \cD _{4b} ^{(n+4)}$
and on the operator $\bar X_v ^\rho$ in the last line of $\bar \cD _{4b} ^{(n+4)}$.

\subsubsection{Avoiding the danger of double counting}

In particular, there is a danger of double counting contributions that must be
addressed.
\begin{enumerate}
\item The $x_+$ field in $\bar X^\rho _{t_n}$ is to be contracted with $x_+$
in the exponential factor of $Z_0$ at the points $r,s,u,v, t_\beta  \cdots, t_{n-1}$,
and only at those points. No contractions at the points $t_1 , \cdots , t_{\beta -1}$
are to be included. From the point of view of $x_+$-contractions, the points
$t_1, \cdots, t_{\beta -1}$ are new points, and these contributions are already
included in the terms with $\beta \to \beta -1$.
\item The term $i k^\rho _{t_n}$ in  $\bar X^\rho _{t_n}$ must be evaluated
at all points except $r,s,u$, including  when $t_n$ coincides with the fermionic
insertion points $t_1 , \cdots , t_{\beta -1}$.
\item The $x_+$ field in $X^\rho _v$ is to be contracted with $x_+$
in the exponential of $Z_0$ at the points $r,s,u,v$,
and only at those points. No contractions at the points $t_1 , \cdots , t_n$
are to be included. From the point of view of $x_+$-contractions, the points
$t_1, \cdots, t_n$ are new points, and these contributions are already
included in the terms with $n \to n+1$.
\item The term $i k^\rho _v$ in  $\bar X^\rho _v$ must be evaluated
at all points different from $r,s,t$, including when $v$ coincides with the fermionic
insertion points $t_1 , \cdots , t_n$.
\end{enumerate}
It will be convenient to reorganize these instructions according
to the following pattern.
\begin{itemize}
\item
The $i k^\rho _{t_n}$ contribution to $\bar X^\rho _{t_n}$ for the $\beta=n$
term, on the first line of $\bar \cD _{4b} ^{(n+4)}$, is identical to the
terms we would obtain by contracting the $x_+$ field in $X^\rho _v$
in the last line of $\bar \cD _{4b} ^{(n+4)}$ with the fermion contraction points
$t_1,\cdots, t_n$. Of course, in the last line of $\bar \cD _{4b} ^{(n+4)}$, such
contractions were not to be included according to the instruction of point 3. above.
\item
Our reorganization will consist in obtaining the $i k^\rho _{t_n}$ contribution to
$\bar X^\rho _{t_n}$ for the $\beta=n$ term from the last line in which we
now instruct the $x_+$ field in $X^\rho _v$ to also be contracted with the
points $t_1, \cdots , t_n$. In addition, we add the instruction that the
$i k^\rho _{t_n}$ contribution to $\bar X^\rho _{t_n}$ for the $\beta=n$ term
is to be included only for points $t_n$ that are different from the points
$r,s,u,t_1, \cdots t_n$, as in fact the instruction $[rsuvt_1 \cdots t_n]$ under
the sum indicates.
\item
Similarly, the $i k^\rho _{t_n}$ terms coincident with the fermion contraction points
$t_1, \cdots , t_{\beta -1}$ in the summand for $\beta$ are reproduced exactly by
the $x_+$ field contractions of $X^\rho _{t_n}$ in the summand for $\beta +1$.
\item This leaves only a single source of $i k^\rho $ terms that will survive,
namely for $v = t_1, t_2, \cdots, t_n$ in the last line of $\bar \cD _{4b} ^{(n+4)}$.
\item Note that the differential block $\cD_{3f}$ is exactly of the form of the
$i k^\rho$ term with $v=r$.
\end{itemize}
The reorganized instructions for the blocks $\bar \cD _{4b} ^{(n+4)}$
are as follows,
\bea
\bar \cD_{4b} ^{(n+2)}
\! & \!  = \!  & \!
\sum _{\beta =1} ^{n-2} \sum _{[rst_1 \cdots t_n]}
d_s \bigg ( Z_0 ^{rs t_1 \cdots t_n}
\ep ^\mu _s \theta _s K^{\mu \rho} _{[rt_1\cdots  \cdots t_{n-1}]}
 \bar X^\rho _{t_n} \FF (r,s)
\\ && \hskip 1.1in \times
S^\beta  (r,t_1, \cdots ,t_\beta ) \FF (t_\beta ,t_{\beta +1} )
G^{n-\beta-1}(t_{\beta +1} , t_{\beta +2} , \cdots , t_n)  \bigg )
\no \\ &&  +
\sum _{[rst_1 \cdots t_n]} \! \!
d_s \bigg ( Z_0 ^{rs t_1 \cdots t_n}
\ep ^\mu _s \theta _s K^{\mu \rho} _{[rt_1\cdots t_n  ]}
 \bar X^\rho _{t_n}
\FF (r,s) S^{n-1} (r,t_1, \cdots ,t_{n-1})  \FF (t_{n-1},t_n)    \bigg )
\no \\ &&
+
\sum _{[rst_1 \cdots t_n]}
d_s \bigg ( Z_0 ^{rs t_1 \cdots t_n}
\ep ^\mu _s \theta _s K^{\mu \rho} _{[rt_1\cdots t_n  ]}
\sum _{a=1}^{n-2} ik ^\rho _{t_a} \delta _{t_a,t_n} \FF (r,s)
\no \\ && \hskip 1in \times
S^{n-1} (r,t_1, \cdots ,t_{n-1})  \FF (t_{n-1},t_n)    \bigg )
\no
\eea
with the following instructions,
\begin{description}
\item[(I)] The point $t_n$ is genuinely distinct from all points $rst_1\cdots t_{n-1}$.
(Thus no $i k^\rho$ terms arise for $t_n$ coincident with those points.)
\item[(II)] The $x_+$ field in $\bar X^\rho _{t_n}$, in both lines, must be
contracted with the $x_+$
field in the exponential of $Z_0$ for all points $r,s,t_1 , \cdots , t_{n-1}$,
namely including all the fermion contraction points.
\end{description}
Finally, the second summation may be safely lumped with the
first summation as $\beta =n-1$, so that our final formula is
\bea
\bar \cD_{4b} ^{(n+2)}
& = &
\sum _{\beta =1} ^{n-1} ~ \sum _{[rst_1 \cdots t_n]}
d_s \bigg ( Z_0 ^{rs t_1 \cdots t_n}
\ep ^\mu _s \theta _s K^{\mu \rho} _{[rt_1\cdots  \cdots t_{n-1}]}
 \bar X^\rho _{t_n} \FF (r,s)
\\ && \hskip 1.1in \times
S ^\beta (r,t_1, \cdots ,t_\beta ) \FF (t_\beta ,t_{\beta +1} )
G^{n-\beta-1}(t_{\beta +1} , t_{\beta +2} , \cdots , t_n)  \bigg )
\no \\ &&
+
\sum _{[rst_1 \cdots t_n]}
d_s \bigg ( Z_0 ^{rs t_1 \cdots t_n}
\ep ^\mu _s \theta _s K^{\mu \rho} _{[rt_1\cdots t_n  ]}
\sum _{a=1}^{n-2} ik ^\rho _{t_a} \delta _{t_a,t_n}
\no \\ && \hskip 1in \times
\FF (r,s) S^{n-1} (r,t_1, \cdots  ,t_{n-1})  \FF (t_{n-1},t_n)    \bigg )
\no
\eea
subject to the same instructions (I) and (II). Note that these blocks are defined for
$n \geq 2$, and that the second sum only contributes starting at $n=3$.

\subsection{Partial fermionic Wick contractions in the remaining blocks}

In order to properly group together the various components of the blocks
$\Pi, \Pi _I$, and $\Pi _\pm$, we need to carry out the Wick contractions
of the worldsheet fermion fields in order to complete the chain blocks, including
their linking. We shall organize these contributions according to the number $n$
of Szeg\"o kernels that need to be inserted to complete the chain.

\sm

To organize this procedure effectively, it will be useful to recognize first that
certain blocks $\cD$ naturally fit in the same sequence of chains.
For example, the block $\cD _{2c}$ may really be viewed as an extension
of the block $\cD _{3a}$ in which we allow the point $t$ to coincide with $r$;
similarly the block $\cD_{2d}$ is an extension of $\cD_{3c}$ with $t=r$,
and $\cD _{3d}$ is an extension of $\cD_{4a}$ with $u=t$ and $r$ and $t$
interchanged. Thus, we have the following expansions in the number $n$
of Szeg\"o kernel insertions,
\bea
\cD_{2c} + \cD _{3a} & = & \sum _{n=0} \cD_{3a}^{(n+2)}
\hskip 1in \cD _{3a} ^{(2)} \equiv \cD_{2c}
\no \\
\cD _{3b} & = & \sum _{n=1} \cD_{3b}^{(n+2)}
\no \\
\cD_{2d} + \cD _{3c} & = & \sum _{n=0} \cD_{3c}^{(n+2)}
\hskip 1in \cD _{3c} ^{(2)} \equiv \cD_{2d}
\no \\
\cD _{3f} & = & \sum _{n=1} \cD_{3f}^{(n+2)}
\no \\
\cD_{3d} + \cD _{4a} & = & \sum _{n=1} \cD_{4a}^{(n+2)}
\hskip 1in \cD _{4a} ^{(3)} \equiv \cD_{3d}
\eea
with
\bea
\label{master7}
\cD_{3a} ^{(n+2)} & = &
 \sum _{[rst_1 \cdots t_n]} d_s \biggl (
Z_0 ^{rst_1 \cdots t_n}  \ep ^\mu _s \theta _s K^{\mu \rho} _{[rt_1 \cdots t_n]} p^\rho _I
 \FF (r,s) S^n(r,t_1 \cdots ,t_n)    \hat \omega _{I0} (t_n) \biggr )
\\
\cD_{3b} ^{(n+2)} & = &
 \sum _{[rst_1 \cdots t_n]} d_s \biggl (
Z_0 ^{rst_1 \cdots t_n} \half \ep ^\mu _s \theta _s
K^{\mu \rho} _{[r t_1\cdots t_n]} i k^\rho _s
 \FF (r,s)  S^n(r,t_1 \cdots ,t_n) \FF (t_n,s) \biggr )
\nonumber \\
\cD_{3c} ^{(n+2)} & = &
- \sum _{[rst_1 \cdots t_n]}  d_s \biggl (Z_0 ^{rst_1 \cdots t_n} {1 \over 2}
K^\rho _s K^\mu _s (ds)^{-1}   K^{\mu \rho} _{[r t_1 \cdots t_n]}
S^n(r,t_1 \cdots ,t_n) Q_B (s;r,t_n) \biggr )
\no \\
\cD_{3f} ^{(n+2)} & = &
 \sum _{[rst_1 \cdots t_n]}  d_s \bigg (
Z^{rst_1 \cdots t_n} _0   \ep ^\mu _s \theta _s ik ^\rho _r
K^{\mu \rho} _{[rt_1 \cdots t_n]}
\FF (r,s) S^n(r,t_1 \cdots ,t_n) \FF (t_n,r) \bigg )
\no \\
\cD_{4a} ^{(n+2)} & = &
\sum _{[rst_1 \cdots t_n]}  d_r d_s \bigg (
Z^{rst_1 \cdots t_n} _0 \half \ep ^\mu _r \theta _r
K^{\mu \rho} _{[ t_1 \cdots t_n]}
 \ep ^\rho _s \theta _s \FF (t_1,r) S^{n-1} (t_1 \cdots ,t_n)  \FF (t_n,s) \bigg )
\no
\eea
To obtain the last line, we have taken $\cD_{4a}$ from (\ref{master2}),
changed variables from $(r,s,u) $ to $(u,r,s)$ and recast the resulting formula
for $n+3 \to n+2$ vertex insertion points. We use the convention whereby
$\cD_{3e\p} ^{(n+2)}$ and $\cD_{4a} ^{(n+2)}$ vanish for $n=0$,
and $\cD_{4b\p} ^{(n+2)}$ vanishes for both values $n=0,1$.

\subsection{Contributions to
$\ep_r^\mu\theta_r K_{[t_1\cdots t_n]}^{\mu\nu}\ep_s^\nu\theta_s$}

These contributions precisely coincide with the terms that have two
differentials $d_rd_s$; they combine as follows,
\bea
 \cD_{2b\p} ^{(n+2)} +  \cD_{3e\p} ^{(n+2)} +  \cD_{4b\p} ^{(n+2)} + \cD_{4a} ^{(n+2)}
=  \sum _{[rs]} d_r d_s \cS_{rs} ^{(n+2)}
\eea
where $\cS_{rs} ^{(n+2)}$ was given explicitly in (\ref{S}). This result is
up to terms with two exact $(1,0)$ differentials $\p_r \p_s$, which will
combine with similar  terms in $\cZ$.

\subsection{Contributions to $\ep_s^\mu\theta_s K_{[t_1\cdots t_nr]}^{\mu\nu}p_I^\nu$}

These contributions precisely coincide with the terms that have a single
differential and are linear in $p$ (we do not count the $p$-dependence
inside $Z_0$ here). The dependence
of the terms linear in $p$ arises from the blocks $\cD_1$, $\cD_{2c}$, $\cD_{3a}$
as well as from the Wick contraction terms $\bar \cD_{2b} ^{(n+2)}$,
$\bar \cD_{3e} ^{(n+2)}$, and $\bar \cD _{4b} ^{(n+2)}$. To make the
$p$-dependence explicit, we split each of the contraction terms as follows,
\bea
\bar \cD_{2b} ^{(n+2)} & = & \hat \cD_{2b} ^{(n+2)} + \cD_{2bp} ^{(n+2)}
\hskip 1in n \geq 0
\no \\
\bar \cD_{3e} ^{(n+2)} & = & \hat \cD_{3e} ^{(n+2)} + \cD_{3ep} ^{(n+2)}
\hskip 1in n \geq 1
\no \\
\bar \cD_{4b} ^{(n+2)} & = & \hat \cD_{4b} ^{(n+2)} + \cD_{4bp} ^{(n+2)}
\hskip 1in n \geq 2
\eea
Here, the blocks denoted by $\hat \cD$ are given by the same formulas as
the blocks $\bar \cD$, but now with $\bar X$ replaced by $\hat X$.
The explicit expressions for the $p$-dependent terms are easily derived, and we have,
\bea
\cD_{2bp} ^{(n+2)}& = &
 \sum _{[rst_1 \cdots t_n]}
d_s \bigg ( Z_0 ^{rs t_1 \cdots t_n}
\ep ^\mu _s \theta _s K^{\mu \nu}_{[rt_1\cdots t_n]} p^\nu _I
\ff (r,s) G^n(r,t_1, \cdots, t_n)  \o_I (t_n)  \bigg )
\no \\
\cD_{3a} ^{(n+2)} & = &
\sum _{[rst_1 \cdots t_n]} d_s \biggl (
Z_0 ^{rst_1 \cdots t_n}  \ep ^\mu _s \theta _s K^{\mu \nu} _{[rt_1 \cdots t_n]} p^\nu _I
 \FF (r,s) S^n (r,t_1, \cdots ,t_n)    \hat \omega _{I0} (t_n) \biggr )
\no \\
\cD_{3ep} ^{(n+2)}& = &
- \sum _{[rst_1 \cdots t_n]}
d_s \bigg ( Z_0 ^{rs t_1 \cdots t_n}
\ep ^\mu _s \theta _s K^{\mu \nu} _{[rt_1 \cdots t_n]} p^\nu _I
\FF (r,s) \FF (r,t_1)  G^n(t_1,\cdots,t_n) \o_I(t_n) \bigg )
\no \\
\cD_{4bp} ^{(n+2)}& = &
-  \sum _{[rst_1 \cdots t_n]}
d_s \bigg ( Z_0 ^{rs t_1 \cdots t_n}
\ep ^\mu _s \theta _s K^{\mu \nu} _{[rt_1\cdots  t_n]} p^\nu _I
\sum _{\beta =1} ^{n-1} \FF (r,s) S^\beta  (r,t_1, \cdots ,t_\beta)
\no \\ && \hskip 1in \times
\FF (t_\beta,t_{\beta +1} )
G^{n-\beta-1}(t_{\beta +1}, \cdots t_n)  \o_I(t_n) \bigg )
\eea
Therefore, the sum of all contributions linear in $p$ is given by,
\bea
\delta _{n,0} \cD_1 + \cD_{2bp} ^{(n+2)} + \cD_{2c} + \cD_{3a} ^{(n+2)}
+ \cD_{3ep} ^{(n+2)} + \cD_{4bp} ^{(n+2)}
= \sum _s d_s \cS _s ^{(n+1)}
\eea
where $\cS_s ^{(n+1)}$ was given explicitly in (\ref{Sp}). This result is up to terms
with one exact $(1,0)$ differential $\p_s$, which will combine with similar terms in $\cZ$.

\newpage

\section{Extraction of  Exact Differentials\\
{\it Part II ~ Singly linked Chain blocks} }
\setcounter{equation}{0}

In the preceding section, we have carried out the bosonic and fermionic Wick
contractions needed to identify the chiral blocks, and we have explicitly
accounted for the chiral differential blocks that involve $\Pi$ and $\Pi _I$.
The entire block functions $\cD_1$, $\cD_{2a}$, $\cD_{2bp}$, $\cD_{2b\p}$, $\cD_{2c}$,
$\cD_{3a}$, $\cD_{3ep}$, $\cD_{3e\p}$, $\cD_{4a}$, $\cD_{4bp}$, and $\cD_{4b\p}$
were consumed in this process.
In the present section, we shall collect all the remaining contributions
to $\cD$, which are $\hat \cD_{2b}$, $\cD_{2d}$ (included in $\cD_{3c}$),
$\cD_{3b}$, $\cD_{3c}$,
$\hat \cD _{3e}$, $\cD_{3f}$ and $\hat \cD_{4b}$, and derive their representation
in terms of the  singly linked chain blocks $\Pi _\pm$.

\subsection{Remaining differential blocks}

We begin by exhibiting the explicit forms of the remaining differentials,
\bea
\hat \cD_{2b} ^{(n+2)}& = & - \sum _{[rst_1 \cdots t_n]}
d_s \bigg ( Z_0 ^{rs t_1 \cdots t_n}
\ep ^\mu _s \theta _s K^{\mu \rho}_{[rt_1\cdots t_{n-1}]}
\ff (r,s) G^n(r,t_1, \cdots, t_n)  \hat X^\rho _{t_n}  \bigg )
\\
\cD_{3b} ^{(n+2)} & = &
+ \sum _{[rst_1 \cdots t_n]} d_s \biggl (
Z_0 ^{rst_1 \cdots t_n} \half \ep ^\mu _s \theta _s
K^{\mu \rho} _{[r t_1\cdots t_n]} i k^\rho _s
 \FF (r,s)  S^n (r,t_1, \cdots ,t_n) \FF (t_n,s) \biggr )
\no \\
\cD_{3c} ^{(n+2)} & = &
- \sum _{[rst_1 \cdots t_n]}  d_s \biggl (Z_0 ^{rst_1 \cdots t_n} {1 \over 2}
K^\rho _s K^\mu _s (ds)^{-1}   K^{\mu \rho} _{[r t_1 \cdots t_n]}
S^n (r,t_1, \cdots ,t_n) Q_B (s;r,t_n) \biggr )
\no \\
\hat \cD_{3e} ^{(n+2)}& = & + \sum _{[rst_1 \cdots t_n]}
d_s \bigg ( Z_0 ^{rs t_1 \cdots t_n}
\ep ^\mu _s \theta _s K^{\mu \rho} _{[rt_1 \cdots t_{n-1}]}
\FF (r,s) \FF (r,t_1)  G^n(t_1,\cdots,t_n)  \hat X^\rho _{t_n}  \bigg )
\no \\
\cD_{3f} ^{(n+2)} & = &
+ \sum _{[rst_1 \cdots t_n]}  d_s \bigg (
Z^{rst_1 \cdots t_n} _0   \ep ^\mu _s \theta _s ik ^\rho _r
K^{\mu \rho} _{[rt_1 \cdots t_n]}
\FF (r,s) S^n (r,t_1, \cdots , t_n) \FF (t_n,r) \bigg )
\no
\eea
and finally the most involved block,
\bea
\hat \cD_{4b} ^{(n+2)}
& = &
\sum _{\beta =1} ^{n-1} \sum _{[rst_1 \cdots t_n]}
d_s \bigg ( Z_0 ^{rs t_1 \cdots t_n}
\ep ^\mu _s \theta _s K^{\mu \rho} _{[rt_1\cdots  \cdots t_{n-1}]}
 \hat X^\rho _{t_n} \FF (r,s)
\no \\ && \hskip 1.1in \times
S^\beta  (r,t_1, \cdots ,t_\beta ) \FF (t_\beta ,t_{\beta +1} )
G^{n-\beta-1}(t_{\beta +1} , t_{\beta +2} , \cdots , t_n)  \bigg )
\no \\ &&
+
\sum _{[rst_1 \cdots t_n]}
d_s \bigg ( Z_0 ^{rs t_1 \cdots t_n}
\ep ^\mu _s \theta _s K^{\mu \rho} _{[rt_1\cdots t_n  ]}
\sum _{a=1}^{n-2} ik ^\rho _{t_a} \delta _{t_a,t_n}
\no \\ && \hskip 1.1in \times
\FF (r,s) S^{n-1}  (r,t_1, \cdots  ,t_{n-1})  \FF (t_{n-1},t_n)    \bigg )
\eea
Note that, upon the identification $t_0\equiv r$, the block $\cD_{3f}$ corresponds
to the term $a=0$ in the sum over $a$ on the third line of the above expression
for $\hat \cD_{4b} ^{(n+2)}$. Henceforth, we shall assume that this reformulation
has been carried out.

\medskip

The instructions on the contractions of the composite $\hat X^\rho _{t_n}$ are
the same as the ones given in section 9.5; we repeat them here for completeness.
\begin{description}
\item[(I)] The point $t_n$ is genuinely distinct from all points $rst_1\cdots t_{n-1}$.
(Thus no $i k^\rho$ terms arise for $t_n$ coincident with those points.)
\item[(II)] The $x_+$ field in $\hat X^\rho _{t_n}$, must be contracted with the $x_+$
field in the exponential of $Z_0$ for all points $r,s,t_1 , \cdots , t_{n-1}$,
namely including all the fermion contraction points.
\end{description}

\sm

The strategy adopted here will be to first isolate the terms associated with the
bosonic Wick contraction between the points $t_n$ and $s$.
These contractions produce two distinctive kinematical factors, namely
$K_{[rt_1\cdots t_ns]}^{\mu\mu}$ from anti-symmetrization in
$\ep_s^\mu \theta_s ik_s^\nu$, or the factor $\ep _s ^{\{ \mu} \theta _s k_s ^{\nu \} }
K^{\mu \nu} _{[r t_1\cdots t_n]}$ from symmetrization.
They will produce the blocks $\Pi ^{(\ell+1)}_\pm$, and will completely consume the
term $\cD_{3b}^{(n+2)}$ above. The remaining blocks are then organized
according to the symmetry properties of $\Pi ^{(m+1|\ell+1)}_+$
or $\Pi ^{(m+1|\ell+1)}_-$ respectively.

\subsection{Symmetric contraction between $t_n$ with $s$}

The symmetric bosonic Wick contractions have the following characteristic
kinematical factor,
$\ep _s ^{\{ \mu} \theta _s k_s ^{\nu \} }K^{\mu \nu} _{[r t_1\cdots t_n]}$,
contributions to which we shall now isolate, and indicate with a subscript $+s$.
From assembling the $s+$ parts of $\hat \cD _{2b} $, $\cD _{3b} $,
$\hat \cD _{4b} $, with the entire contribution of $\hat \cD _{3e} $,
we find,
\bea
\label{Piplus2}
\sum _s d_s \cS _{+s} ^{(n+2)}
& \equiv &
\hat \cD _{2b+s} ^{(n+2)} +  \cD _{3b} ^{(n+2)}
+ \hat \cD _{3e+s} ^{(n+2)} + \hat \cD _{4b+s} ^{(n+2)}
\\
\cS _{+s} ^{(n+2)} &=&
- \sum _{s \not \in [rt_1 \cdots t_n]}
Z_0 ^{rst_1 \cdots t_n} \half i  \ep _s ^{\{ \mu} \theta _s k_s ^{\nu \} }
K^{\mu \nu} _{[r t_1\cdots t_n]} \Pi_+ ^{(n+2)} (s;r,t_1,\cdots , t_n)
\no
\eea
The contributions in the block $\Pi_+ ^{(n+2)}(s;r,t_1, \cdots, t_n)$
derived directly from $\hat \cD _{2b} ^{(n+2)}$, $\cD _{3b} ^{(n+2)}$,
$\hat \cD _{4b} ^{(n+2)}$, and $\hat \cD _{3e} ^{(n+2)}$ are given by
\bea
 &  &
- 2 \ff (r,s) G^{n+1} (r,t_1,\cdots, t_n, s)
+ \FF (r,s) S^n (r,t_1, \cdots, t_n) \FF (t_n,s)
\no \\ &&
+ 2 \sum _{\beta =0} ^{n-1} \FF (r,s) S^\beta (r,t_1, \cdots , t_\beta)
\FF (t_\beta, t_{\beta +1} ) G^{n-\beta} (t_{\beta +1}, \cdots , t_n,s)
\eea
where $t_0 \equiv r$, and $S(t_{-1}, t_0)\equiv 1$ in the above sum.
Actually, in view of the symmetry properties of the kinematic factor,
we may symmetrize the above contributions to $\Pi_+ ^{(n+2)}$,
without loss of generality. The symmetry operation will divide the
above contributions by a factor of 2, and add $(-)^{n+1}$ times
those terms with $[r t_1 t_2 \cdots t_{n-1} t_n] \to [t_n t_{n-1}  \cdots t_2 t_1r]$
interchanged.
One checks that
\bea
\FF (r,s) S^n (r,t_1, \cdots , t_n) \FF (t_n,s)
= (-)^{n+1} \FF (t_n,s) S^n (t_n, \cdots ,t_1, r) \FF (r,s)
\eea
As a result, we obtain the symmetrized block $\Pi _+ ^{(n+2)}(s;r,t_1,\cdots, t_n)$,
given by
\bea
&&
-  \ff (r,s) G^{n+1} (r,t_1,\cdots, t_n, s)
+ (-)^n G^{n+1} (t_n, \cdots , t_1,r) \ff (t_n,s)
\no \\ &&
+ \FF (r,s) S^n (r,t_1, \cdots , t_n) \FF (t_n,s)
\no \\ &&
+  \sum _{\beta =0} ^{n-1} \FF (r,s) S^\beta (r,t_1, \cdots, t_\beta)
\FF (t_\beta, t_{\beta +1} ) G^{n-\beta} (t_{\beta +1}, \cdots , t_n,s)
\no \\ &&
+ (-)^{n+1} \sum _{\beta =0} ^{n-1} \FF (t_n,s) S(t_n,t_{n-1}) \cdots S(t_{n+1-\beta }, t_{n-\beta})
\no \\ && \hskip 1in  \times
\FF (t_{n- \beta}, t_{n-1-\beta} ) G^{n-\beta} (t_{n-\beta -1}, \cdots , r,s)
\eea
This expression coincides with $- \Pi ^{(n+3)} (s;r,t_1,\cdots, t_n;s)$ up to terms
which are exact holomorphic differentials in $s$, and which will combine with similar
contributions arising from contractions of $X$ in $\cZ$. Using now also the definition
of the block $\Pi_+ ^{(n+2)}$ in terms of $\Pi ^{(n+3)}$, we obtain the result
announced in (\ref{Piplus2}).

\subsection{Anti-Symmetric contraction between $t_n$ with $s$}

The anti-symmetric bosonic Wick contractions have the following characteristic
kinematical factor, $K^{\mu \mu} _{[r t_1\cdots t_ns]}$, contributions to which
we shall now isolate, and indicate with a subscript $-s$.
From assembling the $-s$ parts of $\hat \cD _{2b} $, $\cD _{3e} $,
$\hat \cD _{4b} $, with the entire contribution of $\hat \cD _{3c} $,
we find,
\bea
\label{Bminus4}
\sum _s d_s \cS ^{(n+2)} _{-s} & \equiv &
\hat \cD _{2b-s} ^{(n+2)} +  \cD _{3c} ^{(n+2)}
+ \hat \cD _{3e-s} ^{(n+2)} + \hat \cD _{4b-s} ^{(n+2)}
\\
\cS ^{(n+2)} _{-s} &=&
- \sum _{s \not \in [r t_1 \cdots t_n]}
Z_0 ^{rst_1 \cdots t_n} \half
K^{\mu \mu} _{[r t_1\cdots t_ns]} (ds)^{-1} \Pi _- ^{(n+2)} (s;r,t_1,\cdots , t_n)
\no
\eea
The contributions in the block $\Pi_- ^{(n+2)}(s;r,t_1, \cdots, t_n)$
derived directly from $\hat \cD _{2b} ^{(n+2)}$, $\cD _{3c} ^{(n+2)}$,
$\hat \cD _{4b} ^{(n+2)}$, and $\hat \cD _{3e} ^{(n+2)}$ are given by
\bea
\label{B-}
&&
\ff (r,s) G^{n+1} (r,t_1,\cdots, t_n, s)
-   Q_B(s;r, t_n)S^n (r,t_1, \cdots , t_n)
\no \\ &&
- \sum _{\beta =0} ^{n-1} \FF (r,s) S^\beta (r,t_1, \cdots, t_\beta)
\FF (t_\beta, t_{\beta +1} ) G^{n-\beta} (t_{\beta +1}, \cdots , t_n,s)
\quad
\eea
The kinematical factor has the following reflection and cyclic permutation symmetries,
\bea
K^{\mu \mu} _{[rt_1\cdots t_n s]} & = & (-)^n K^{\mu \mu} _{[st_n\cdots t_1 r]}
\no \\
K^{\mu \mu} _{[st_n\cdots t_1 r]} & = &  K^{\mu \mu} _{[t_n\cdots t_1 rs]}
\eea
which combine to yield the mixed reflection and cyclic permutation symmetry,
\bea
K^{\mu \mu} _{[rt_1\cdots t_n s]} = (-)^n K^{\mu \mu} _{[t_n\cdots t_1 rs]}
\eea
It is under this symmetry that we can ``symmetrize" the contributions of
(\ref{B-}) by averaging (\ref{B-}) with $(-)^n$ times the expression with
$[rt_1\cdots t_n s] \to [t_n\cdots t_1 rs]$. The resulting symmetrized contributions
to  $\Pi _- ^{(n+2)}(s;r,t_1,\cdots, t_n)$ will be denoted $B_- ^{(n+2)}(s;r,t_1,\cdots, t_n)$,
and are  given as follows,
\bea
\label{BB}
B_- ^{(n+2)} &=&
-   Q_B(s;r, t_n)S^n (r,t_1, \cdots, t_n)
+ \half  \ff (r,s) G^{n+1} (r,t_1,\cdots, t_n, s)
\\ &&
+ \half (-)^n \ff (t_n,s) G^{n+1} (t_n,t_{n-1},\cdots, t_1,r, s)
\no \\ &&
- \half \sum _{\beta =0} ^{n-1} \FF (r,s) S^\beta (r,t_1, \cdots, t_\beta)
\FF (t_\beta, t_{\beta +1} ) G^{n-\beta} (t_{\beta +1}, \cdots , t_n,s)
\no \\ &&
- \half (-)^n \sum _{\alpha =1} ^n \FF (t_n,s) S^{n-\a} (t_n,  \cdots , t_\alpha)
\FF (t_\alpha, t_{\alpha -1} ) G^\alpha (t_{\alpha -1}, \cdots , t_1,r,s)
\no
\eea
In view of the symmetrization, the blocks
automatically have the symmetry,
\bea
\label{Bsym}
B_- ^{(n+2)}(s;r,t_1,\cdots, t_n) = (-)^n B_- ^{(n+2)}(s;t_n,t_{n-1},\cdots, t_1,r)
\eea
This definition applies to all $n\geq 1$. When $n=0$,
we can still define the block $B _-^{(2)}(s;r)$
by the above expression by setting $t_n=r$ and $n=0$.
Thus the lowest order blocks are
\bea
B_- ^{(2)} (s;r)& = & - Q_B (s;r,r) + \ff (r,s) G(r,s)
\no \\
B_- ^{(3)} (s;r,t) & = & - Q_B (s;r,t) S(r,t)
\no \\ &&
+ \half \ff (r,s) G(r,t) G(t,s) - \half \ff (t,s) G(t,r) G(r,s)
\no \\ &&
+ \half \FF (t,s) \FF  (t,r) G(r,s) - \half \FF (r,s) \FF (r,t) G(t,s)
\eea
The presence of the cubic vertex function $Q_B$ alerts us to the possibility
that $B_- ^{(n+2)}$ is related to the blocks $\Pi _- ^{(n+2)}$ constructed
in section 8. We recall here the lowest cases, adapted to the present notation,
\bea
\Pi _- ^{(2)} (s;r) & = & Q_B(s;r,r) - 2 Q_\ff (s;r,r) - G(r,s) \ff (r,r)
\no \\
\Pi _- ^{(3)} (s;r,t) & = & Q_B(s;r,t) S(r,t) + Q_0 (s;r,t) \ff (r,t)
\no \\ &&
-  Q_\ff (s;r,t) G(r,t) + Q_\ff (s;t,r) G(t,r)
\no \\ &&
+ Q_F (s;t,r) \FF (r,t) - Q_F (s;r,t) \FF( t,r)
\no \\ &&
- G(r,s) \Pi _- ^{(2)} (r;t) + G(t,s) \Pi _- ^{(2)} (t;r)
\eea
To compare, we compute
\bea
\label{derB2minus}
\p_{\bar s} \bigg ( B^{(2)} (s;r) + \Pi ^{(2)} _- (s;r) \bigg )
& = &
\p_{\bar s} \bigg ( \ff (r,s) G (r,s) - 2 Q_\ff (s;r,r) - G(r,s) P(r,r) \bigg )
\no \\
& = & \p_{\bar s} G(r,s) \bigg ( \ff (r,s) - \ff (r,r) \bigg ) =0
\eea
Here, the term $Q_B(s;r,r)$, which arises in the expressions of both
$B_- ^{(2)}$ and $\Pi ^{(2)}_-$ cancels out. The final result of
(\ref{derB2minus}) vanishes, from which we conclude that
$B_- ^{(2)}$ equals $\Pi ^{(2)}_-$, up to terms holomorphic in $s$.
As usual, in the full chiral amplitude, such exact holomorphic
differentials in $s$ will combine with similar terms in $\cZ$.

\sm

The case of $B^{(3)}_- (s;r,t)$ is similar, and we calculate
\bea
\label{derB3minus}
\p_{\bar s} \bigg ( B^{(3)} (s;r,t) + \Pi ^{(3)} _- (s;r,t) \bigg )
& = &
\p_{\bar s} \bigg (
 \half \ff (r,s) G(r,t) G(t,s) - \half \ff (t,s) G(t,r) G(r,s)
\no \\ &&
+ \half \FF (t,s) \FF  (t,r) G(r,s) - \half \FF (r,s) \FF (r,t) G(t,s)
\no \\ &&
-  Q_\ff (s;r,t) G(r,t) + Q_\ff (s;t,r) G(t,r)
\no \\ &&
+ Q_F (s;t,r) \FF (r,t) - Q_F (s;r,t) \FF( t,r) \bigg )
\eea
Here, we have readily omitted the terms that produce $\delta$-functions
at coincident vertex insertion points only, such as the term in $Q_0$,
and the terms  in $\Pi ^{(2)}_-$ that arise in $\Pi ^{(3)}_-$.
Using on the right hand side of  (\ref{derB3minus}) the defining
$\p_{\bar s}$ differentials of $Q_\ff$, and $Q_\FF$ in terms of
$\p_{\bar s} \ff$ and $\p_{\bar s} \FF$, neglecting $\delta$-functions
at coincident points immediately shows that (\ref{derB3minus}) vanishes.
The proof to all orders is readily obtained by generalizing the above
case to all $n$, and we obtain,

\begin{lemma}
The blocks $B_-^{(n+2)}(s;r,t_1,\cdots,t_n)$ and ~
$- \Pi_-^{(n+2)}(s;r,t_1,\cdots,t_n)$ differ only by terms which are
holomorphic in $s$,
\bea
\p_{\bar s}\bigg ( B_-^{(n+2)}(s;r,t_1,\cdots,t_n)
+ \Pi_-^{(n+2)}(s;r,t_1,\cdots,t_n)\bigg)=0.
\eea
\end{lemma}
This completes the proof of (\ref{Bminus4}).

\subsection{Blocks remaining after contractions of $t_n$ with $s$}

Having identified and extracted the differential blocks $d_s\cS^{(n+2)}_{+s}$ and
$d_s \cS^{(n+2)}_{-s}$, we are left with only  the following differential blocks,
\bea
\tilde \cD_{2b} ^{(n+2)}& = & - \sum _{[rst_1 \cdots t_n]}
d_s \bigg ( Z_0 ^{rs t_1 \cdots t_n}
\ep ^\mu _s \theta _s K^{\mu \rho}_{[rt_1\cdots t_{n-1}]}
\ff (r,s) G^n(r,t_1, \cdots, t_n)  \tilde X^\rho _{t_n}  \bigg )
\no \\
\tilde \cD_{4b} ^{(n+2)}
& = &
\sum _{\beta =0} ^{n-1} \sum _{[rst_1 \cdots t_n]}
d_s \bigg ( Z_0 ^{rs t_1 \cdots t_n}
\ep ^\mu _s \theta _s K^{\mu \rho} _{[rt_1\cdots  \cdots t_{n-1}]}
 \tilde X^\rho _{t_n} \FF (r,s)
\no \\ && \hskip 1.1in \times
S^\beta  (r,t_1, \cdots,t_\beta ) \FF (t_\beta ,t_{\beta +1} )
G^{n-\beta-1}(t_{\beta +1}  , \cdots , t_n)  \bigg )
\no \\ &&
+
\sum _{[rst_1 \cdots t_n]}
d_s \bigg ( Z_0 ^{rs t_1 \cdots t_n}
\ep ^\mu _s \theta _s K^{\mu \rho} _{[rt_1\cdots t_n  ]}
\sum _{a=0}^{n-2} ik ^\rho _{t_a} \delta _{t_a,t_n} \FF (r,s)
\no \\ && \hskip 1.3in \times
 S^{n-1} (r,t_1, \cdots ,t_{n-1})  \FF (t_{n-1},t_n)    \bigg )
\eea
Here, we have denoted by $\tilde X^\rho _{t_n}$ {\sl the instruction that
$\hat X_{t_n}^\rho$ is not to be contracted with $s$} any more (since those
contractions were already isolated in sections 10.2 and 10.3),
and the corresponding blocks also by $\tilde \cD$.
In $\tilde \cD_{4b} ^{(n+2)}$, we have now also included  $\hat \cD _{3e} ^{(n+2)}$ as
the contribution with $\beta=0$ in the first sum, as well as $\cD _{3f}^{(n+2)}$
as the contribution with $a=0$ in the second sum (where $t_0 \equiv r$).
For later use, we shall relabel $r \to t_1$ and $t_i \to t_{i+1}$ when $i=1, \cdots, n-1$,
and finally $t_{n+1} \to r$, so that
\bea
\tilde \cD_{2b} ^{(n+2)}& = & - \sum _{[st_1 \cdots t_n r]}
d_s \bigg ( Z_0 ^{s t_1 \cdots t_n r}
\ep ^\mu _s \theta _s K^{\mu \rho}_{[t_1\cdots t_n]}
\ff (s,t_1) G^n(t_1, \cdots, t_n,r)  \tilde X^\rho _r  \bigg )
\no \\
\tilde \cD_{4b} ^{(n+2)}
& = &
\sum _{\beta =1} ^m \sum _{[st_1 \cdots t_n r]}
d_s \bigg ( Z_0 ^{rs t_1 \cdots t_n r}
\ep ^\mu _s \theta _s K^{\mu \rho} _{[t_1\cdots  \cdots t_n]}
 \tilde X^\rho _r \FF (t_1,s)
\no \\ && \hskip 1.1in \times
S^{\b -1}  (t_1, \cdots ,t_\beta ) \FF (t_\beta ,t_{\beta +1} )
G^{n-\beta}(t_{\beta +1} ,  \cdots , t_n,r)  \bigg )
\no \\ &&
+
\sum _{[st_1 \cdots t_n,r]}
d_s \bigg ( Z_0 ^{rs t_1 \cdots t_n r}
\ep ^\mu _s \theta _s K^{\mu \rho} _{[t_1\cdots t_n r  ]}
\sum _{a=1}^{n-1} ik ^\rho _{t_a} \delta _{t_a,r}
\no \\ && \hskip 1in \times
\FF (t_1,s) S^{n-1}  (t_1, \cdots  ,t_m)  \FF (t_n,r)    \bigg )
\eea

\subsection{The single-link structure of the remaining blocks}

The blocks $\tilde \cD_{2b} ^{(n+2)}$ and $\tilde \cD_{4b} ^{(n+2)}$ involve
the composite $\tilde X^\rho _r$, which has two parts,
\bea
\tilde X^\rho _r
=
i k^\rho _r  + K^\nu _r K^ \rho  _r \p_r x_+ ^\nu (r)
\eea
with the instruction that $x_+ ^\nu (r)$ is to be contracted only with the points
$t_1, \cdots, t_n$, but not with $s$. The first term will necessarily be contracted
with a fermion loop. This splits each block into a part with $x$-contraction and
a part with $\psi$-contraction,
and for $\tilde \cD_{4b} ^{(n+2)}$ also a part arising from the second sum,
which did not involve any further contractions,
\bea
\tilde \cD_{2b} ^{(n+2)} & = & \tilde \cD_{2bx} ^{(n+2)} + \tilde \cD_{2b\psi } ^{(n+2)}
\no \\
\tilde \cD_{4b} ^{(n+2)} & = & \tilde \cD_{4bx} ^{(n+2)} + \tilde \cD_{4b\psi } ^{(n+2)}
+ \tilde \cD _{4f} ^{(n+2)}
\eea
The contractions are organized as follows,
\bea
\tilde \cD_{2b\psi } ^{(n+2)} = \sum _{\ell =1}   \cD_{2b\psi } ^{(n + 1 | \ell +1)}
& \hskip 1in &
\tilde \cD_{2bx} ^{(n+2)} = \sum _{\ell =1} ^n  \cD_{2bx} ^{(n-\ell + 1 | \ell +1)}
\no \\
\tilde \cD_{4b\psi } ^{(n+2)} = \sum _{\ell =1}   \cD_{4b\psi } ^{(n + 1 | \ell +1)}
& \hskip 1in &
\tilde \cD_{4bx} ^{(n+2)} = \sum _{\ell =1} ^n  \cD_{4bx} ^{(n-\ell + 1 | \ell +1)}
\no \\
& \hskip 1in &
\tilde \cD_{4f} ^{(n+2)} = \sum _{\ell =1} ^n  \cD_{4f} ^{(n-\ell + 1 | \ell +1)}
\eea
Each of these contributions has the connectivity of a closed loop with $\ell+1$
points, attached at the point $r$ to a linear chain.
We shall label the points on the closed loop by $r, u_1, \cdots , u_\ell$ and the
points on the linear chain by $s, t_1, \cdots, t_m, r$.

\smallskip

\noindent $\bullet$
The fermion loop contributions may be expressed as follows\footnote{Note that an
extra $-$ sign has been included for the closed fermion loop. In our formalism,
this comes about because $K^\mu _r S(r,u_1) K^\mu _{u_1} \cdots  K^\nu _{u_\ell}
S(u_\ell, r) K^\nu _r = - K^\mu_{u_1} K^\rho _{u_1} \cdots K^\nu _{u_\ell} K^\nu _r K^\mu _r
S(r,u_1) \cdots S(u_\ell, r)$.}
\bea
\cD_{2b\psi } ^{(m + 1 | \ell +1)}
& = &
+ \sum _{[st_*ru_*]} d_s \bigg ( Z_0 ^{st_* r u_*}
\ep ^\mu _s \theta _s
K^{\mu \rho} _{[t_1 \cdots t_m]} i k^\rho _r K^{\sigma \sigma } _{[r u_1 \cdots u_\ell]}
\ff (s,t_1)
 \\ && \hskip 0.9in
\times G^m (t_1 , \cdots , t_m,r) S^{\ell +1} (r,u_1, \cdots , u_\ell, r) \bigg )
\no \\
\cD_{4b\psi } ^{(m + 1 | \ell +1 )} & = &
- \sum _{[st_*ru_*]} d_s \bigg ( Z_0 ^{st_* r u_*}
\ep ^\mu _s \theta _s
K^{\mu \rho} _{[t_1 \cdots t_m]} i k^\rho _r K^{\sigma \sigma } _{[r u_1 \cdots u_\ell]}
\no \\ && \hskip 0.8in
\sum _{\beta =1} ^m  \FF (t_1,s) S^{\b -1} (t_1, \cdots , t_\beta)
\FF (t_\beta , t_{\beta +1})
\no \\ && \hskip 0.9in
\times G^{m-\beta}  (t_{\beta +1} , \cdots , t_m,r)
S^\ell (r,u_1, \cdots , u_\ell, r) \bigg )
\no
\eea
Here and below, we use the notation $t_*$ as an abbreviation for
$t_1\cdots t_m$, and $u_*$ as an abbreviation for $u_1 \cdots u_\ell$.

\sm

\noindent $ \bullet$
The bosonic contributions $\cD _{2bx} ^{(m+1|\ell+1)}$  may be expressed as follows,
\bea
\cD _{2bx} ^{(m+1|\ell +1)}
& = &
- \sum _{[st_*ru_*]} d_s \bigg ( Z_0 ^{st_* r u_*}
\ep ^\mu _s \theta _s
K^{\mu \rho} _{[t_1 \cdots t_m r u_1 \cdots u_\ell]} i k^\rho _r
\ff (s,t_1)
\no \\ && \hskip 0.9in \times
G^{m+\ell +1} (t_1,\cdots, t_m, r, u_1 , \cdots, u_\ell, r) \bigg )
\eea
The bosonic contributions $\cD _{4bx} ^{(m+1|\ell+1)}$ are given by
\bea
\cD _{4bx} ^{(m+1|\ell +1)} \! & \!  = \!  & \!
 \sum _{[st_* r u_*]}
d_s \bigg ( Z_0 ^{s t_* r u_*}
\ep ^\mu _s \theta _s K^{\mu \rho} _{[t_1\cdots  t_m r u_1 \cdots u_\ell ]} i k^\rho _r
 \sum _{\beta =1} ^{n} \FF (t_1,s) S^{\b-1}  (t_1, \cdots ,t_\beta)
\no \\ && \hskip 1in \times
\FF (t_\beta,t_{\beta +1})
G^{n-\beta+1}(t_{\beta +1} , \cdots , t_{n+1},r)  \bigg ) \delta _{r,t_{m+1}}
\eea
In the formula above, we have used the notations,
\bea
n & = & m + \ell
\no \\
r & = & t_{m+1}
\no \\
u_i & = & t_{m+i+1} \hskip 1in i = 1, \cdots, \ell
\eea
and we continue to use this notation below.
The contributions $\cD_{4f} ^{(m + 1 | \ell +1)}$ are given by
\bea
\cD _{4f} ^{(m+1|\ell+1)} & = &
 \sum _{[st_* r u_*]}
d_s \bigg ( Z_0 ^{s t_* r u_*}
\ep ^\mu _s \theta _s K^{\mu \rho} _{[t_1\cdots  t_m r u_1 \cdots u_\ell ]} i k^\rho _r
\no \\ && \hskip 1in \times
\FF (t_1,s) S^n (t_1, \cdots ,t_{n+1})
 \FF (t_{n+1},r)   \bigg ) \delta _{r,t_{m+1}}
\no
\eea
Note that $\cD_{4f} ^{(m + 1 | \ell +1)}$ coincides with the $\beta =n+1$ contribution
of $\cD _{4bx} ^{(m+1|\ell +1)}$, so that, more succinctly,
\bea
&&
\cD _{4f} ^{(m+1|\ell +1)} + \cD _{4bx} ^{(m+1|\ell +1)}
\no \\ && \quad =
 \sum _{[st_* r u_*]}
d_s \bigg ( Z_0 ^{s t_* r u_*}
\ep ^\mu _s \theta _s K^{\mu \rho} _{[t_1\cdots  t_m r u_1 \cdots u_\ell ]} i k^\rho _r
 \sum _{\beta =1} ^{n+1} \FF (t_1,s) S ^{\b -1} (t_1, \cdots ,t_\beta)
\no \\ && \hskip 1in \times
\FF (t_\beta,t_{\beta +1})
G^{n-\beta+1}(t_{\beta +1} , \cdots , t_{n+1},r)  \bigg ) \delta _{r,t_{m+1}}.
\no
\eea
We are now ready to isolate the contributions of the blocks $\Pi ^{(m+1|\ell+1)}_\pm$
that arise from these terms. To do so, we separate the $+$ and $-$ cases,
according to the structure of the kinematic factors.

\subsection{Contributions with kinematic factor
$K^{\mu \rho } _{[t_1 \cdots t_m r \{ u_1 \cdots u_\ell \} ]}  k^\rho _r$}

The blocks $\cD _{2b \psi} ^{(m+1|\ell+1)}$ and
$\cD _{4b \psi} ^{(m+2|\ell)}$ do not contribute to this
kinematic factor. The contribution to this kinematic factor from the remaining blocks
will be denoted with the subscript $+$, and  is given by
\bea
\sum _s d_s \cS ^{(m+1|\ell+1)} _{+s}
& = &
\cD _{2bx+} ^{(m+1|\ell+1)} + \cD _{4bx+} ^{(m+1|\ell+1)} + \cD _{4f+} ^{(m+1|\ell+1)}
\\
\cS ^{(m+1|\ell+1)} _{+s} & = &
- \! \sum _{s \not \in [t_* r u_*]} \! \!
Z_0 ^{s t_* r u_*}
\ep ^\mu _s \theta _s K^{\mu \rho} _{[t_1\cdots  t_m r \{ u_1 \cdots u_\ell  \} ]} i k^\rho _r
\Pi_+ ^{(m+1 | \ell+1)} (s;t_*, r, u_*) \quad
\no
\eea
The contributions to $\Pi_+ ^{(m+1 | \ell+1)}$ arising directly from
$\cD _{2bx+} ^{(m+1|\ell+1)}$, $\cD _{4bx+} ^{(m+1|\ell+1)}$, and
$\cD _{4f+} ^{(m+1|\ell+1)}$ are given as follows by $B^{(m+1|\ell+1)} _+
(s;t_*,r,u_*)$,
\bea
B^{(m+1|\ell+1)} _+ & = &
- \half \ff (s,t_1) G^{m+\ell +1} (t_1,\cdots, t_m, r, u_1 , \cdots, u_\ell, r)
\no \\ &&
+ \half \sum _{\beta =1} ^{n+1} \FF (t_1,s) S^{\b -1} (t_1, \cdots ,t_\beta)
\FF (t_\beta,t_{\beta +1})
G^{n-\beta+1}(t_{\beta +1} , \cdots , t_{n+1},r)  \delta _{r,t_{m+1}}
\no \\ &&
+ (-)^\ell \bigg ( [u_1 \cdots u_\ell ] \to [u_\ell \cdots u_1 ] \bigg )
\eea
with the identification
\bea
n & = & m + \ell
\no \\
r & = & t_{m+1}
\no\\
u_i & = & t_{m+i+1} \hskip 1in i = 1, \cdots, \ell
\eea
It is straightforward to see that, up to terms which are holomorphic in $s$,
$B_+ ^{(m+1|\ell+1)}$ coincides with $- \Pi _+ ^{(m+1|\ell+1)}$. More precisely,
we have

\begin{lemma}
The blocks $B_+^{(m+1|\ell+1)}$ and ~ $-\Pi_+^{(m+1|\ell+1)}$
differ only by terms which are holomorphic in $s$,
\bea
\p_{\bar s}
\bigg(B_+^{(m+1|\ell+1)}(s;t_*,r,u_*)
+
\Pi_+^{(m+1|\ell+1)}(s;t_*,r,u_*)\bigg)=0
\eea
\end{lemma}
As usual, we omit $\delta$-function contributions at coincident
vertex insertion points for such considerations.
We have continued to use the notations $t_*$  and $u_*$
as an abbreviation for $t_1\cdots t_m$,   and $u_1 \cdots u_\ell$ respectively.

\subsection{Contributions with kinematic factor
$K^{\mu \rho} _{[t_1 \cdots t_m]}  k^\rho _r
K^{\sigma \sigma} _{[ r u_1 \cdots u_\ell]}$}

This part occurs in all blocks. This contribution will be denoted with the
subscript $-$, and  is given by
\bea
\label{Bminus5}
\sum _s d_s \cS _{-s} ^{(m+1 |\ell+1)}
& = &
\cD _{2bx-} ^{(m+1|\ell+1)} + \cD _{4bx-} ^{(m+1|\ell+1)} + \cD _{4f-} ^{(m+1|\ell+1)}
+ \cD _{2b\psi} ^{(m+1|\ell+1)} + \cD _{4b\psi} ^{(m+1|\ell+1)}
\no \\
\cS _{-s} ^{(m+1 |\ell+1)}
& = & \! \!
- \! \! \! \sum _{s \not \in [t_* r u_*]} \! \! \! \!
Z_0 ^{s t_* r u_*}
\ep ^\mu _s \theta _s K^{\mu \rho} _{[t_1 \cdots t_m]} i k^\rho _r
K^{\sigma \sigma} _{[ r u_1 \cdots u_\ell]}
\Pi_- ^{(m+1 | \ell+1)} (s;t_*, r, u_*) \qquad
\eea
The contributions to $\Pi_- ^{(m+1 | \ell+1)}$ arising directly from
$\cD _{2bx-} ^{(m+1|\ell+1)}$, $\cD _{4bx-} ^{(m+1|\ell+1)}$,
$\cD _{4f-} ^{(m+1|\ell+1)}$,
$\cD _{2b\psi} ^{(m+1|\ell+1)}$, and $ \cD _{4b\psi} ^{(m+1|\ell+1)}$
are given as follows by $B^{(m+1|\ell+1)} _- (s;t_*,r,u_*)$,
\bea
\label{Bminus6}
B_- ^{(m+1|\ell+1)}
& = &
+ \half \ff (s,t_1) G^m (t_1, \cdots, t_m,r) S^{\ell+1} (r,u_1, \cdots ,u_\ell, r)
\no \\ &&
- \half \ff (s,t_1) G^{m+\ell+1} (t_1,\cdots, t_m, r, u_1 , \cdots, u_\ell, r)
\no \\ &&
- \half \sum _{\beta =1} ^m  \FF (t_1,s) S^{\b -1} (t_1, \cdots , t_\beta)
\FF (t_\beta , t_{\beta +1})
\no \\ && \hskip 0.6in
\times G^{m-\beta}  (t_{\beta +1} , \cdots , t_m,r)
S^{\ell+1} (r,u_1, \cdots , u_\ell, r)
\no \\ &&
+ \half\sum _{\beta =1} ^{n +1} \FF (t_1,s) S^{\b-1}  (t_1, \cdots ,t_\beta)
\FF (t_\beta,t_{\beta +1})
G^{n -\beta+1}(t_{\beta +1} , \cdots , t_{n +1},r) \delta _{r,t_{m+1}}
\no \\ &&
- (-)^\ell \bigg ( [u_1 \cdots u_\ell ] \to [u_\ell \cdots u_1 ] \bigg )
\eea
with the identification
\bea
n & = & m + \ell
\no \\
r & = & t_{m+1}
\no\\
u_i & = & t_{m+i+1} \hskip 1in i = 1, \cdots, \ell
\eea
One may show by inspection that, up to terms which are holomorphic in $s$,
$B_- ^{(m+1|\ell+1)}$ coincides with $\Pi _- ^{(m+1|\ell+1)}$.

\sm

Some of the crucial steps in this identification are as follows. The  first and third
terms of $B_-^{(m+1|\ell+1)} $ in (\ref{Bminus6}) sum to produce a term
proportional to $\Pi ^{(m+2)} (s;t_1,\cdots,t_m;r)$
up to terms that are holomorphic in $s$. The remaining sum may be split
up into the parts $1\leq \beta \leq m$ and the part $m+1 \leq \beta \leq m + \ell +1$.
The first sum combines with the second term in $B_-^{(m+1|\ell+1)} $ to produce
another term  proportional to $\Pi ^{(m+2)} (s;t_1,\cdots,;r)$
up to terms that are holomorphic in $s$.
The remaining sum may be recast in the following form,
(changing notation for the summation variables,  $\beta = \alpha + m +1$),
\bea
\FF (t_1,s) S^m (t_1, \cdots , t_m,r) \sum _{\alpha=0}^\ell
S^\a (r,u_1, \cdots , u_{\alpha -1}, t_\alpha) \FF (u_\alpha, u _{\alpha +1})
G^{\ell - \alpha } (u_{\alpha +1} , \cdots , u _\ell,r)
\no
\eea
The sum over $\a$ may be recognized as $\Pi _F ^{(\ell +2)} (r;u_1,\cdots , u_\ell;r)$,
while the prefactor equals the expression
$(-)^m \Pi _F ^{(m +2)} (r;t_m,\cdots , t_1;s)$,
up to terms holomorphic in $s$.
Assembling all pieces, we have our final lemma,
\begin{lemma}
The blocks $B_-^{(m+1|\ell+1)}$ and ~
$- \Pi_-^{(m+1|\ell+1)}$ differ only by terms which are holomorphic in $s$,
\bea
\p_{\bar s}\bigg(  B_-^{(m+1|\ell+1)}(s;t_*,r,u_*)
+ \Pi_-^{(m+1|\ell+1)}(s;t_*,r,u_*) \bigg ) =0
\eea
\end{lemma}
which completes the proof of (\ref{Bminus5}).
As usual, we omit $\delta$-function contributions at coincident
vertex insertion points for such considerations.
We have continued to use the notations $t_*$  and $u_*$
as an abbreviation for $t_1\cdots t_m$,   and $u_1 \cdots u_\ell$ respectively.

\newpage

\section{Summary and Additional Remarks}
\setcounter{equation}{0}

In this last section, we shall present a brief summary of the results obtained
in this paper for the structure of the differential blocks $\cD$. We shall
also give a brief discussion of the fate of the various blocks, which are
holomorphic $\p_s$ differentials purely of type $(1,0)$ in all insertion
points, and which will combine with similar terms in $\cZ$, and which we
have systematically omitted in the previous sections. This recombination
in simple examples will set the stage for the next companion paper
in which the complete structure of the holomorphic blocks $\cZ$ will be derived
and studied. We close with an analysis of the monodromy structure of
the blocks $\cD$ and $\cZ$.

\subsection{Summary of Differential blocks}

The various results in this paper on the structure of the differential
blocks may be summarized as follows,
\bea
\cD [\delta ] & = & \sum _{n=0} ~ \sum _{[rs]} d_r d _s \cS_{rs} ^{(n+2)}
+ \sum _{n=0} ~ \sum _r d _s  \cS_s ^{(n+1)}
\no \\ &&
+ \sum _{m=-1} ~ \sum _{\ell =1} ~ \sum _s
d_s \bigg ( \cS _{+s} ^{(m+1|\ell+1)} + \cS _{-s} ^{(m+1|\ell+1)} \bigg )
\eea
up to terms which are exact holomorphic $\p_s$ differentials, purely
of type $(1,0)$ in all insertion points. Here, the linear chain blocks $\cS_{rs}^{(n+2)}$
and $ \cS _s ^{(n+2)}$ are given by
\bea
S^{(n+2)} _{rs} & = &
- \sum _{r,s \not \in [t_1 \cdots t_n ]}  \! \!
Z_0 ^{rst_1 \cdots t_n} \half \ep_r ^\mu \theta _r
 \ep ^\nu _s \theta _s  K^{\mu \nu}  _{[t_1 \cdots  t_n]}
 \Pi ^{(n+2)} (r;t_1,\cdots , t_n;s)
\no \\
\cS ^{(n+1)} _s & = &
- \sum _{s \not \in [t_1 \cdots t_n ]}
Z_0 ^{st_1 \cdots t_n} p^\mu _I  K^{\mu \nu} _{[t_1 \cdots t_n]}
 \ep ^\nu _s \theta _s   \Pi ^{(n+1)} _I (t_1;t_2,\cdots , t_n;s)
\eea
while the singly linked blocks $ \cS _{\pm s} ^{(m+1|\ell+1)}$ are given by
\bea
\cS ^{(m+1|\ell+1)} _{+s}
& = &
- \sum _{s \not \in [t_* r u_*]}
Z_0 ^{s t_* r u_*}
\ep ^\mu _s \theta _s K^{\mu \rho} _{[t_1\cdots  t_m r \{ u_1 \cdots u_\ell  \} ]} i k^\rho _r
\Pi_+ ^{(m+1 | \ell+1)} (s;t_*, r, u_*)
\no \\
\cS _{-s} ^{(m+1 |\ell+1)}
& = & \! \!
- \sum _{s \not \in [t_* r u_*]} \! \!
Z_0 ^{s t_* r u_*}
\ep ^\mu _s \theta _s K^{\mu \rho} _{[t_1 \cdots t_m]} i k^\rho _r
K^{\sigma \sigma} _{[ r u_1 \cdots u_\ell]}
\Pi_- ^{(m+1 | \ell+1)} (s;t_*, r, u_*) \quad
\eea
We use the notations $t_*$  and $u_*$  as abbreviations for $t_1\cdots t_m$,
and $u_1 \cdots u_\ell$ respectively.

\sm

We note that the contributions from $m=-1$ correspond to the contributions
from the blocks $\Pi _\pm ^{(\ell+1)} (s;u_1, \cdots, u_\ell) =
\Pi _\pm ^{(0|\ell+1)} (s;u_1, \cdots, u_\ell)$, in which no variables $t_i$
occur, and in which we set $r=s$.

\subsection{Monodromy  of $\cD [\delta]$}

All blocks have trivial monodromy under $A$-cycles.
Using the monodromy operator around $B$-cycles, defined earlier by
$\Delta _K ^{(s)} f(s) \equiv (f(s+B_K) - f(s) )/(2 \pi i)$,
the transformation property of the internal loop momentum under this
operation, $\Delta _K ^{(s)} p_I ^\mu = -i k_s ^\mu \delta _{IK}$, and the
monodromy properties of the elementary functions $\l_I$, $\hat \o _{I0}$,
$\FF$, $\ff$, and $Q_B$, we readily compute the variations under $B_K$ monodromy of
the differential blocks,
\bea
\Delta _K ^{(s)} \cD [\delta ]   = \ep ^\mu _s \theta _s \p_s \cD_K ^{(s)} [\delta]
\eea
where the variation $\cD_K ^{(s)} [\delta]$ is given by,
\bea
\cD_K ^{(s)} [\delta]
& = &
-   \sum _ {r\not= s} Z_0 ^{rs}  \Big ( X^\mu _r  \lambda _K(r) -
p^\mu _I K^\mu _r K^\nu _r  \hat \omega _{I0}(r) \hat \omega _{K0}(r) \Big )
\\ &&
+   \sum _{s \not= [rt]}  Z_0 ^{rst}
K^\mu _r K^\nu _t  \Big ( p^\nu _I  \hat \o_{I0}(r)  \hat \omega _{K0} (t) +
\hat \o_{K0}(r) i k^\nu _r \FF (t,r) \Big )
\no \\ &&
-  \sum _{s \not= [rt]} Z_0 ^{rst}  K^\mu _r
 K^\nu _r X^\nu _t \hat \o_{K0}(r)  \FF (r,t)
\no \\ &&
 -  \sum _{s \not= [rtu]}  Z_0 ^{rstu}  K^\mu _r
 K^\nu _t  \hat \o_{K0}(t)  \FF (r,u)
\no \\ &&
+ \pi  \sum _{s \not = [rt]} Z_0 ^{rst}  K^\mu _r
 K^\nu _t  k_s ^\nu \hat \o_{K0}(r)  \hat \o_{K0}(t)
\eea
The key result is that $\Delta _K ^{(s)} \cD [\delta ]$ turns out to be
a holomorphic exact differential block. This result, together with the
associated monodromy transformation laws of the $\cZ$ blocks will be studied
in detail in the next paper.

\subsection{Non-triviality of the differential blocks}

Under changes of gauge slice, the blocks $\cF [\delta]$ change by
the addition of an exact differential, as was already shown in \cite{V},
Thus, the question naturally arises as to whether it is possible to choose
a gauge in which the sum of all the differential blocks $\cD [\delta]$  vanishes.
We shall argue here that there exists no such gauge choice.

\sm

A natural candidate for such a choice might be the {\sl split gauge},
introduced in \cite{IV} as a calculational tool to evaluate the superstring
chiral measure in terms of $\vartheta$-functions. Split gauge is defined
by taking the worldsheet gravitino $\chi $ to be supported at two points
$q_1$ and $q_2$, such that $\chi$ is parametrized by the two odd moduli
$\zeta ^1, \zeta ^2$ in the following manner,
\bea
\chi (z) = \zeta ^1 \delta (z,q_1) + \zeta ^2 \delta (z,q_2)
\eea
and the points $q_1$ and $q_2$ are related to one another by a spin structure
$\delta$-dependent equation,
\bea
S_\delta (q_1,q_2)=0
\eea
In split gauge, the super period matrix coincides with the bosonic
period matrix $\hat \Omega _{IJ} = \Omega _{IJ}$, and we may choose
$\hat \mu=0$. In split gauge, one also has
\bea
\l _I (s) = \ff (r,s) =0
\eea
As a result, the blocks $\Pi _I ^{(1)} (s) = \lambda _I (s)$ and $\Pi ^{(2)} (r,s)=
\ff (r,s)$ vanish identically, so that $\cS^{(1)} _s = \cS^{(2)} _{rs}=0$.
This cancellation is only a property of the low order at which we consider
the question, and higher blocks will not vanish. For example, in split
gauge, we have the following simplified expression for the block
$\Pi ^{(3)} (r;s,t) = - \FF (t,r)F(t,s)$, but this quantity is non-vanishing.

\sm

More generally, the blocks $\Pi ^{(n+2)}$ can never vanish identically as
soon as $n >0$. This may be seen directly by considering its double
$(0,1)$ derivatives at both end points, already mentioned in the introduction,
\bea
\p_{\bar r} \p_{\bar s} \Pi ^{(n+2)} (r;t_1, \cdots , t_n;s)
= {1 \over 4} \chi (r) S^{n+1} (r,t_1, \cdots, t_n, s)  \chi (s)
\eea
From inspection of the expression on the right hand side, it is clear
that for $n>0$, it cannot vanish for all $t_1, \cdots , t_n$, and thus the
block $\Pi ^{(n+2)}$ itself can also not vanish.

\subsection{Recombination of holomorphic $\p_s$ differentials and $\cZ$}

We illustrate here, for the simplest blocks treated in section 9.2, how
the exact holomorphic $\p_s$ differentials that arise at intermediate
stages from the bosonic Wick contractions by Lemma 1, recombine with
similar terms in $\cZ$. We follow closely the notations of section 9.2,
and refer to the various cases with the numerals used there. In
$(1)$ and $ (2)$, no exact holomorphic $\p_s$ differentials arise for these blocks.
In $(3)$, combining $\cD_{2bp}$ with $\cD_{2c}$ in (\ref{D2}), produces the
exact holomorphic differential,
\bea
\label{D2holo}
\sum _{[rs]} \p_s \bigg ( Z_0 ^{rs} p^\nu _I \ep ^\mu _s \theta _s K^\nu _r K^\mu_r
\lambda _I (r) G(r,s) \bigg )
\eea
This contribution compensates an identical and opposite term arising from the
contractions of certain $\cZ$-blocks. To see how this works, we use Lemma 1 to
carry out partial bosonic Wick contractions on $\cZ_{1a}$, as follows,
\bea
\label{Z2}
\cZ_{1a} & = &
\sum _r Z_0 ^r p^\mu_I
\bigg (i k^\mu _r + K^\nu _r K^\mu _r p_J \o_J(r) \bigg ) \lambda _I (r)
+ \sum _{[rs]} Z_0 ^{rs} p^\mu _I K^\mu_r K^\nu _r X^\nu _s G(r,s) \lambda _I(r)
\no \\ &&
+ \sum _{[rs]} \p_s \bigg ( Z_0 ^{rs} p^\mu _I \ep ^\nu _s \theta _s K^\nu_r K^\mu _r G(r,s) \lambda _I(r) \bigg )
\eea
The total $\p_s$ derivative terms in (\ref{D2}) and (\ref{Z2}) precisely
cancel one another in view of the anti-symmetry of $K_r^\nu K_r ^\mu$ under
interchange in $\mu$ and $\nu$. Using also the definition of the block
$\Pi ^{(1)} _{IJ}(r)$, we may assemble these simplest blocks as follows,
\bea
\cZ_{1a} + \cZ_{1b} + \cD_{2b}^p + \cD_{2c} & = &
\cH^{(1)} _{pp} + d_r \cS^{(2)} _r
+ \sum _r Z_0 ^r p^\mu_I  i k^\mu _r  \lambda _I (r) + \cZ_{1a} ^{(2)}
\no \\
\cZ_{1a} ^{(2)} & = & \sum _{[rs]} Z_0 ^{rs} p^\mu _I K^\mu_r K^\nu _r X^\nu _s
G(r,s) \lambda _I(r)
\no \\
\cH^{(1)}_{pp} & \equiv & - \sum _r Z_0 ^r \half p^\mu_I  K^\mu _r K^\nu_r p^\nu _J
\Pi ^{(1)} _{IJ} (r)
\eea
The $p$-parts in $X^\nu _s$ of $\cZ_{1a}^{(2)}$ and $\cZ_{2d}$ combine with the
the 2-point fermionic contraction of $\cZ_{2a}$ and the double $p$-part of
$\cZ_{2b}$ to produce the holomorphic block
\bea
\cH^{(2)}_{pp} \equiv - \sum _{[rs]} Z_0 ^{rs}
\half p_I ^\mu  K^\mu _r K^\rho _r K^\rho _s K_s ^\nu p^\nu _J \Pi ^{(2)} _{IJ}(r,s)
\eea
From the above pattern, we may deduce all the linear chain blocks
which are bilinear in the internal loop momenta.
We shall prove these formulas systematically in the next paper.
\bea
\label{Hpp}
\cH^{(n)}_{pp} \equiv - \sum _{[t_1 \cdots t_n]} Z_0 ^{t_1\cdots t_n}
\half p_I ^\mu  K^\mu _{t_1}  K^{\mu_1} _{t_1}  \cdots K^{\mu_{n-1}} _{t_n}
K_{t_n} ^\nu p^\nu _J \Pi ^{(n)} _{IJ}(t_1; t_2, \cdots, t_{n-1};t_n )
\eea
The appearance of the blocks $\Pi_{IJ} ^{(n+2)}$ in the $\cZ$ blocks
confirms the general structure announced in the Introduction, and will
be studied in detail in the next paper.

\bigskip \bigskip

\noindent
{\large \bf Acknowledgements}

\medskip

This project was initiated in 2001.
We have benefited from conversations with many colleagues, including
Costas Bachas, Zvi Bern, Sam Grushevsky, Michael Gutperle, Shamit Kachru,
Per Kraus, Igor Krichever, Boris Pioline, John Schwarz, Eva Silverstein, Tomasz Taylor,
Pierre Vanhove, Edward Witten, and Chuan-Ji Zhu.

\sm

We have also benefited from the hospitality of
many institutions where part of this work was discussed and or carried out:
E.D. is happy to acknowledge the kind hospitality of
the Aspen Center for Physics,
the Columbia University Mathematics Department,
the Ecole Normale Sup\'erieure (Paris),
the Institut Henri Poincar\'e (Paris),
the Institute for Pure and Applied Mathematics (UCLA),
and the Kavli Institute for Theoretical Physics (UCSB).
D.H.P. would like to thank the Centro di Ricerca Matematica Ennio De Georgi in Pisa,
the Mathematical Sciences Research Institute in Berkeley,
the American Institute of Mathematics in Palo Alto,
the Centre International de Recherches Mathematiques in Luminy,
the International Center for Mathematical Sciences in Hsin Chu, Taiwan,
and the Johns Hopkins University
for their kind hospitality.

\end{document}